\documentclass[fleqn,usenatbib]{mnras}
\usepackage[T1]{fontenc}
\usepackage{ae,aecompl}
\usepackage{multirow}
\usepackage{graphicx}	
\usepackage{amsmath}	
\usepackage{amssymb}	

\newcommand{\eps}{$\epsilon$ Lupi}

\usepackage{booktabs,caption}
\usepackage[flushleft]{threeparttable}
\usepackage{newtxtext,newtxmath}
\usepackage{color}
\usepackage{xcolor}
\usepackage{soul}

\definecolor{vero}{rgb}{0.82, 0.25, 0.95}

\title[Variable X-ray from $\epsilon$ Lupi]{Discovery of extraordinary X-ray emission from magnetospheric interaction in the unique binary stellar system $\epsilon$ Lupi}

\author[B. Das et al.]{
B. Das,$^{1}$\thanks{E-mail: barnali@udel.edu}
V. Petit,$^{1}$
Y. Naz\'e,$^2$\thanks{F.R.S.-FNRS Senior Research Associate}
M. F. Corcoran,$^{3,4}$
D. H. Cohen,$^{5}$
A. Biswas,$^{6,7}$
\newauthor
P. Chandra,$^{8}$ 
A. David-Uraz,$^{9,3}$ 
M. A. Leutenegger,$^{10}$ 
C. Neiner,$^{11}$ 
H. Pablo,$^{12}$
\newauthor
E. Paunzen,$^{13}$ 
M. E. Shultz,$^{1}$
A. ud-Doula$^{14}$ 
and G. A. Wade$^{7}$ 
\\
$^{1}$Department of Physics and Astronomy, Bartol Research Institute, University of Delaware, 217 Sharp Lab, Newark, DE 19716, USA\\
$^{2}$Groupe d'Astrophysique des Hautes Energies, STAR, Universit\'e de Li\`ege, Quartier Agora (B5c, Institut d'Astrophysique et de G\'eophysique), \\
All\'ee du 6 Ao\^ut 19c, B-4000 Sart Tilman, Li\`ege, Belgium\\
$^{3}$Center for Research and Exploration in Space Science and Technology, and X-ray Astrophysics Laboratory, NASA/GSFC, Greenbelt, MD 20771, USA\\
$^{4}$Institute for Astrophysics and Computational Sciences, The Catholic University of America, 620 Michigan Avenue, N.E. Washington, DC 20064, USA\\
$^{5}$Swarthmore College, Department of Physics and Astronomy, Swarthmore, PA 19081, USA\\
$^{6}$Department of Physics, Engineering Physics \& Astronomy, Queen’s University, Kingston, Ontario K7L 3N6, Canada\\
$^{7}$Department of Physics and Space Science, Royal Military College of Canada, PO Box 17000, Station Forces, Kingston, ON, K7K 7B4\\
$^{8}$National Radio Astronomy Observatory, 520 Edgemont Road, Charlottesville VA 22903, USA\\
$^{9}$Department of Physics and Astronomy, Howard University, Washington, DC 20059, USA\\
$^{10}$X-ray Astrophysics Laboratory, NASA/GSFC, Greenbelt, MD 20771, USA\\
$^{11}$LESIA, Paris Observatory, PSL University, CNRS, Sorbonne University, Université Paris Cit\'e, 5 place Jules Janssen, 92195 Meudon, France\\
$^{12}$American Association of Variable Star Observers, 185 Alewife Brook Pkwy, Cambridge, MA 02138, USA\\
$^{13}$Department of Theoretical Physics and Astrophysics, Masaryk University, Kotl\'a\v{r}sk\'a 2, 611\,37 Brno, Czechia\\
$^{14}$Penn State Scranton, 120 Ridge View Drive, Dunmore, PA 18512, USA\\
}

\date{Accepted XXX. Received YYY; in original form ZZZ}

\pubyear{2022}

\begin{document}
\label{firstpage}
\pagerange{\pageref{firstpage}--\pageref{lastpage}}
\maketitle

\begin{abstract}
We report detailed X-ray observations of the unique binary system \eps, the only known short-period binary consisting of two magnetic early-type stars. The components have comparably strong, but anti-aligned magnetic fields. The orbital and magnetic properties of the system imply that the magnetospheres overlap at all orbital phases, suggesting the possibility of variable inter-star magnetospheric interaction due to the non-negligible eccentricity of the orbit. To investigate this effect, we observed the X-ray emission from \eps~both near and away from periastron passage, using the Neutron Star Interior Composition Explorer mission (NICER) X-ray Telescope. We find that the system produces excess X-ray emission at the periastron phase, suggesting the presence of variable inter-star magnetospheric interaction. 
We also discover that the enhancement at periastron is confined to a very narrow orbital phase range ($\approx 5\%$ of the orbital period), but the X-ray properties close to periastron phase are similar to those observed away from periastron.
From these observations, we infer that the underlying cause is magnetic reconnection heating the stellar wind plasma, rather than shocks produced by wind-wind collision. Finally, by comparing the behavior of \eps~with that observed for cooler magnetic binary systems, we propose that elevated X-ray flux at periastron phase is likely a general characteristic of interacting magnetospheres irrespective of the spectral types of the constituent stars. 
\end{abstract}




\section{Introduction}\label{sec:intro}
Roughly 10\% of early-type stars harbour large-scale, highly stable, kG strength surface magnetic fields \citep{grunhut2017,sikora2019}.  
The consequences of such magnetic fields on hot stars have been studied extensively. It is now well-established that magnetic OBA stars are surrounded by enormous, co-rotating magnetospheres (resulting from stellar wind-magnetic field interaction) that often extend up to several tens of stellar radii \citep[e.g.][]{shultz2019c}. The magnetospheres lead to different kinds of phenomena, such as the generation of variable $\mathrm{H\alpha}$, X-ray and non-thermal radio emission \citep[e.g.][]{drake1987,trigilio2000,gagne2005,oksala2012,petit2013,ud-doula2014,naze2014,shultz2020,owocki2020, leto2021,shultz2022,owocki2022}, and also have profound impacts on stellar evolution due to magnetospheric braking and wind confinement \citep[e.g.][]{petit2017,keszthelyi2019,keszthelyi2020,keszthelyi2021,keszthelyi2022}. 
However, not much is known about the effect(s) of binarity on hot magnetic stars, even though binarity is known to be an important ingredient of stellar evolution \citep{sana2012}. 
\citet{naze2017} explored the role of binary interaction on the generation of magnetic fields in massive stars, but did not find any observational evidence in support of that idea. 
\citet{vidal2018} adopted a theoretical approach to examine the role of tidal interaction in binaries in magnetic field generation, and inferred that tidally generated dynamos can lead to weak (up to several Gauss) surface magnetic fields.
In 2019, \citeauthor{schneider2019} showed by performing magnetohydrodynamic (MHD) simulations that mergers of two massive stars can give rise to a single magnetic massive star, and speculated that the observed small fraction of magnetic massive star binaries \citep[the `Binarity and Magnetic Interactions in various classes of stars, or, BinaMIcS project;][]{alecian2015} is a consequence of the fact that the magnetic massive stars are merger products of binary systems.
The opposite possibility, i.e. tidal interactions leading to the \textit{destruction} of magnetic fields, has been proposed by \citet{vidal2019},
who suggested that tidal interactions in binary systems with non-circular orbits would erase fossil magnetic fields over timescales of a few million years. 

As described above, so far the effect of binarity on magnetic massive stars has primarily been investigated in the context of understanding the magnetic incidence fraction among massive star binaries.
However, there are few observational constraints on the effect of binary interactions on the characteristics of massive star magnetospheres and the associated emission \citep[e.g.][]{shultz2018b}.
Existing studies have been mostly limited to pre-main sequence late-type stars \citep[e.g.][]{massi2006,salter2008,salter2010,adams2011,getman2011,getman2016}. In such systems, it has been proposed that colliding magnetospheres lead to inter-star magnetic reconnection that manifests as enhancement of radio and/or X-ray flux. While this idea was proposed for individual cases, the only attempt towards investigating the validity of this proposition for the general population was performed by \citet{getman2016} by considering a sample of four binary systems consisting of pre-main sequence stars of spectral types F and later. 
They found the X-ray flux at periastron to be on average higher than that at a phase away from the periastron, with a significance of ~2.5$\sigma$.
The low statistical significance of their result, however, hindered them from drawing a firm conclusion.

Compared to the late-type pre-main sequence stars, the magnetic early-type stars are a more attractive test-bed to understand the interplay between binarity and magnetism, since these stars have extremely stable and relatively simple (usually dipolar) magnetic fields. 
These fields are fundamentally different from the convective-dynamo generated magnetic fields seen in cool stars \citep[e.g.][]{donati2009} in the sense that the former is either of fossil origin \citep[e.g.][]{braithwaite2004}, or a product of mergers \citep{schneider2019}, and as mentioned already, stable in time unlike the case for dynamo-generated fields.
This implies that for a binary system consisting of two magnetic early-type stars, the inter-magnetospheric interactions will be free from time-variability induced by the change in the magnetic fields themselves, and probably will vary only at the orbital and/or rotational timescales of the constituent stars. However, early-type magnetic binaries are scarce. 
Only around 2\% of all massive stars in close binaries (orbital period smaller than 20 days, where mutual interactions are expected) have been found to host a detectable magnetic field \citep{alecian2015}. Among them,
there are only three known doubly magnetic massive star systems \citep{elkin1999,semenko2011,shultz2015b,shultz2021}. 
Out of these three, $\epsilon$ Lupi stands out as the only short-period binary system, the other two being wide binaries (hence non-interacting). In $\epsilon$ Lupi, the constituent stars are nearly identical (B2/B3), but have anti-aligned magnetic fields \citep{shultz2015b,pablo2019}. 

In this paper, we report X-ray observations of the unique magnetic binary system $\epsilon$ Lupi acquired with the Neutron Star Interior Composition Explorer (NICER) mission X-ray Telescope. Our data provide evidence of inter-star magnetospheric interaction, and also shed light on the nature of the interaction.


This paper is structured as follows: in the next section (\S\ref{sec:X-ray}), we provide a brief description regarding different channels of X-ray production from magnetic massive stars, followed by an introduction to our target of interest \eps, and a summary of the results from past X-ray observation of the system (\S\ref{sec:target}). We then describe our observations (\S\ref{sec:obs}) and data analysis (\S\ref{sec:data_analysis}), followed by the results (\S\ref{sec:results}). We discuss our key findings in \S\ref{sec:discussion}, and then present our conclusions in \S\ref{sec:conclusion}.

\section{X-ray emission from magnetic massive stars}\label{sec:X-ray}
X-ray emission from solitary magnetic massive stars is explained in the framework of the `Magnetically Confined Wind Shock' model \citep[MCWS,][]{babel1997}. It was further refined thanks to MHD simulations by \citet{ud-doula2014}. 
In these stars, the radiatively driven stellar wind materials are channeled by the predominantly dipolar magnetic field from the magnetic poles towards the magnetic equatorial regions. The collision between the wind flows from the two magnetic hemispheres gives rise to shocks that can heat the gas up to $10^7-10^8\,\mathrm{K}$ \citep[e.g.][]{ud-doula2022} leading to X-ray emission. If the star's rotation axis is not aligned with the line of sight, and the magnetic dipole axis is also misaligned with respect to the rotation axis, variability can occur. In the X-ray domain, the emission can exhibit modulation with rotational phase since the visibility of the sites of emission, as well as the amount of magnetospheric absorption (for sufficiently dense magnetospheres) vary as the star rotates. The extent of the modulation is determined by several parameters such as the stellar geometry (angles made by line of sight and magnetic dipole axis with the stellar rotation axis), magnetic confinement, density and size of the magnetosphere \citep[e.g.][]{naze2014}. For O-stars that have high mass-loss rates, another channel for X-ray production is embedded wind shocks that produce relatively soft X-rays \citep[e.g.][]{lucy1982,feldmeier1997,berghoefer1997}. Finally, \citet{leto2017,leto2020} proposed that X-ray emission from magnetic massive stars can also have a non-thermal component, produced at the surface magnetic polar caps when energetic electrons bombard the stellar surface (`Auroral X-ray Emission' or AXE) although an unambiguous signature of non-thermal X-ray emission has not yet been reported from hot magnetic stars
\citep[the only unambiguous case is the non-magnetic colliding wind binary system $\eta$ Car, ][]{leyder2008}.

\citet{naze2014} performed a population study of magnetic massive stars regarding their X-ray properties and discovered that a few of these stars are overluminous with respect to the X-ray luminosity expected from the MCWS scenario.
Most recently, \citet{shultz2020} proposed that this excess luminosity might originate from magnetic reconnection triggered by continuous ejections of magnetically confined plasma from the stellar magnetosphere \citep[centrifugal breakout or CBO,][]{townsend2005,shultz2020,owocki2020}. 
This possibility was discussed in a greater detail by \citet{owocki2022}, who showed (theoretically) that indeed CBO events can lead to significant enhancement in X-ray production.

For massive stars in binary systems, X-ray emission can sometimes also be produced due to shocks resulting from collision between the wind from the two stars \citep[e.g.][]{corcoran2007,gosset2016, naze2017b,naze2018, rauw2022}. 
For a system like \eps, where both components are magnetic, and mass-loss rates are not very high, a more relevant scenario is heating resulting from magnetic reconnection. We will discuss this scenario in greater detail in subsequent sections.

\section{$\epsilon$ Lupi: The system and results from past X-ray observations}\label{sec:target}
The system $\epsilon$ Lupi consists of three stars, with the inner pair forming a double-lined spectroscopic binary consisting of two early-type stars with $T_\mathrm{eff}\approx 20.5$ kK and $18$ kK \citep{pablo2019}, referred to as $\epsilon$ Lupi A; and another relatively distant component known as $\epsilon$ Lupi B, also an early-type star \citep[$T_\mathrm{eff}\approx 18$ kK,][]{pablo2019}. The binary system $\epsilon$ Lupi A has an orbital period of $\approx 4.6$ days \citep{thackeray1970,uytterhoeven2005,pablo2019}; on the other hand, the orbital period of the binary system formed of $\epsilon$ Lupi A and $\epsilon$ Lupi B is estimated to be 740 years \citep{zirm2007}. Hence the interactions between A and B components are of no importance in the context of the work presented here. Henceforth, we will refer to the $\epsilon$ Lupi A system as simply $\epsilon$ Lupi.

The magnetic field in that system was first reported by \citet{hubrig2009} with low-resolution spectropolarimetric data, and confirmed by \citet{shultz2012} using high-resolution spectropolarimetric data. The fact that both components of \eps~are magnetic was discovered by \citet{shultz2015b}.
The system thus appears to be the first (and only) known close binary consisting of two magnetic massive stars. Due to their proximity, the two stars' magnetospheres overlap at all orbital phases \citep{shultz2015b}. Based on the observed longitudinal magnetic fields, which exhibit little variation with rotational phase, and the known projected rotational velocities, \citet{shultz2015b} estimated the two stars to have anti-aligned magnetic fields with polar strengths of 900 G (primary) and 600 G (secondary). The magnetic axes are assumed to be aligned with the rotation axes (consistent with the observed Stokes V profiles), and the rotation axes are assumed to have the same inclination angle of $21^\circ$ \citep[updated to $18.8^\circ$ by][]{pablo2019} with the line-of-sight as that of the orbital plane.
\citet{pablo2019} performed a detailed study of the system using radial velocity measurements acquired over decades, and also modeling the photometric heartbeat variation detected in data from the BRIght Target Explorer (BRITE) Constellation \citep{weiss2014},
which enabled them to make precise measurements of the orbital parameters, along with direct (but not very precise) measurements of stellar masses and radii. They also refined the eccentricity ($e$) of the system ($e=0.2806^{+0.0059}_{-0.0047}$), and the apsidal motion ($1.1\pm 0.1^\circ$/yr) from their previously reported values by \citet{uytterhoeven2005} and \citet{thackeray1970}.


X-ray emission from \eps~was reported by \citet{naze2014}. These data were acquired with the XMM-Newton telescope for an exposure time of 5 ks (0.01 orbital phase ranges) as part of a survey of magnetic massive stars. \citet{naze2014} fitted the spectrum with an absorbed optically thin thermal plasma model assuming solar abundances. They considered two sources of X-ray absorption: the absorption in the interstellar medium (ISM), and the absorption in the stellar magnetosphere. The ISM absorption is determined by the column density of hydrogen along the line of sight, and this was fixed at $0.03\times 10^{22}\,\mathrm{cm^{-2}}$ \citep[based on excess color,][]{petit2013,naze2014}. The emission was modelled using two strategies. In the first, the hot plasma is assumed to be made of two thermal components, each modeled using the Astrophysical Plasma Emission Code \citep[`\textit{apec}',][]{smith2001}. The individual plasma temperatures and the corresponding emission measures (EMs), equivalent to the normalization factors or `norms'\footnote{The relation between norm and EM is the following:\\
$\mathrm{EM} = 4\pi d^2\times 10^{14}\times\mathrm{norm}$, where $d$ is the distance to the star; all quantities are in CGS units.
} of the \textit{apec} model, were kept as free parameters. 
This method yielded the two temperatures as $kT=0.3$ and $3.0$ keV ($T=3.5$ MK and 34.8 MK respectively), with the latter having a higher emission measure. In their second strategy, they assumed four thermal components, the temperatures of which were fixed at 0.2, 0.6, 1.0 and 4.0 keV. The corresponding norms were kept as free parameters. This strategy also showed that the system has its highest emission measure for the hottest plasma component (4 keV or 46.4 MK). Within the sample studied by \citet{naze2014},
which consisted of 40 stars with 28 of them well-detected in X-rays,
only three stars (including \eps) exhibit this property for both strategies. The other two stars are HD\,57682 and HD\,182180 \citep{naze2014}. For most of the magnetic B stars, including \eps, \citet{naze2014} found the absorption by the stellar magnetospheres to be negligible, whatever the adopted spectral fitting strategy (but to the limits of the usual temperature/absorption trade-off). The highest absorption was observed for NGC 1624-2, a magnetic O star with extreme conditions (a very strong magnetic field and mass-loss rate, making it the O-star with the largest magnetosphere), where 70--95\% of the X-ray emission gets absorbed \citep{petit2015}.  \eps~is neither known nor expected to have such extreme magnetospheric conditions. 

The snapshot observation of \eps~reported by \citet{naze2014} thus showed the system to have a harder X-ray spectrum than that of typical magnetic massive stars. 
The system's X-ray luminosity, compared to its bolometric luminosity, was, however, found to be consistent with that of other stars in their sample \citep[$\log(L_\mathrm{X}/L_\mathrm{BOL})\approx -7.2$,][]{naze2014}.

It is worth mentioning that the orbital phase corresponding to these observations happened to lie close to the periastron phase (Figure \ref{fig:obs_strategy}).

\section{Observations}\label{sec:obs}
\begin{table*}
\begin{threeparttable}
{
\caption{Log for our observations of \eps~with NICER. The first column lists the observation IDs, assigned by the observatory; the second column lists the IDs that \textit{we} assign to the different observations for convenience; the third, fourth and fifth column show the exposure times for the unfiltered, filtered with \texttt{nicerl2} with default inputs, and filtered following manual filtering of events (see \S\ref{subsec:data_filter}) in addition to that done by \texttt{nicerl2} respectively; the sixth column shows the range of Heliocentric Julian Days (HJDs) spanned by each observation; and the seventh column shows the corresponding range of orbital phases (apsidal motion corrected).  \label{tab:obs}}
\begin{tabular}{lc|cccrc}
\hline
Obs ID & ID & \multicolumn{3}{c}{Exposure Time (ks)} & HJD & Orbital phase\\
 &  & Unfiltered & Default filtering & Default$+$Manual filtering & $-2459255$& \\
\hline\hline
3627010101 & 1 & 2.296 & 1.916 & 1.818 & 0.91605--0.99237 & 0.953--0.970\\
3627010201 & 2 & 2.710 & 1.856 & 1.856 & 3.25038--3.39004 & 0.465--0.496\\
3627010301 & 3 & 2.603 & 1.966  & 1.897 & 5.51470--5.59070 & 0.962--0.978\\
3627010401$^*$ & 4 & 1.142 & 1.034 & 0.932 & 7.45142--7.46347 & 0.386--0.389\\
3627010402$^*$ & 4 & 1.356 & 1.238 & 1.023 & 7.51489--7.52799& 0.400-0.403\\
3627010501 & 5 & 2.795 & 2.570 & 2.570 & 10.16385--10.30294 & 0.981--1.012\\ 
3627010601 & 6 & 2.300 & 0.714 & 0.571 & 12.22154--12.28901& 0.432--0.447\\
3627010701 & 7 & 2.200 & 1.582 & 0.972 & 14.79680--14.87256 & 0.997--1.014\\
3627010801 & 8 & 2.889 & 2.082 & 1.814 & 16.80092--17.06744& 0.437--0.495\\
3627010901 & 9  & 2.852 & 1.448 & 0.795 & 19.32231--19.45561 & 0.990--1.019\\
3627011001 & 10 & 4.088 & 2.138 & 1.819 & 20.93592--21.39237& 0.344--0.444\\
\hline
\end{tabular}
\begin{tablenotes}
       \item $^*$3627010401 and 3627010402 were later combined as they were separated by only 1.2 hours (see column 6), and hence were assigned the same ID (column 2). They were assigned different Obs Ids as the `NICER day' changed during the observation. The `NICER day' is defined with respect to UTC midnight.
\end{tablenotes}
}
\end{threeparttable}
\end{table*}

\begin{figure}
    \centering
    \includegraphics[trim={1cm 8cm 11cm 1cm},clip,width=0.48\textwidth]{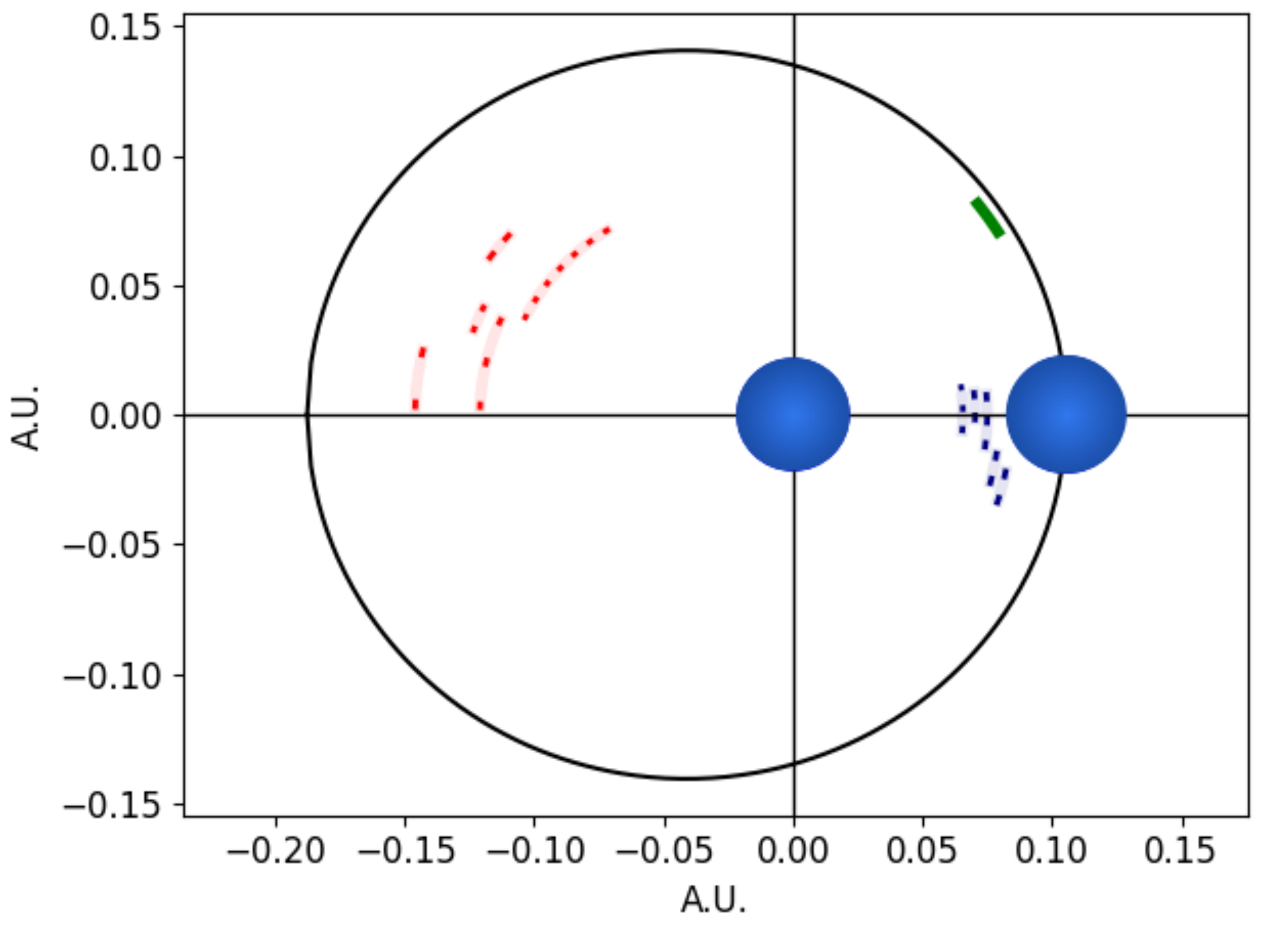}
    \includegraphics[trim={1cm 14.5cm 1.5cm 2cm},clip,width=0.48\textwidth]{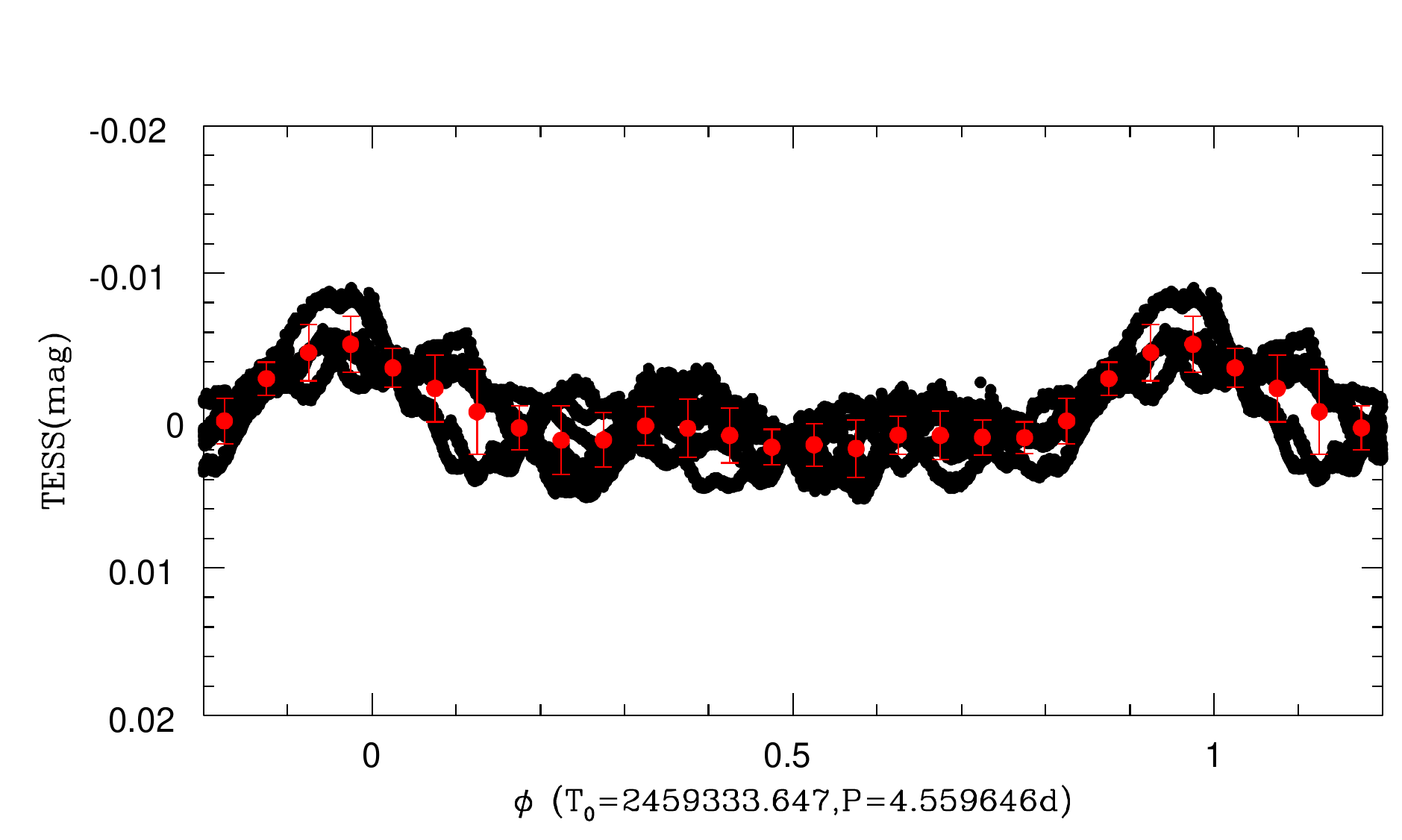}
    \caption{\textbf{Top:} The strategy adopted for our NICER observations of \eps. This cartoon diagram shows the two stars during the periastron phase in the stationary frame of reference of the primary star. The sizes of the stars, and the orbit are drawn in accordance with the values of stellar radii and orbital parameters reported by \citet{pablo2019}. The orbital phase increases from 0 (periastron) in the anti-clockwise direction. 
    We conducted five observations close to periastron (blue lines), and another five away from periastron (red lines) using the NICER telescope. As described in \S\ref{sec:obs}, each observation session consists of multiple snapshot observations (shown by darker shades). The distances of the arcs from the orbit (black solid line) used to mark the NICER observations (red and blue arcs) increase with increasing observation IDs (see Table \ref{tab:obs}).
    Also shown is the orbital phase range (green line) spanned by the XMM-Newton observation of \eps~reported by \citet{naze2014}.
    \textbf{Bottom:} The most recent TESS light curve of \eps~phased with the ephemeris derived from \citet{pablo2019}. It clearly shows the enhanced amplitude at periastron due to the ``heartbeat effect''.
    }
    \label{fig:obs_strategy}
\end{figure}

In order to study the effect of binarity on magnetospheric X-ray production, we observed \eps~using the NICER telescope on the International Space Station \citep[ISS,][]{gendreau2012,arzoumanian2014}. NICER is an X-ray timing instrument (not an imaging instrument) with a spectral band of 0.2-12.0 keV. The angular resolution is 6 arcmin (equivalently, the field of view is $\approx 30\,\mathrm{arcmin^2}$). \eps~is the brightest source in the field of view. The second brightest source is approximately three times fainter than \eps, located $\approx 5$ arcmin away, where the NICER response drops to 5\% of its maximum value at the pointing center\footnote{\url{https://heasarc.gsfc.nasa.gov/docs/nicer/data_analysis/workshops/NICER-CalStatus-Markwardt-2021.pdf}}.

We acquired observations of \eps~during five consecutive stellar orbits. At each cycle, one observation was taken when the system was at periastron, and another was taken when it was `out-of-periastron' (top panel of Figure \ref{fig:obs_strategy}). Thus, a total of ten observation sessions were conducted in this campaign. The exposure times for individual observations varied between 2.2 ks to 4.1 ks, with a total exposure time of 27.2 ks. The details of individual observations are given in Table \ref{tab:obs}.

Since NICER is attached to the ISS, its ability to observe a source of interest is governed by the ISS orbit (in addition to the instrument's own visibility constraints). As a result, the desired exposure time for the target is usually obtained via multiple `snapshot' observations of the target. If those `snapshots' are acquired on the same day (days are defined with respect to the UTC midnight), they are combined to form a single observation `segment', which is assigned a unique observation ID (column 1 of Table \ref{tab:obs}). In our case, we intended to obtain ten observation segments, five at near-periastron phases, and another five at out-of-periastron phases. However, one of the observation sessions was conducted around UTC midnight, and following the NICER convention, it resulted in two different segments with observation IDs 3627010401 and 3627010402. In the post-processing of the data, we combined these two segments (see \S\ref{sec:data_analysis}).

For the orbital parameters, \citet{pablo2019} found an orbital period $P_\mathrm{orb}=4.559646^{+5\times 10^{-6}}_{-8\times 10^{-6}}\,\mathrm{days}$, a reference Heliocentric Julian Day (corresponding to periastron) $\mathrm{HJD_0}=2439379.875^{+0.024}_{-0.019}$, and a rate of periastron advance $\dot \omega=1.1\pm 0.1\,\mathrm{^\circ/year}$. In view of the apsidal motion, and since the period provided in \citet{pablo2019} is not the anomalistic
one, the periastron reference time $\mathrm{HJD_0^{new}}$ at any other epoch should be calculated using the following equation:
\begin{align}
  \mathrm{HJD}_0^\mathrm{new} = \mathrm{HJD}_0+k\times P_\mathrm{orb}+ \frac{\dot \omega}{360^{\circ}}\left(\frac{\mathrm{HJD}-\mathrm{HJD}_0}{365.25}\right)\times P_\mathrm{orb}  \label{eq:hjd_ref}
\end{align}
where $k$ is the number of elapsed cycles since $\mathrm{HJD_0}$. For NICER data, $k$=4359 and $\mathrm{HJD}_0^\mathrm{new}$ is then 2459256.130 (2459333.647 for TESS data and 2456356.085 for XMM-Newton data). This ensures that $\phi=0$ always corresponds to periastron passage. 
Using the uncertainties in orbital period, reference HJD and the rate of periastron advance, we estimate (with the help of a Monte Carlo analysis) the uncertainty in the orbital phases to be $\approx 0.009$ cycles, which is significantly smaller than the duration of the variation discussed later.

\citet{pablo2019} used BRITE data acquired between 2014 March and 2015 August. The NICER data were however acquired in the year 2021. To check the validity of the ephemeris for the NICER data, we extracted the most recent photometric light curve acquired by the Transiting Exoplanet Survey Satellite \citep[TESS,][]{ricker2015}. These data were acquired in Sector 38, between April--May, 2021 while the NICER data were acquired in February 2021 (see Table \ref{tab:obs}). The bottom panel of Figure \ref{fig:obs_strategy} shows the TESS light curve phased with the above $\mathrm{HJD_0^{new}}$: it clearly shows the increased luminosity at periastron, confirming the heartbeat effect and thereby validating the ephemeris for the NICER data.


\section{Data analysis}\label{sec:data_analysis}
The default strategy to analyze NICER data is to use the pipeline \texttt{nicerl2}\footnote{\url{https://heasarc.gsfc.nasa.gov/lheasoft/ftools/headas/nicerl2.html}} included in the HEASoft package (version 6.29). This pipeline performs standard calibration, screening and filtering of events. The next step is to use the `cleaned' events to extract the light curves and spectra using `\textsc{xselect}'\footnote{\url{https://heasarc.gsfc.nasa.gov/ftools/xselect/}}. This is followed by generation of background spectra using either the 3C50 model \citep{remillard2022} or the space weather model\footnote{\url{https://heasarc.gsfc.nasa.gov/docs/nicer/data_analysis/workshops/environmental_bkg_model.pdf}} (since NICER is a non-imaging instrument), and the generation of the response matrices. The spectra for the events and the backgrounds, as well as the response matrices are then analyzed with `PyXspec', which is a python interface to the spectral fitting program `\textsc{xspec}'\footnote{\url{https://heasarc.gsfc.nasa.gov/xanadu/xspec/}} (version 12.12.0). Below we describe each step in detail.

\subsection{NICER data filtering, and extraction of light curves and spectra}\label{subsec:data_filter}

NICER has two main sources of background: a high energy particle background originating from cosmic rays as well as local energetic particles,
and optical loading. The latter usually affects only the low energy portion of the spectrum ($\lesssim 0.25$ keV). In our analysis, we do not use the spectrum below 0.3 keV, and hence the optical loading component is not relevant to us. Furthermore, the `underonly counts', which reflect the extent of contamination due to optical loading, is always well below the default NICER threshold for filtering.
The particle background, on the other hand, dominates the total background contribution at higher energies. For identifying time intervals 
that are highly contaminated by particle background, one may calculate the count rates over the 12--15 keV energy band, over which the effective area of NICER is basically zero.
Any high value of that rate then indicates background contamination in the corresponding time interval. 

\begin{figure*}
    \centering
    \textbf{Default filtering\hspace{6cm} Default$+$Manual filtering}\\
    \includegraphics[width=0.49\textwidth]{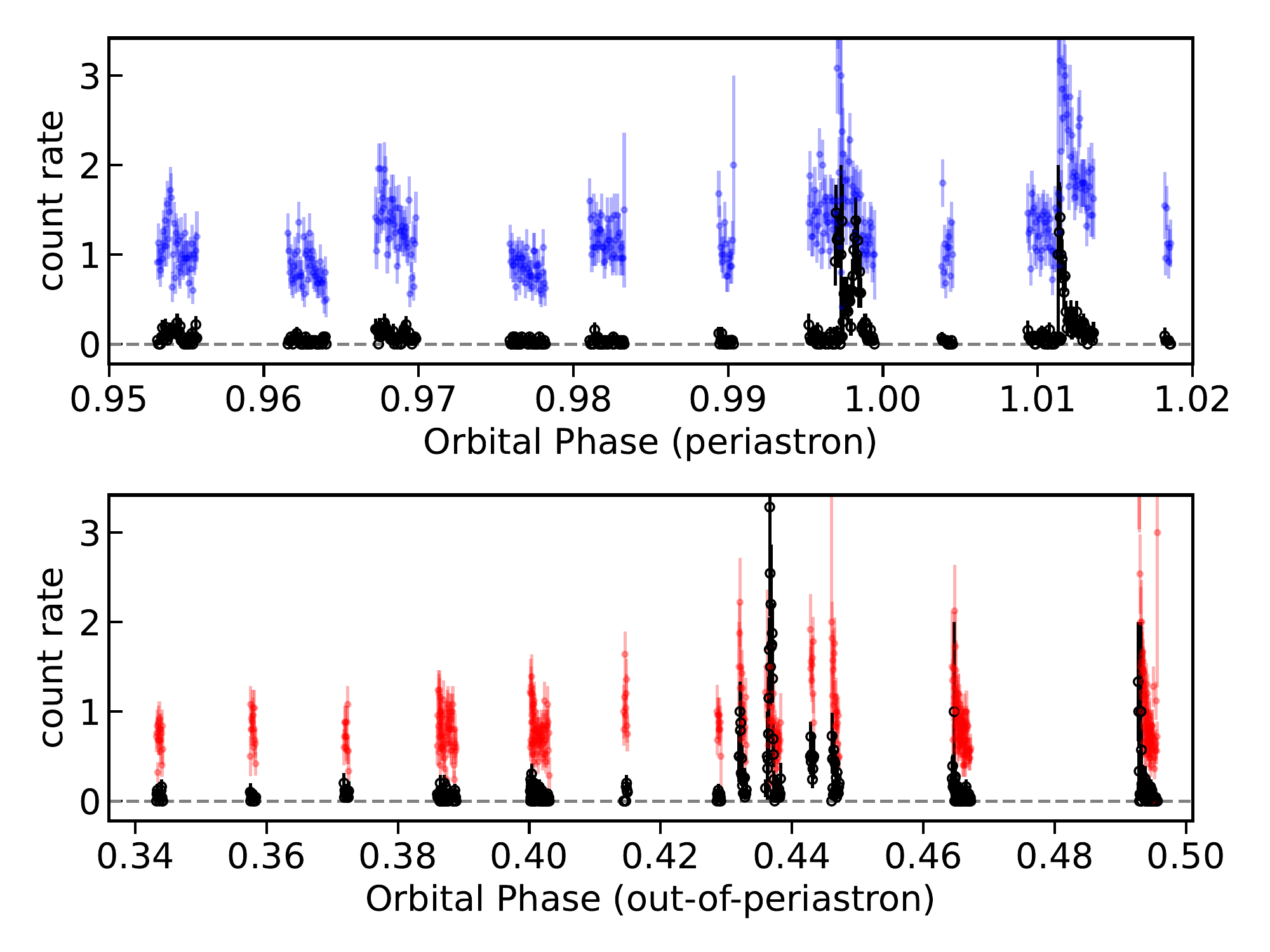}
    \includegraphics[width=0.49\textwidth]{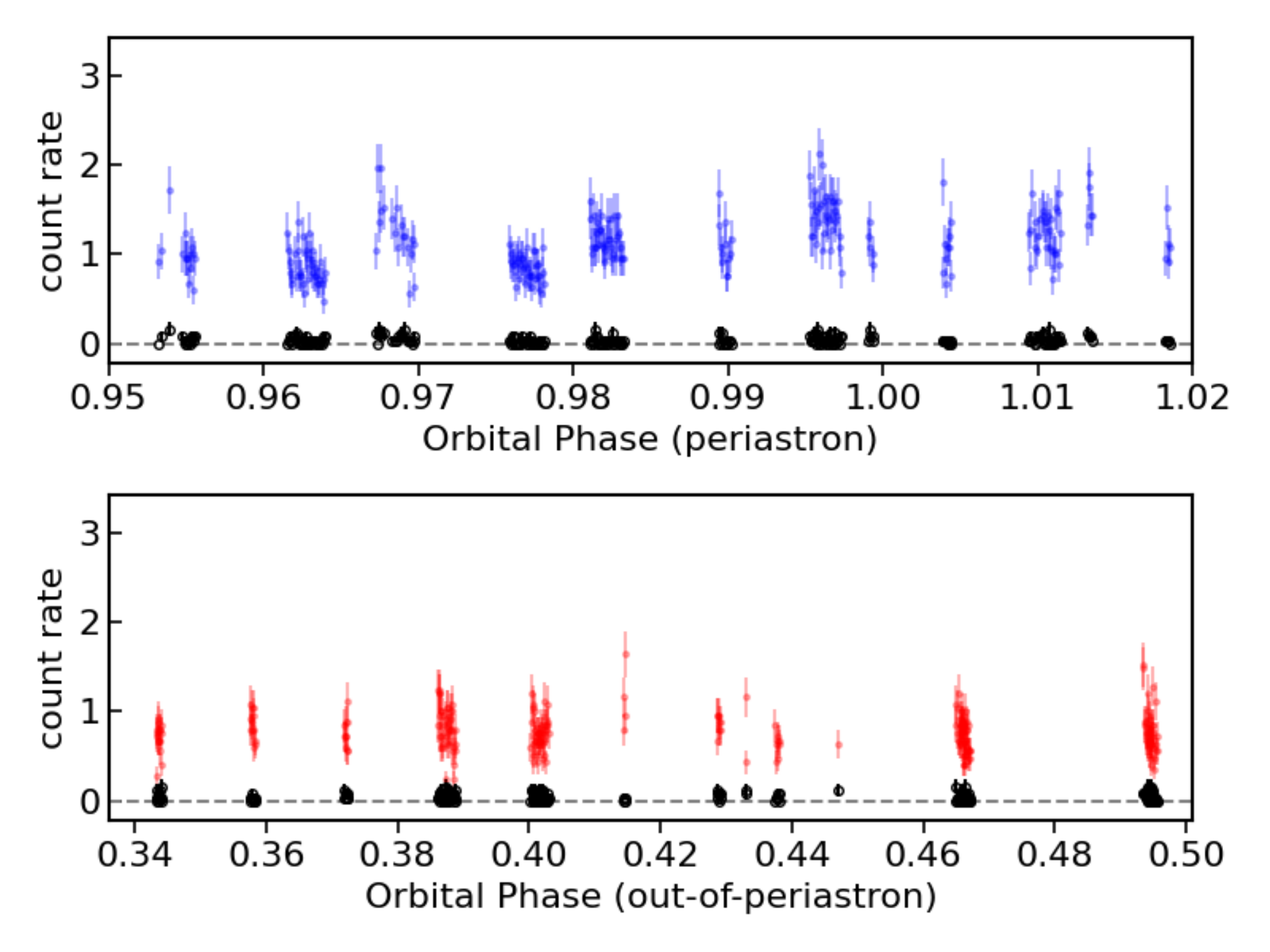}
    \caption{The observed variation of count rates for the different observations after using \texttt{nicerl2} with default input parameters ($\textbf{left}$), and after manually filtering GTIs with high count rates over 12--15 keV (\textbf{right}). The black markers represent data for the energy band 12--15 keV (attributed to background), and the blue (top panels) and red (bottom panels) markers respectively represent data for periastron and out-of-periastron observations respectively, over the energy range of 0.4--2.0 keV.}
    \label{fig:raw_lc}
\end{figure*}

After running \texttt{nicerl2} with default inputs (but with \texttt{nicersaafilt=NO saafilt=YES}, a more conservative approach to minimize the background contribution) \footnote{`saa' or `SAA' stands for `South Atlantic Anomaly', which is a geographic location where the Earth's inner Van Allen radiation belt is closest to the Earth's surface, resulting in a greater flux of high-energy particles.} 
for the ten observations, we extracted the light curves over 0.4--2.0 keV and 12.0--15.0 keV from the cleaned files using \textsc{xselect}. We discovered that despite the filtering imposed by \texttt{nicerl2}, there were still a few `Good Time Intervals' (GTIs) where the count rates over 12--15 keV were high (see left of Figure \ref{fig:raw_lc}).
We identified these GTIs by first using an absolute threshold of 0.3 $\mathrm{counts\,s^{-1}}$ for the rates in the 12--15 keV range. The absolute threshold was chosen based on visual inspection.
Next we calculate the median and the median absolute deviation (MAD) over the full time-series
of count rates at 12–15 keV spanning all ten observations. We then mark a time interval as contaminated if the corresponding rate at 12–15 keV is above the median$+2\times \mathrm{MAD}$ (after considering the error bars of the rates). Note that this strategy is devised
specifically for the data on \eps, and need not be applicable for other NICER data. Finally, we provided the list of contaminated GTIs to \texttt{nicerl2} to exclude them from further consideration, with the help of the \texttt{nimaketime}\footnote{\url{https://heasarc.gsfc.nasa.gov/lheasoft/ftools/headas/nimaketime.html}} tool. We also reduced the overonly\_range parameter (a large value of `overonly' indicates a higher amount of energetic background events) to 0--0.8 (the default is 0--1). This strategy successfully removed the GTIs with high count rates above 12 keV (see right hand panel of Figure \ref{fig:raw_lc}).

The final cleaned observations were then passed to \textsc{xselect} to extract light curves (with a time resolution of 25 s) and spectra.

\subsection{Merging observations}\label{subsec:merge_event}
Since the observations represented by Obs. IDs 3627010401 and 3627010402 were originally intended to be a single event list, we merged the two after producing the cleaned event lists.
This was done using the tools \texttt{nimpumerge}\footnote{\url{https://heasarc.gsfc.nasa.gov/lheasoft/ftools/headas/nimpumerge.html}} (that merges the event files) and \texttt{nimkfmerge}\footnote{\url{https://heasarc.gsfc.nasa.gov/lheasoft/ftools/headas/nimkfmerge.html}} (that merges the filter files). The resulting event list was assigned the ID `4' (Table \ref{tab:obs}).

\subsection{Background spectra generation}\label{subsec:bkg}
As mentioned previously, there are two ways to generate background models for NICER observations. The 3C50 model uses certain proxies for the background from the X-ray observations themselves to predict their background spectra (using a pre-built background database). The `space weather' model, as the name suggests, relies on the available information regarding space weather at the time of the observation, such as the planetary Kennziffer Index \citep[Kp index,][]{bartels1939}, and the magnetic cut-off rigidity, and is independent of the observed X-ray event list.

For our observations, we found that the space weather model predicts background rates that are sometimes higher than the total observed count rates (indicating negative count rates for the target, which is unphysical). The 3C50 model, however, predicts background rates that do not exceed the total count rates. Because of that, we prefer using the 3C50 model (tool \texttt{nibackgen3C50}\footnote{\url{https://heasarc.gsfc.nasa.gov/docs/nicer/tools/README_nibackgen3C50_v7b.txt}}) to generate the background spectra for our observations. 

\subsection{Spectral analysis}\label{subsec:spec_analysis}
We generated the response matrices using the \texttt{nicerarf}\footnote{\url{https://heasarc.gsfc.nasa.gov/lheasoft/ftools/headas/nicerarf.html}} and \texttt{nicerrmf}\footnote{\url{https://heasarc.gsfc.nasa.gov/lheasoft/ftools/headas/nicerrmf.html}} tools included in the HEASoft package. Before performing any fit to the spectra, we grouped each spectrum so that the minimum count per bin was 15. This step is needed as \textsc{xspec}, by default, uses Gaussian statistics for performing the fit. Additionally, this helps to avoid negative values of count rates following background subtraction.
The tool used for this purpose is \texttt{ftgrouppha}\footnote{\url{https://heasarc.gsfc.nasa.gov/lheasoft/help/ftgrouppha.html}}.

For spectral fitting, we used the energy range 0.3--10.0 keV.
Following \citet{naze2014}, we used absorbed optically thin thermal plasma models for the spectra. Our model ($tbabs\times \sum apec$) uses the `Tuebingen-Boulder ISM absorption' model (\textit{tbabs}\footnote{\url{https://heasarc.gsfc.nasa.gov/xanadu/xspec/manual/node268.html}}) for absorption. It also does not consider a second absorption component to account for absorption in the stellar magnetosphere.
This is motivated by the fact that \citet{naze2014} estimated the contribution of the stellar magnetosphere towards absorption to be negligible as compared to that of the ISM. Nevertheless, we attempted to fit the spectra by both fixing the neutral hydrogen column density \citep[to the value used by][]{naze2014}, and by keeping it as a free parameter. 

Similar to \citet{naze2014}, we considered two models for fitting our observations. The first one is given by $tbabs\times (apec+apec)$, where both temperatures (expressed in keV) and the norms for the $apec$ components are free parameters. We will refer to this model as the 2T model. The second model is expressed as $tbabs\times (apec+apec+apec+apec)$, where the four temperatures are kept fixed at 0.2, 0.6, 1.0 and 4.0 keV; and only the norms are fitted. 
This model will be referred as the 4T model. 
These four temperatures reasonably span the emissivities of the spectral lines within the energy range of observation. 
For both 2T and 4T models, we use solar abundances taken from \citet{anders1989}.
Note that \citet{cohen2021} provided a set of six fixed temperatures between 0.11 and 1.56 keV to approximate the X-ray spectra from continuous temperature distributions in the magnetospheres of O stars.
In the case of \eps, however, we find that a hotter plasma component is necessary to reproduce the observed X-ray spectra. We therefore prefer the 4T model of \citet{naze2014}, which has a smaller number of free parameters, and contains a high temperature component, for modelling the X-ray spectra of \eps.

In order to obtain a better understanding regarding the uncertainty associated with the fitted parameters, we performed a Markov Chain Monte Carlo (MCMC) analysis using the \texttt{chain}\footnote{\url{https://heasarc.gsfc.nasa.gov/xanadu/xspec/python/html/chain.html}} command in PyXspec.
The results of our spectral analysis are given in the next section.

\section{Results}\label{sec:results}
Between the 2T and 4T models, we find the latter to be more effective in terms of reproducing the observed behavior of the data without the need to consider unusually high plasma temperatures, and also for constraining spectral parameters. The 4T model also makes it easier to compare the differential emission measures (DEMs) of the individual observations. We hence present the results for only the 4T model here. The results obtained with the 2T model are given in Appendix \S\ref{app_sec:2T} and \S\ref{sec:mcmc_results}.

As a first step, we investigate whether there is any difference between the average spectrum at periastron and that at the out-of-periastron phases. This is followed by a detailed examination of individual observations. We present the results obtained from the two exercises in \S\ref{subsec:result_merged_events} and \S\ref{subsec:result_individual_obs}.

\begin{table*}
{\tiny
\caption{The spectral fitting results for merged observations using the 4T model with $n_\mathrm{H}$ (in the \textit{tbabs} model, in units of $10^{22}\,\mathrm{cm^{-2}}$) allowed to vary between 0--0.3 (\S\ref{subsec:result_merged_events}). The energy range used during the fitting process is 0.3--10.0 keV. For each observation number, the first row gives the median values. The second row gives the 68\% ($1\sigma$) confidence interval from the posterior probability distribution marginalized for a given parameter (from the MCMC analysis). For the ${\chi}_\mathrm{red}^2$, the values within parentheses give the numbers of degrees of freedom for each fit.\label{tab:4T_result_merged_varying_nH}}
\begin{tabular}{cc|cccc||c|cccccc}
\hline
ID & $n_\mathrm{H}$ & \multicolumn{4}{c}{Norm $\left(\times 10^{-5}\,\mathrm{cm^{-5}}\right)$} & $\chi^2_\mathrm{red}$ & \multicolumn{6}{c}{Flux $\left(\times 10^{-13}\,\mathrm{erg\,cm^{-2}\,s^{-1}}\right)$}\\
&  & &  &  & & (dof) & \multicolumn{3}{c}{Observed} & \multicolumn{3}{c}{ISM corrected}\\
 & $\left(\times 10^{22}\,\mathrm{cm^{-2}}\right)$  & 0.2 keV & 0.6 keV & 1.0 keV & 4.0 keV & & 0.5--1 keV& 1--2 keV & 2--10 keV & 0.5--1 keV & 1--2 keV& 2--10 keV\\
\hline\hline
$1+3+5+7+9$ & 0.024 & $13.0$ & $6.8$ & $1.8$ & $51.2$ & 2.5(108) & 4.3 & 2.8  & $5.1$  & 4.9 & 2.9 & 5.1  \\
(periastron)& $(0.020-0.029)$ & $(11.9-14.2)$ & $(6.4-7.3)$ & $(1.1-2.4)$ & $(49.0-53.4)$ &  & (4.2--4.4) & (2.7--2.9)  &$(4.9-5.3)$   & (4.8--5.1) & (2.8--3.0) & $(4.9-5.3)$ \\
$2+4+6+8+10$ & 0.032 & $10.9$ & $3.4$ & $0.9$ & $31.2$ & 3.2(106) & 2.7 & 1.6 & $3.1$& 3.2 & 1.7 & 3.1 \\
(out-of-periastron) & $(0.026-0.039)$ & $(9.8-12.2)$ & $(3.0-3.8)$ & $(0.4-1.5)$ & $(29.2-33.1)$ &  & (2.6--2.7) & (1.6--1.7) & $(2.9-3.3)$ & (3.1--3.4) & (1.6--1.8) & $(3.0-3.3)$\\
\hline
$1+3+5+9$ & 0.024 & $13.5$ & $6.8$ & $2.8 $ & $42.5$ & 1.6(105) & 4.3 & 2.5 & $4.2$ & 4.9 & 2.6  & 4.3\\
(periastron)& $(0.020-0.029)$ & $(12.4-14.8)$ & $(6.2-7.3)$ & $(2.1-3.6)$ & $(40.2-44.7)$ &  & (4.2--4.4) & (2.4--2.6) & $(4.0-4.5)$ & (4.7--5.0) & (2.5--2.7) & $(4.1-4.5)$ \\
$2+4+6+8$ & 0.033 & $12.0$ & $3.3$ & $2.3$ & $19.0$ & 1.5(103) & 2.6 & 1.2 & $1.9$ & 3.2 & 1.3 & 1.9\\
(out-of-periastron) & $(0.025-0.041)$& $(10.7-13.4)$ & $(2.8-3.8)$ & $(1.6-2.9)$ & $(16.8-21.1)$ &  & (2.6--2.7) & (1.2--1.3) & $(1.7-2.1)$ & (3.0--3.4) & (1.2--1.4)  & $(1.7-2.1)$ \\
\hline
\end{tabular}
}
\end{table*}

\begin{table*}
{\tiny
\caption{Same as Table \ref{tab:4T_result_merged_varying_nH}, but with $n_\mathrm{H}$ fixed at the interstellar value of $0.03\times 10^{22}$\,cm$^{-2}$.\label{tab:4T_result_merged_fixed_nH}}
\begin{tabular}{c|cccc||c|cccccc}
\hline
ID & \multicolumn{4}{c}{Norm $\left(\times 10^{-5}\,\mathrm{cm^{-5}}\right)$} & $\chi^2_\mathrm{red}$ & \multicolumn{6}{c}{Flux $\left(\times 10^{-13}\,\mathrm{erg\,cm^{-2}\,s^{-1}}\right)$}\\
&  &  &  & & (dof) & \multicolumn{3}{c}{Observed} & \multicolumn{3}{c}{ISM corrected}\\
 & 0.2 keV & 0.6 keV & 1.0 keV & 4.0 keV & & 0.5--1.0 keV & 1.0--2.0 keV & 2.0--10.0 keV & 0.5--1.0 keV & 1.0--2.0 keV & 2.0--10.0 keV\\
\hline\hline
$1+3+5+7+9$ & $14.2$ & $6.9$ & $1.8$ & $51.6$ & 2.5(109) & 4.3 & 2.8  & $5.2$ & 5.1&2.9 & 5.2\\
(periastron)& $(13.5-14.9)$ & $(6.4-7.4)$ & $(1.1-2.5)$ & $(49.3-53.8)$ &  & (4.3--4.4) & (2.7--2.9) & $(4.9-5.4)$ & (5.0--5.2) & (2.8--3.0) & $(5.0-5.4)$ \\
$2+4+6+8+10$ & $10.5$ & $3.4$ & $0.9$ & $31.1$ & 3.2(107) & 2.6 & 1.6 & $3.1$ & 3.1 & 1.7 & 3.1\\
(out-of-periastron) & $(10.0-11.1)$ & $(3.0-3.8)$ & $(0.4-1.5)$ & $(29.2-32.9)$ &  & (2.6--2.7) & (1.6--1.7) & $(2.9-3.3)$ & (3.1--3.2) & (1.6--1.8) & $(2.9-3.3)$ \\
\hline
$1+3+5+9$ & $14.8$ & $6.8$ & $2.9$ & $42.7$ & 1.6(106) & 4.3& 2.5& $4.3$ &5.1 & 2.6 & 4.3\\
(periastron)& $(14.1-15.5)$ & $(6.3-7.3)$ & $(2.2-3.6)$ & $(40.5-45.0)$ &  & (4.2--4.4) & (2.5--2.6) & $(4.1-4.5)$ & (5.0--5.2) & (2.6--2.7) & $(4.1-4.6)$ \\
$2+4+6+8$ & $11.6$ & $3.3$ & $2.2$ & $19.0$ & 1.4(104) & 2.6 &1.2 & $1.9$ &3.1 &1.3 & 1.9\\
(out-of-periastron) & $(10.9-12.3)$ & $(2.8-3.8)$ & $(1.6-2.9)$ & $(17.0-21.1)$ &  & (2.5--2.7) & (1.2--1.3) & $(1.7-2.1)$ & (3.0--3.2) & (1.2--1.4) & $(1.8-2.1)$\\
\hline
\end{tabular}
}
\end{table*}

\subsection{Spectral analysis of merged spectra}\label{subsec:result_merged_events}
We merge the observations taken at periastron to obtain an average spectrum corresponding to the periastron phase following the same tools as in \S\ref{subsec:merge_event}.
The merged event lists were used to get the predicted background over all of the GTIs involved.
Similarly, we obtained an average spectrum for the out-of-periastron phases, as well as the corresponding background spectrum. The exposure times for the merged periastron and merged out-of-periastron observations are 8.05 and 8.02 ks respectively.

\begin{figure*}
    \centering
    \textbf{All observations\hspace{7cm} Excluding observations 7 \& 10}\\
    \includegraphics[trim={0.5cm 0.5cm 12cm 0.7cm},clip,width=0.45\textwidth]{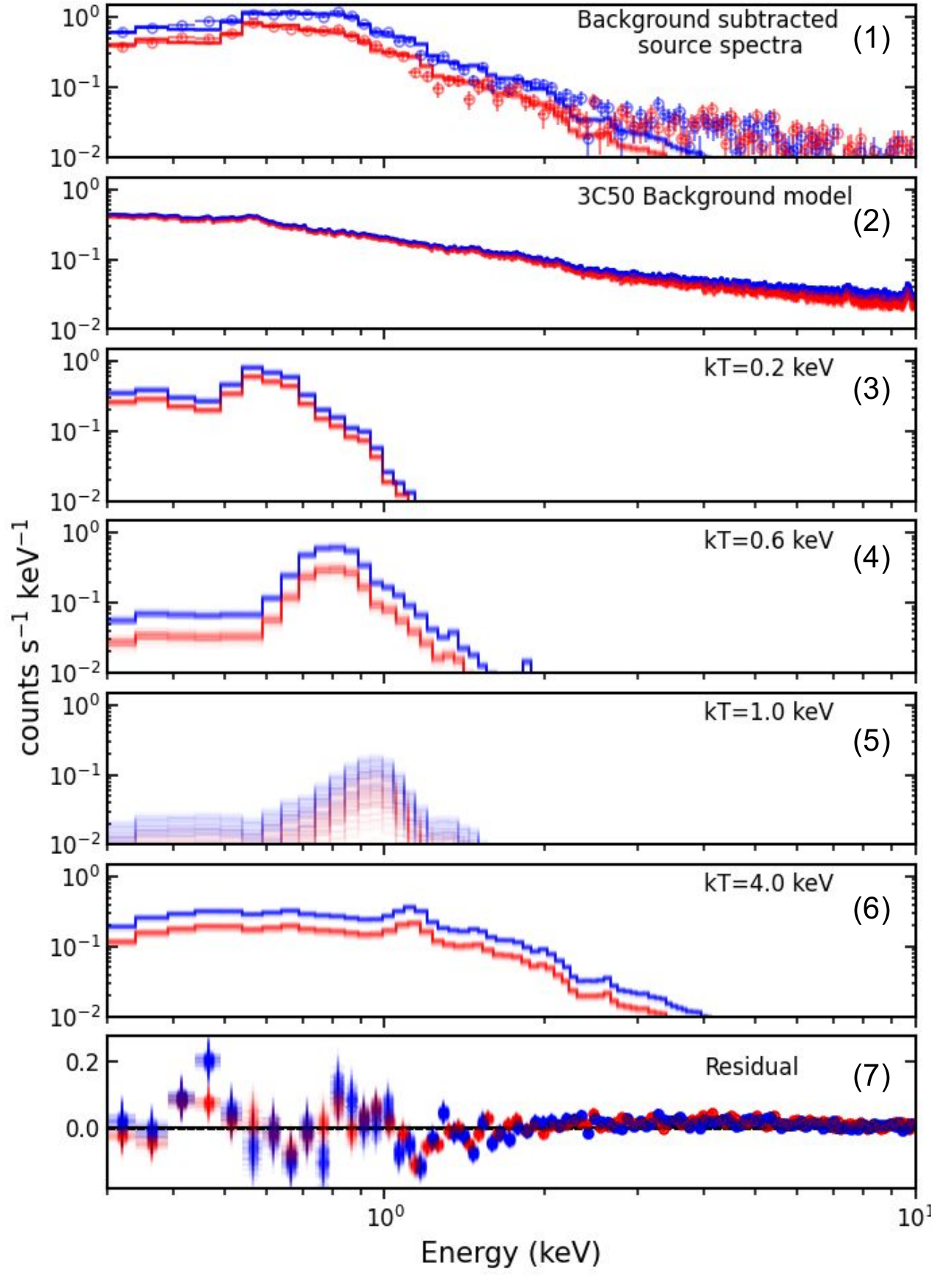}
    \hspace{1cm}
    \includegraphics[trim={0.5cm 0.5cm 12cm 0.7cm},clip,width=0.45\textwidth]{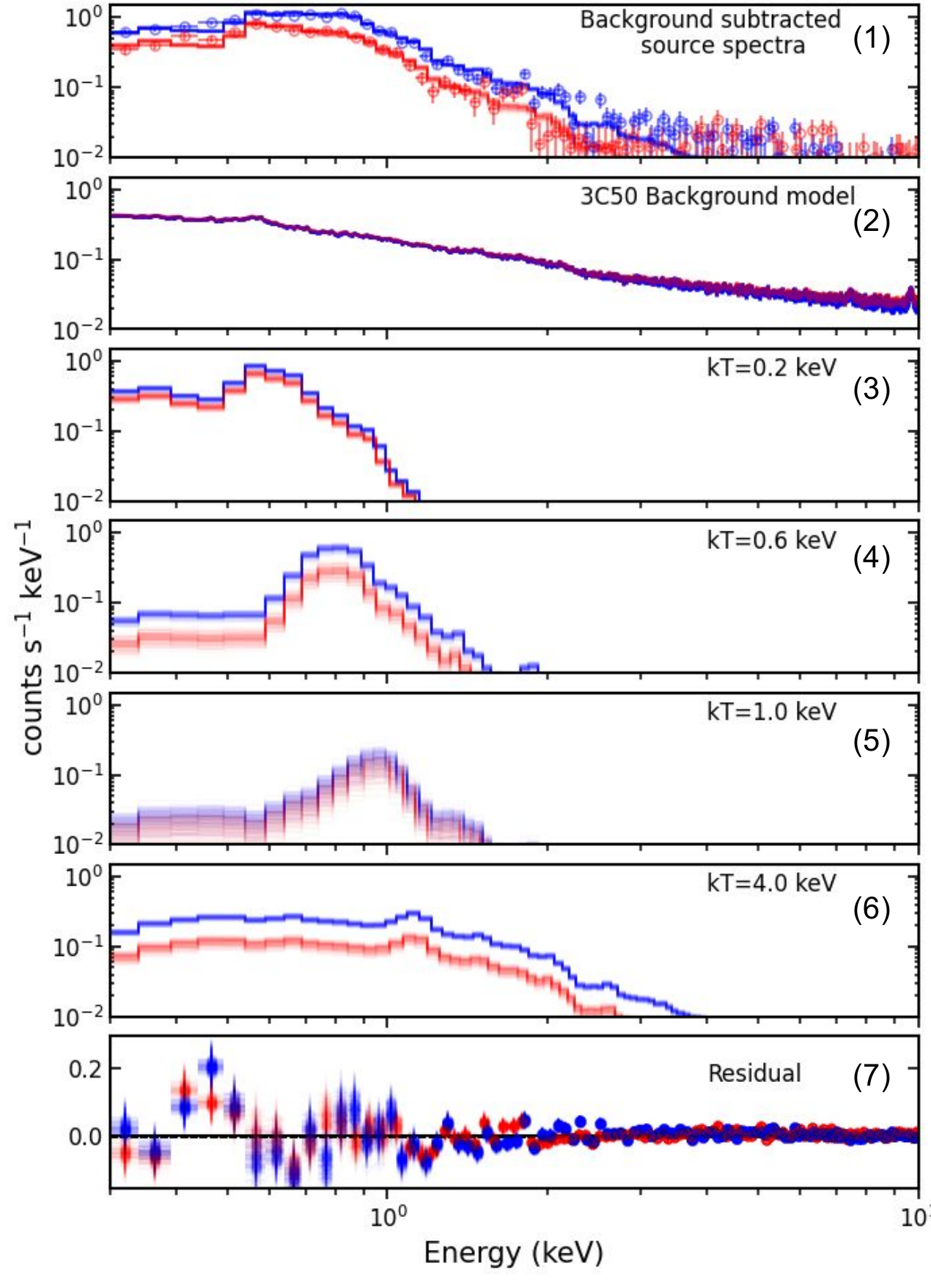}
    %
    \caption{Spectral analysis of the merged observations. The left panels show the results for the average periastron (shown in blue) and average out-of-periastron spectra (shown in red) obtained by merging all the periastron observations (observations 1, 3, 5, 7 and 9), and all the out-of-periastron observations (observations 2, 4, 6, 8 and 10) respectively. The right panels show the same but without using observations 7 and 10.
    The background subtracted spectra (both data and the fitted model) are shown in the top panels. Note that we have used the 4T model with $n_\mathrm{H}$ fixed at interstellar value.
    The second panels show the average background spectra predicted by the 3C50 model,
    panels 3--6 show the individual model components,
    finally the bottom panel show the residuals.
    Instead of showing just the best-fit model, we show the models corresponding to each of the parameter combinations sampled according to the joint posterior distribution of all the parameters (obtained from our MCMC analysis, see Figure \ref{fig:merged_spectra_corner_plots}). The opacities of the markers reflect their probability densities. See \S\ref{subsec:result_merged_events} for details. 
    }
    \label{fig:merged_spectra}
\end{figure*}

\begin{figure*}
    \centering
    \textbf{All observations\hspace{6cm} Excluding observations 7 \& 10}\\
    \includegraphics[width=0.45\textwidth]{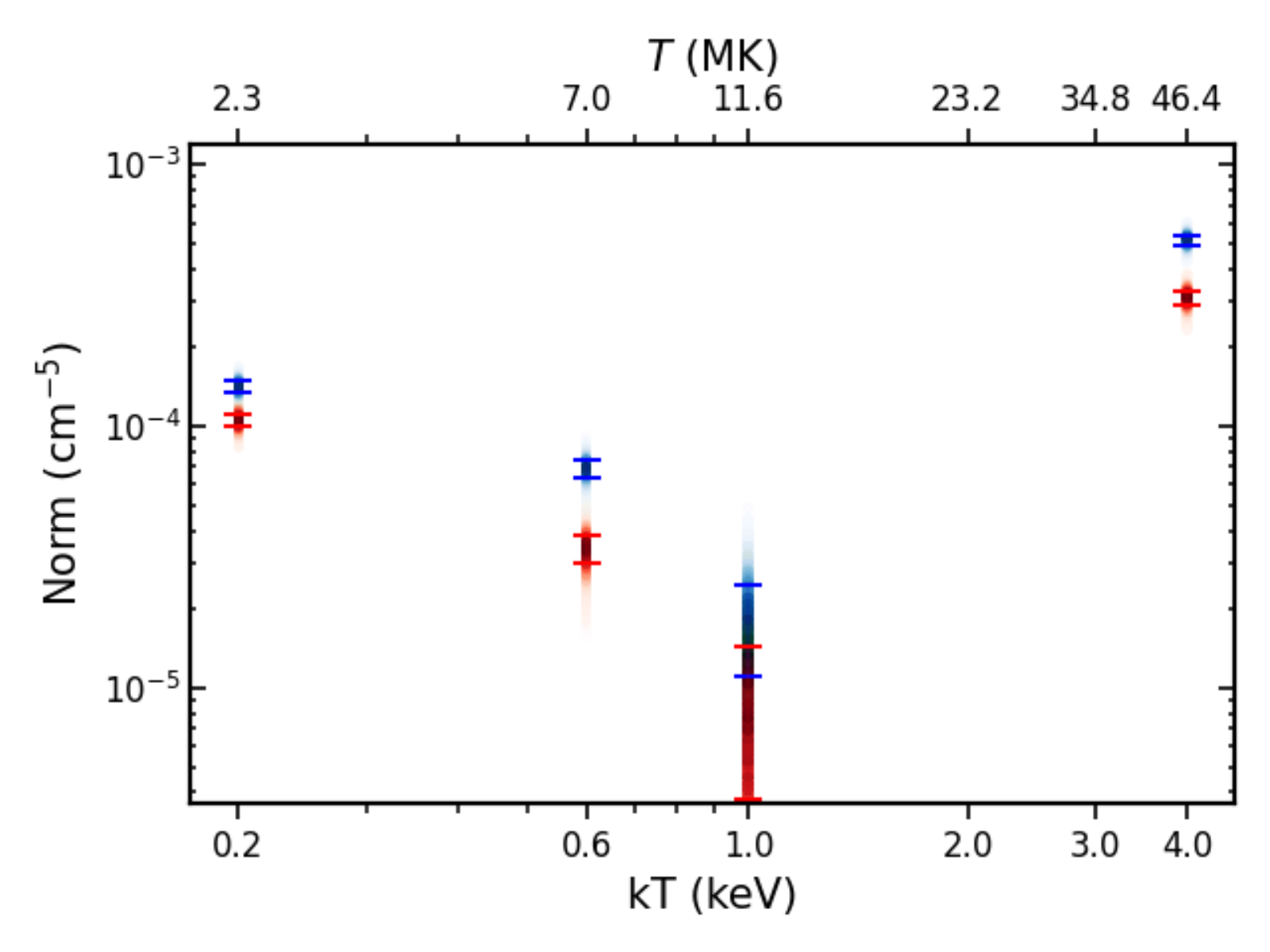}
    \hspace{1cm}
    \includegraphics[width=0.45\textwidth]{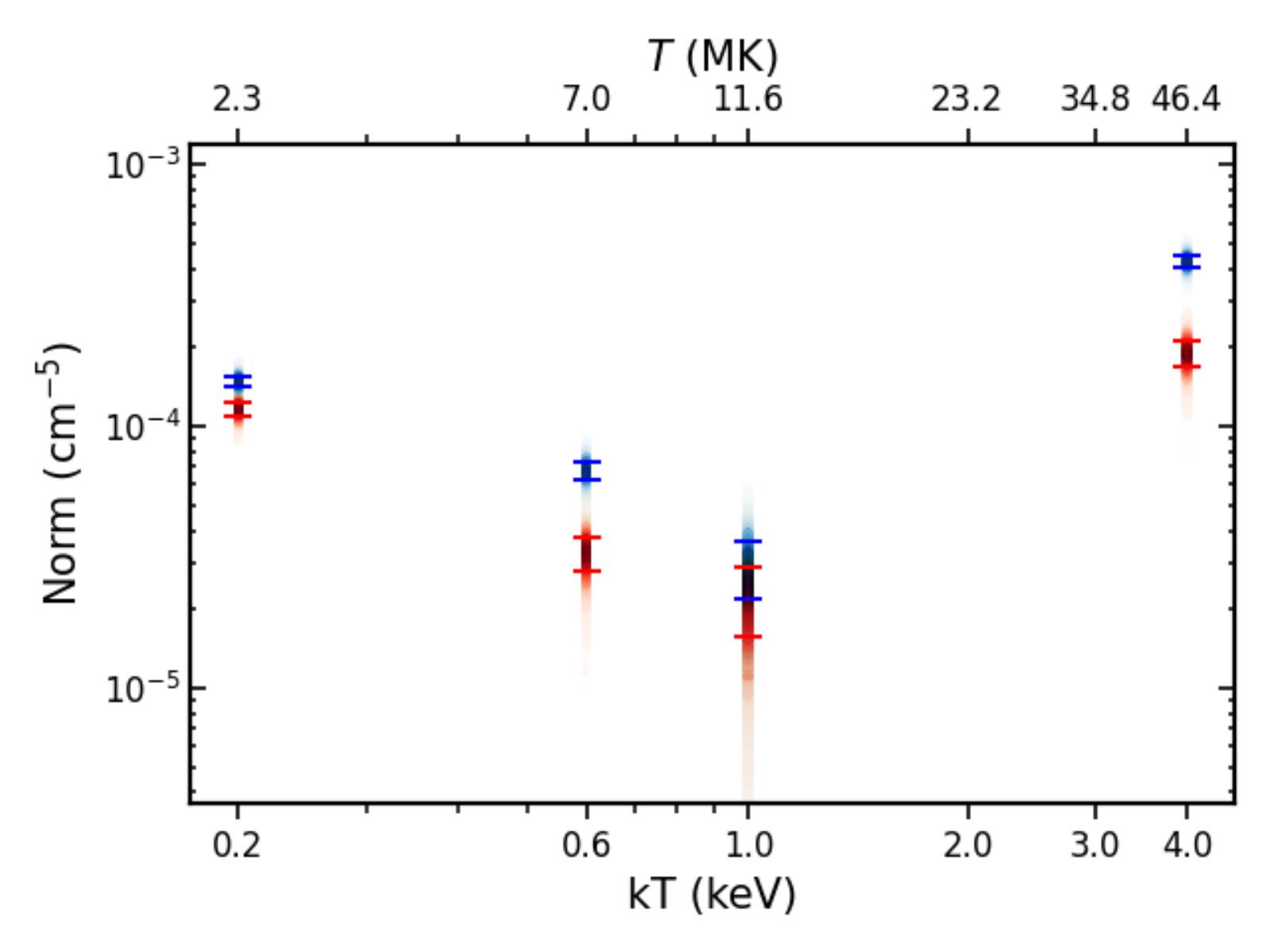}
    \caption{\textbf{Left:} The normalization factors as a function of temperature (expressed in units of keV in the lower axis, and in units of megakelvin in the upper axis) for the merged periastron (blue) and out-of-periastron (red) observations. Instead of plotting a single value for each norm and the associated error bar, we plot all the possible values of the norms corresponding to their respective marginalized posterior probability distributions, with the opacity of each marker being proportional to the probability density of that value. The horizontal bars mark the 68\% ($1\sigma$) confidence intervals. \textbf{Right:} Same as the left figure but without including spectra 7 and 10. See \S\ref{subsec:result_merged_events} for details.
    }
    \label{fig:dem_merged_spectra}
\end{figure*}

In the 4T model, there are five free parameters: the norms corresponding to the four plasma temperatures, and the neutral hydrogen column density $n_\mathrm{H}$.
The $n_\mathrm{H}$ parameter quantifies the absorption in the stellar magnetosphere itself and in the ISM, which generates a difference between the intrinsic and observed X-ray flux. The ISM contribution is already known : $0.03\times 10^{22}$\,cm$^{-2}$ (see \S\ref{sec:target}). \citet{naze2014} considered the possibility of absorption in the magnetosphere, but found the corresponding $n_\mathrm{H}$ to be zero. As already mentioned, this suggests that there is no significant absorption of the emitted X-rays in the stellar magnetosphere, and hence it is sufficient to use a single absorption with $n_\mathrm{H}$ fixed at $0.03\times 10^{22}$\,cm$^{-2}$ during spectral analysis.

The above assumption is strictly valid only when the absorbing material from the stellar magnetosphere(s) on the plane of the sky does not vary with time. For a single star, this is true if either the inclination angle (the angle between the line of sight and the rotation axis) or the obliquity (the angle between the rotation axis and the magnetic dipole axis) is zero. 
In the case of \eps, the obliquity of each star has been estimated to be consistent with zero, and the inclination angles are also small \citep{shultz2015b}. Thus, if we ignore the binarity, the contribution of the magnetospheres of the individual stars towards X-ray absorption will be time-invariant. 
However, if the X-ray emission is related to binarity, and given the fact that the system has non-negligible eccentricity, the net magnetospheric configuration will vary with orbital phase, and hence in principle, the value of $n_\mathrm{H}$ could be a function of orbital phase. Note that the orbital phase corresponding to the observation reported by \citet{naze2014} is $\approx 0.1$, which is close to the periastron phase (see Figure \ref{fig:obs_strategy}). Thus, binarity does not seem to provide an additional channel for X-ray absorption. 
Nevertheless, we examine possible additional absorption using the merged spectra by making $n_\mathrm{H}$ a free parameter\footnote{Note that the \textit{tbabs} model does not account for absorption by ionized materials, whereas the magnetospheric plasma is likely to be in ionized state. Thus, use of \textit{tbabs} will only provide a lower limit to the magnetospheric absorption. However, this caveat does not really impact in our case, since (1) we do not have any strong evidence of magnetospheric absorption and (2) spectra are fitted above 0.5 keV, where the differences between ionized and neutral medium absorptions are small and certainly below the error bars.}.

In the top two rows of Table \ref{tab:4T_result_merged_varying_nH}, we list the values of fitted parameters and their 68\% confidence intervals ($1\sigma$) obtained from MCMC analysis.
This analysis produces a multivariate posterior probability distribution, which is then marginalized for each parameter (by integrating over the other parameters). The best-fit values listed in Table \ref{tab:4T_result_merged_varying_nH} correspond to the medians of these marginalized distributions. The $\chi^2_\mathrm{red}$ values correspond to the model using these median values for all free parameters.
This `median $\chi^2_\mathrm{red}$' is larger than 2 for both spectra with corresponding $p$-values 
of approximately zero. We find that two spectra 7 (periastron) and 10 (out-of-periastron) are primarily responsible for the discrepancy between the data and the model. These two spectra have the highest $\chi^2_\mathrm{red}$ when fitted with the 4T model (see \S\ref{subsec:result_individual_obs} and Appendix \ref{sec:obs_7_10}). 
This motivated us to examine the average spectra without including the spectra for observations 7 and 10. Indeed we find that the fitting improves significantly upon exclusion of the two spectra (see the bottom two rows of Table \ref{tab:4T_result_merged_varying_nH}). The $p$-values still remain small, but become non-negligible in this case (approximately $10^{-4}$ and $10^{-3}$ respectively for the periastron and out-of-periastron spectra).

Table \ref{tab:4T_result_merged_varying_nH} also shows the value of $n_\mathrm{H}$ obtained for the average spectra.
We find the median values of $n_\mathrm{H}$ to be 0.02 and $0.03\times 10^{22}$\,cm$^{-2}$, with the 68\% confidence intervals as 0.02--0.03 and $0.03-0.04\times 10^{22}$\,cm$^{-2}$ for the merged periastron and out-of-periastron observations, respectively. Thus, our observations are consistent with the interstellar value, i.e. without significant X-ray absorption in the stellar magnetsosphere itself at any phase, as expected.
Consequently, we will set $n_\mathrm{H}=0.03\times 10^{22}$\,cm$^{-2}$ in all subsequent analyses (the number of free parameters is thus reduced to 4). In Table \ref{tab:4T_result_merged_fixed_nH}, we list the results of our spectral analysis when $n_\mathrm{H}$ is kept fixed at the interstellar value. As can be seen, the parameter values are nearly identical to those listed in Table \ref{tab:4T_result_merged_varying_nH}.

In Figure \ref{fig:merged_spectra}, we show the background subtracted spectra (panel 1), the 3C50 background spectra (panel 2), contribution from individual model components (panels 3--6) and the residuals (panel 7) obtained from the spectral fitting (4T model with $n_\mathrm{H}$ fixed at interstellar value) of the merged spectra for periastron (blue markers) and out-of-periastron (red markers) observations. On the left, we show the result obtained by using all the observations, and on the right after the exclusion of spectra 7 and 10.
As can be seen, the average background spectra are nearly identical. The contributions of the different plasma components primarily differ at the 0.6 keV and 4 keV energy bins. This becomes more apparent in Figure \ref{fig:dem_merged_spectra}, where we show the variation of normalization factors (the values of the parameter `norm' in the \textit{apec} model, see \S\ref{sec:target}) with temperature for the average spectra. 
The left panel compares the normalization factors (which are proportional to the EMs) for the average periastron (blue) and out-of-periastron (red) observations. The right panel shows the same but without including spectra 7 and 10 in the average periastron and out-of-periastron observations, respectively. 
Both Figures \ref{fig:merged_spectra} and \ref{fig:dem_merged_spectra} clearly show that 
the brightest X-ray observations are obtained at periastron.

Thus, we conclude that there is excess X-ray flux from \eps~at periastron as compared to away from periastron. This suggests an impact of the binarity in the production of X-ray emission from \eps.

\begin{table*}
{\scriptsize
\caption{Same as Table \ref{tab:4T_result_merged_fixed_nH}, but for the individual observations.
\label{tab:4T_result}}
\begin{tabular}{c|cccc||c|cccccc}
\hline
ID & \multicolumn{4}{c}{Norm $\left(\times 10^{-5}\,\mathrm{cm^{-5}}\right)$} & $\chi^2_\mathrm{red}$ & \multicolumn{6}{c}{Flux $\left(\times 10^{-13}\,\mathrm{erg\,cm^{-2}\,s^{-1}}\right)$}\\
 &  &  &  & & (dof)& \multicolumn{3}{c}{Observed} & \multicolumn{3}{c}{ISM corrected}\\
 & 0.2 keV & 0.6 keV & 1.0 keV & 4.0 keV & & 0.5--1.0 keV & 1.0--2.0 keV & 2.0--10.0 keV & 0.5--1.0 keV& 1.0--2.0 keV & 2.0--10.0 keV\\
\hline\hline
1 & $13.1$ & $8.0$ & $2.3 $ & $12.8$ & 1.2(74) & $3.6$ & 1.3 & $1.4$ & 4.3 & 1.3 & 1.4\\
& $(11.7-14.6)$ & $(7.0-9.1)$ & $(1.0-3.7)$ & $(8.3-17.3)$ &  & (3.5--3.8) & (1.1--1.4)& $(0.9-1.8)$ & (4.1--4.4) & (1.2--1.5) & $(0.9-1.8)$\\
2 & $14.5$ & $2.6$ & $3.2$ & $23.8$ & 1.1(49) & 3.0 & 1.5 & $2.4$ & 3.6 & 1.6 & 2.4\\
& $(13.2-15.7)$ & $(1.7-3.4)$ & $(2.0-4.4)$ & $(20.1-27.5)$ &  & (2.8--3.2) & (1.4--1.6) & $(2.1-2.8)$ & (3.4--3.8) & (1.4--1.7) & $(2.1-2.8)$ \\
3 & $17.5$ & $6.4$ & $3.6$ & $16.2$ & 0.9(43) & 4.0 & 1.4 & $1.7$ & 4.7 & 1.5 & 1.7\\
& $(16.3-18.7)$ & $(5.5-7.4)$ & $(2.4-4.7)$ & $(13.0-19.5)$ &  & (3.9--4.1) & (1.4--1.6) & $(1.4-2.0)$ & (4.6--4.9) & (1.4--1.6) & $(1.4-2.0)$\\
4 & $10.9$ & $3.9$ & $0.8$ & $20.5$ & 1.2(64) & 2.5 & 1.2 & $2.0$ & 3.0 & 1.3 & 2.0\\
& $(9.8-12.0)$ & $(3.2-4.6)$ & $(0.2-1.6)$ & $(17.4-23.6)$ &  &(2.4--2.6) &(1.1--1.3)& $(1.7-2.4)$ & (2.8--3.1) & (1.2--1.4) & $(1.8-2.4)$\\
5 & $16.1$ & $6.3$ & $3.5$ & $68.9$ & 1.6(87) & 5.0 & 3.7 & $6.9$ & 5.9 & 3.8 & 6.9\\
& $(14.9-17.4)$ & $(5.4-7.2)$ & $(2.2-4.8)$ & $(64.7-72.9)$ &  & (4.9--5.2) & (3.6--3.8) & $(6.5-7.3)$ & (5.8--6.1) & (3.7--4.0) & $(6.5-7.3)$\\
6 & $11.7$ & $5.0$ & $2.6$ & $8.8$ & 1.3(33) & 2.8 & 1.0 & $1.0$ & 3.4 & 1.0 & 1.0 \\
& $(9.0-14.3)$ & $(3.1-6.8)$ & $(0.9-5.0)$ & $(3.1-16.4)$ &  & (2.5--3.2) & (0.8--1.2) & $(0.4-1.7)$ & (3.0--3.7) & (0.8--1.3) & $(0.4-1.7)$\\
7 & $11.5$ & $5.7$ & $0.7$ & $98.5$ & 2.4(89) & 4.8 & 4.6 & $9.7$ & 5.6 & 4.8 & 9.8 \\
& $(9.1-14.0)$ & $(4.2-7.2)$ & $(0.2-1.7)$ & $(90.2-107.0)$ &  & (4.6--5.0) & (4.4--5.0) & $(8.9-10.6)$ & (5.4--5.9) & (4.5--5.2) & $(9.0-10.6)$\\
8 & $9.6$ & $2.8$ & $3.7$ & $6.8$ & 0.8(68) & 2.2 & 0.8 & $0.8$ & 2.6 & 0.9 & 0.8\\
& $(8.4-10.7)$ & $(1.9-3.7)$ & $(2.4-4.9)$ & $(3.2-10.8)$ &  & (2.1--2.3) & (0.7--1.0) & $(0.4-1.1)$ & (2.4--2.7) & (0.7--1.0) & $(0.4-1.1)$\\
9 & $7.7$ & $5.8$ & $2.4$ & $69.9$ & 2.2(44) & 4.0 & 3.6 & $7.0$ & 4.6 & 3.7 & 7.0\\
& $(5.8-9.7)$ & $(4.3-7.1)$ & $(0.9-4.3)$ & $(63.4-76.6)$ & & (3.8--4.1) & (3.3--3.8) & $(6.4-7.6)$ &  (4.4--4.8) & (3.4--3.9) & $(6.4-7.6)$\\
10 & $8.9$ & $2.5$ & $0.4$ & $54.2$ & 4.1(86) & 2.8 & 2.5 & $5.4$ &  3.3 & 2.6 & 5.4\\
& $(7.7-10.1)$ & $(1.8-3.2)$ & $(0.0-0.9)$ & $(50.2-57.9)$ & & (2.6--2.9) & (2.4--2.7) & $(5.0-5.7)$ & (3.1--3.4) & (2.5--2.8) & $(5.0-5.7)$\\
\hline
\end{tabular}
}
\end{table*}

\begin{figure}
    \centering
    \includegraphics[width=0.45\textwidth]{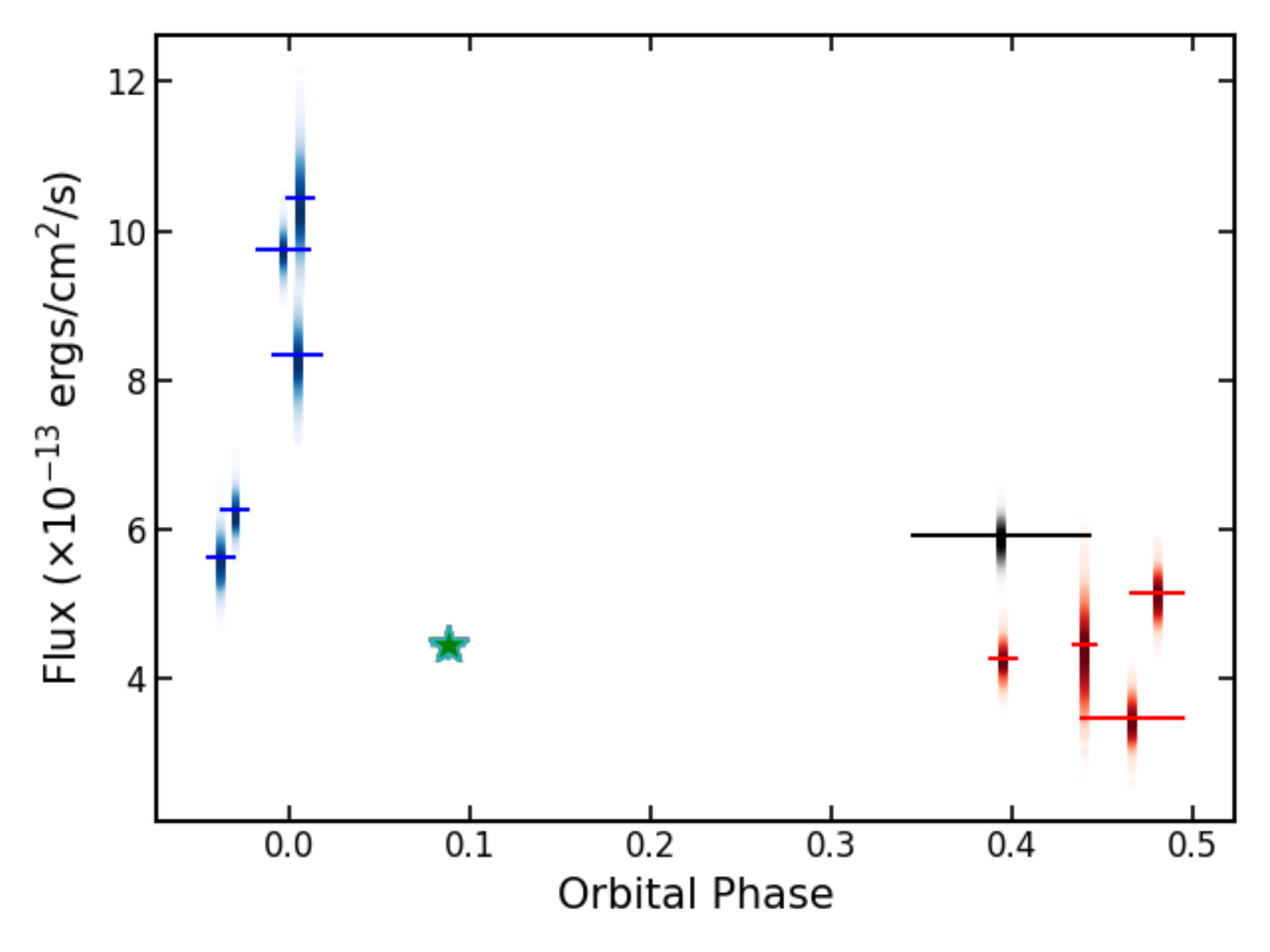}
    \caption{The variation with orbital phases of X-ray flux (ISM corrected) over 0.5--2.0 keV
    obtained using the 4T model with $n_\mathrm{H}$ fixed at the interstellar value. The blue and red points correspond to periastron and out-of-periastron observations respectively. The grey point signifies unreliable estimates (see \S\ref{subsec:result_individual_obs} for details). At a given orbital phase, instead of plotting a single value of the flux and the associated error bars, we plot all the possible values of the flux corresponding to the marginalized posterior probability distributions of the fitted parameters, with the opacity of each marker being proportional to the probability density of that value. The green star represents the corresponding fluxes obtained using XMM-Newton data by \citet{naze2014}.
    }
    \label{fig:lightcurves_4T}
\end{figure}

\begin{figure}
    \centering
    \includegraphics[width=0.45\textwidth]{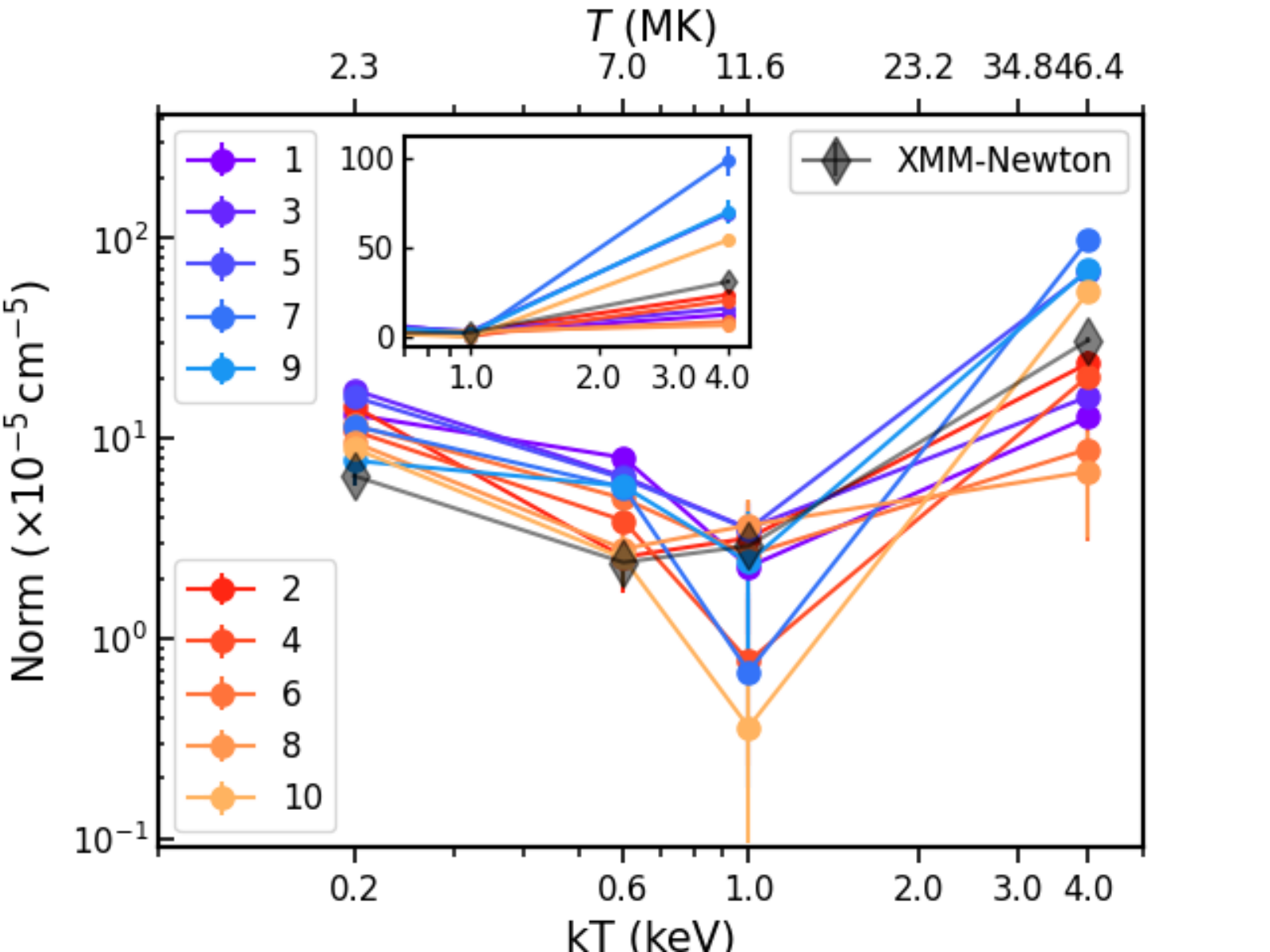}
    \caption{The normalization factors as a function of temperature (expressed in keV here) for the 4T model with $n_\mathrm{H}$ fixed at $0.03\times 10^{22}$\,cm$^{-2}$. The error bars in the normalization factors correspond to 68\% confidence intervals ($1\sigma$). The observations labeled with odd numbers correspond to periastron observations, and those labelled with even numbers correspond to out-of-periastron observations. The black diamonds represent the result obtained by \citet{naze2014} using the XMM-Newton observations. In the inset, we zoom over the region between 0.6 and 4 keV to show the difference between the normalization factors at the highest temperature (4 keV) for the different observations.
    Note that we have used a linear scale on the Y-axis to make the difference more apparent.}
    \label{fig:dem_4T_fixed_nH}
\end{figure}

\subsection{Spectral Analysis of individual observations}\label{subsec:result_individual_obs}
We now perform spectral fitting of the individual observations represented by IDs 1--10 (Table \ref{tab:obs}) so as to investigate potential differences among the observations.
We fixed $n_\mathrm{H}$ at $0.03\times 10^{22}$\,cm$^{-2}$. The resulting parameter values and the associated uncertainties (obtained from MCMC analysis) are listed in Table \ref{tab:4T_result} (also see Figure \ref{fig:4T_fixed_nH_mcmc_spectra} for the MCMC corner plots and the best fits).
As can be seen, the reduced $\chi^2$ is less than 2 except for three spectra: 7, 9 and 10. For the first two, the reduced $\chi^2$ lies below 2 if we evaluate it over the energy range of 0.5--2.0 keV (without re-fitting the spectra). For the last spectrum however (ID 10), although the reduced $\chi^2$ decreases if we only consider the energy range 0.5--2.0 keV, it still remains higher than 2 (2.5). In all three cases, the $\chi^2_\mathrm{red}$ increases if evaluated over the energy range 2.0--10.0 keV making the corresponding flux estimation in this energy range unreliable. 
To summarize, between 0.5 and 2.0 keV, the flux estimated using the 4T model can be trusted for all but spectrum 10, whereas between 2.0 and 10.0 keV, we can trust all flux estimations except for spectra 7, 9 and 10. 
There is however a caveat here, which is that Figure \ref{fig:merged_spectra} clearly shows that over 2--10 keV, the target spectrum is comparable to the predicted (3C50) background spectrum. This, combined with the fact that the background spectrum is not obtained from observation, but is modelled, makes our results over 2-10 keV less robust against any limitation in the predicted background spectra (see Appendix \ref{sec:test}). We will, therefore, not use the flux values obtained for this energy range for drawing any inference about the system.

The orbital variation of X-ray flux over 0.5--2.0 keV is shown in Figure \ref{fig:lightcurves_4T}. 
From the analysis of the previous subsection, we expect to observe higher fluxes at periastron phases than those at out-of-periastron phases. Interestingly, the light curve shown in Figure \ref{fig:lightcurves_4T} reveals that not all periastron observations have identical X-ray properties. The X-ray flux rises sharply around periastron over an orbital phase range of $\approx 0.05$ cycles, and only three (IDs 5, 7 and 9) of the five periastron observations predominantly contribute to the flux enhancement.
The other two periastron observations (IDs 1 and 3), that lie close to periastron, but do not cover phase 0 (top panel of Figure \ref{fig:obs_strategy}), have X-ray properties similar to that of the out-of-periastron observations (also see Table \ref{tab:4T_result}). 
This can also be seen from Figure \ref{fig:dem_4T_fixed_nH} where we plot the normalization factors as a function of temperature for the individual observations.
Unfortunately in this case, the error bars in the normalization factors are too large to investigate the differences between periastron vs out-of-periastron observations, except for the hottest plasma component (4 keV). From the inset of Figure \ref{fig:dem_4T_fixed_nH}, where we zoom in to the normalization factors at 4 keV (using a linear scale),
we find that the normalization factors at 4 keV are higher for observations 5, 7\footnote{Note that we obtained a poor fit for this observation using the 4T model, where our best-fit spectrum has a lower flux than that of the observed spectrum above $\approx 2$ keV (see Figure \ref{fig:4T_fixed_nH_mcmc_spectra}).} and 9 than those for the rest of the observations (see also Figure \ref{fig:4T_fixed_nH_mcmc_spectra}).
The other two observations obtained near (but not at) periastron (IDs 1 and 3) have norms similar to those for out-of-periastron observations. Observation 10 also appears to have an unusually high norm at 4 keV compared to the rest of the out-of-periastron observations. However, this observation is problematic (see preceding paragraph and \S\ref{sec:obs_7_10}). 


\begin{table*}
{\tiny 
\caption{Same as Table \ref{tab:4T_result_merged_fixed_nH}, but for `true periastron' and `approaching periastron' (see \S\ref{subsec:result_individual_obs}) observations.\label{tab:4T_result_merged_peri_fixed_nH}}
\begin{tabular}{c|cccc||c|cccccc}
\hline
ID & \multicolumn{4}{c}{Norm $\left(\times 10^{-5}\,\mathrm{cm^{-5}}\right)$} & $\chi^2_\mathrm{red}$ & \multicolumn{6}{c}{Flux $\left(\times 10^{-13}\,\mathrm{erg\,cm^{-2}\,s^{-1}}\right)$}\\
&  &  &  & & & \multicolumn{3}{c}{Observed} & \multicolumn{3}{c}{ISM corrected}\\
 & 0.2 keV & 0.6 keV & 1.0 keV & 4.0 keV & & 0.5--1.0 keV & 1.0--2.0 keV & 2.0--10.0 keV & 0.5--1.0 keV & 1.0--2.0 keV & 2.0--10.0 keV\\
\hline\hline
$1+3$ & $15.7$ & $7.1$ & $3.3$ & $14.6$ & 0.9(89) & 3.9 &1.4 & 1.5 & 4.6 & 1.4 & 1.5\\
& (14.8--16.6) & (6.3--7.8) & (2.4--4.3) & (11.8--17.3) &  & (3.8--4.0) & (1.3--1.5) & (1.2--1.8)  & (4.4--4.7) & (1.4--1.6) & (1.2--1.8) \\
$5+9$ & $13.7$ & $6.6$ & $2.6$ & $71.7$ & 2.0(96) & 4.8 & 3.7 & 7.1 & 5.6 & 3.8 & 7.2\\
& (12.7--14.7) & (5.8--7.4) & (1.5--3.7) & (68.1--75.1) &  & (4.7--4.9) &(3.6--3.8) & (6.8--7.5) & (5.5--5.8) & (3.7--4.0) & (6.8--7.5)\\
\hline
\end{tabular}
}
\end{table*}

\begin{figure}
    \centering
    \includegraphics[width=0.45\textwidth]{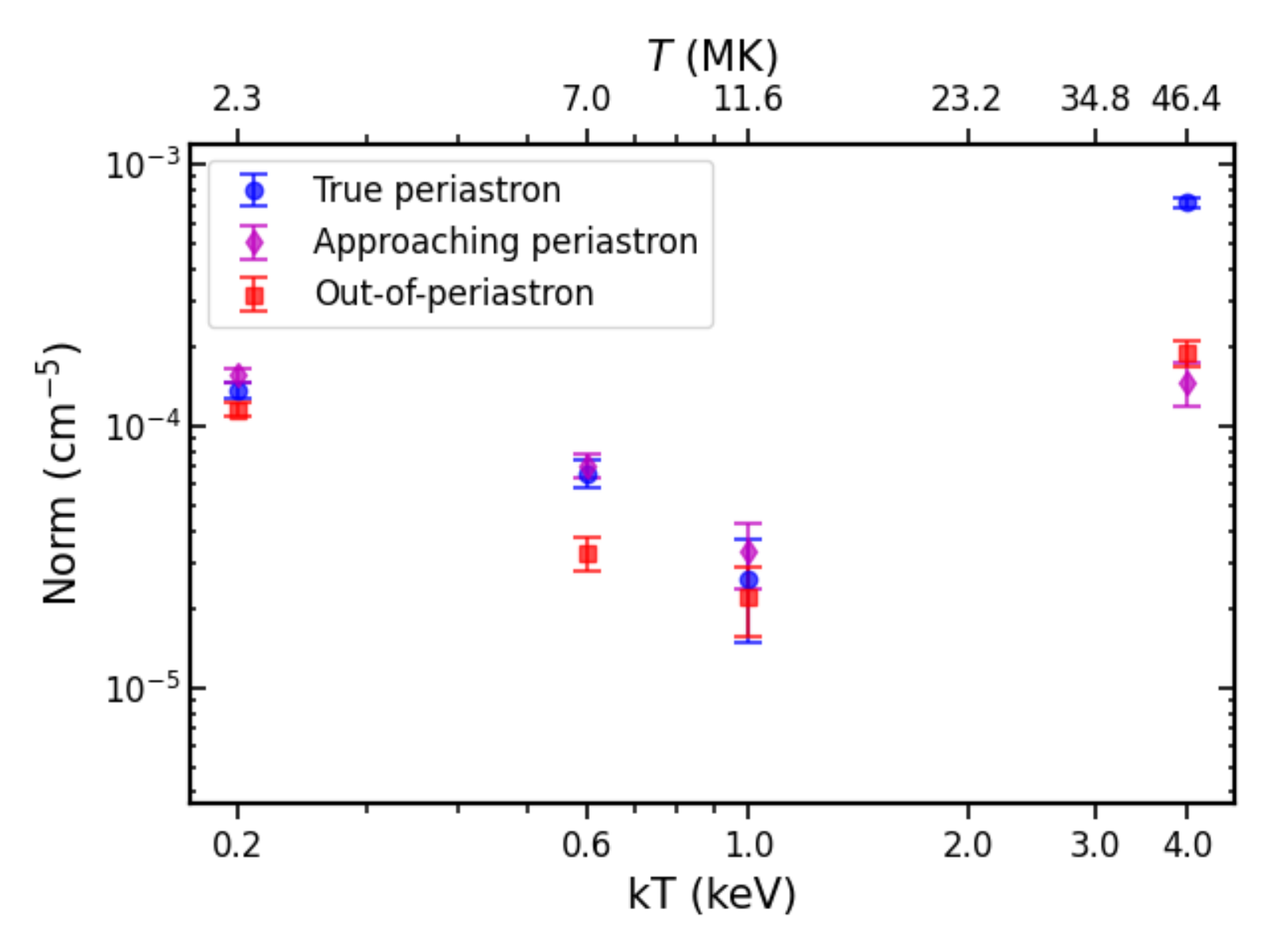}
    \caption{The normalization factors as a function of temperature (expressed in units of keV in the lower axis, and in units of megakelvin in the upper axis), the markers represent the median values obtained from the MCMC analysis, and the errorbars represent the 68\% confidence intervals.
    Magenta diamonds correspond to the spectrum obtained by merging observations 1 and 3 (appoaching periastron); blue circles correspond to the spectrum obtained by merging observations 5 and 9 (true periastron); and the red squares correspond to the spectrum obtained by merging the out-of-periastron observations excluding the observation 10.
    }
    \label{fig:dem_comparison}
\end{figure}

Both Figures \ref{fig:lightcurves_4T} and \ref{fig:dem_4T_fixed_nH} suggest that the merged spectrum obtained by averaging all five periastron observations (\S\ref{subsec:result_merged_events}) does not represent the true X-ray characteristics of the system exactly at periastron. 
To investigate the change in X-ray characteristics as the system approaches periastron, 
we merged spectra 5 and 9, and 1 and 3. We performed spectral analysis using the 4T model with $n_\mathrm{H}$ fixed at the interstellar value (see Table \ref{tab:4T_result_merged_peri_fixed_nH}). Figure \ref{fig:dem_comparison} shows the comparison among the normalization factors for the spectra obtained by merging observations 5 and 9 (representative of the periastron spectrum, shown in blue, will be referred to as `true periastron'), merging observations 1 and 3 (shown in magenta, will be referred to as `approaching periastron'), and the merged out-of-periastron observation (excluding observation 10).
From this figure, we conclude that the primary difference between the exact periastron X-ray characteristics and those away from periastron lies in the contribution of the hottest (4 keV) plasma component. The differential EM (DEM) for the hottest plasma component is much higher at periastron than for the observations away/close to periastron. The common property of the observations acquired around periastron (i.e. both true periastron and approaching periastron) is that both have a higher DEM at 0.6 keV as compared to that obtained for the out-of-periastron observations. Note that inclusion of spectrum 7 while constructing the true periastron spectrum does not change these inferences.


\subsection{Comparison with past X-ray observations}\label{subsec:past_obs}
As mentioned already, \citet{naze2014} reported an X-ray detection from \eps~at a phase close to periastron using the XMM-Newton telescope (ObsID: 0690210201, PI: Naz\'e). 
In Figure \ref{fig:lightcurves_4T}, we compare their flux estimates over the energy ranges 0.5--2.0 keV (4T model, $n_\mathrm{H}=0.03\times 10^{22}$\,cm$^{-2}$) with those obtained for the NICER data. It can be seen that the flux reported by \citet{naze2014} is consistent with that for the out-of-periastron observation, suggesting similar X-ray characteristics at this phase as compared to phases further away from periastron.



\begin{figure}
    \centering
    \includegraphics[width=0.45\textwidth]{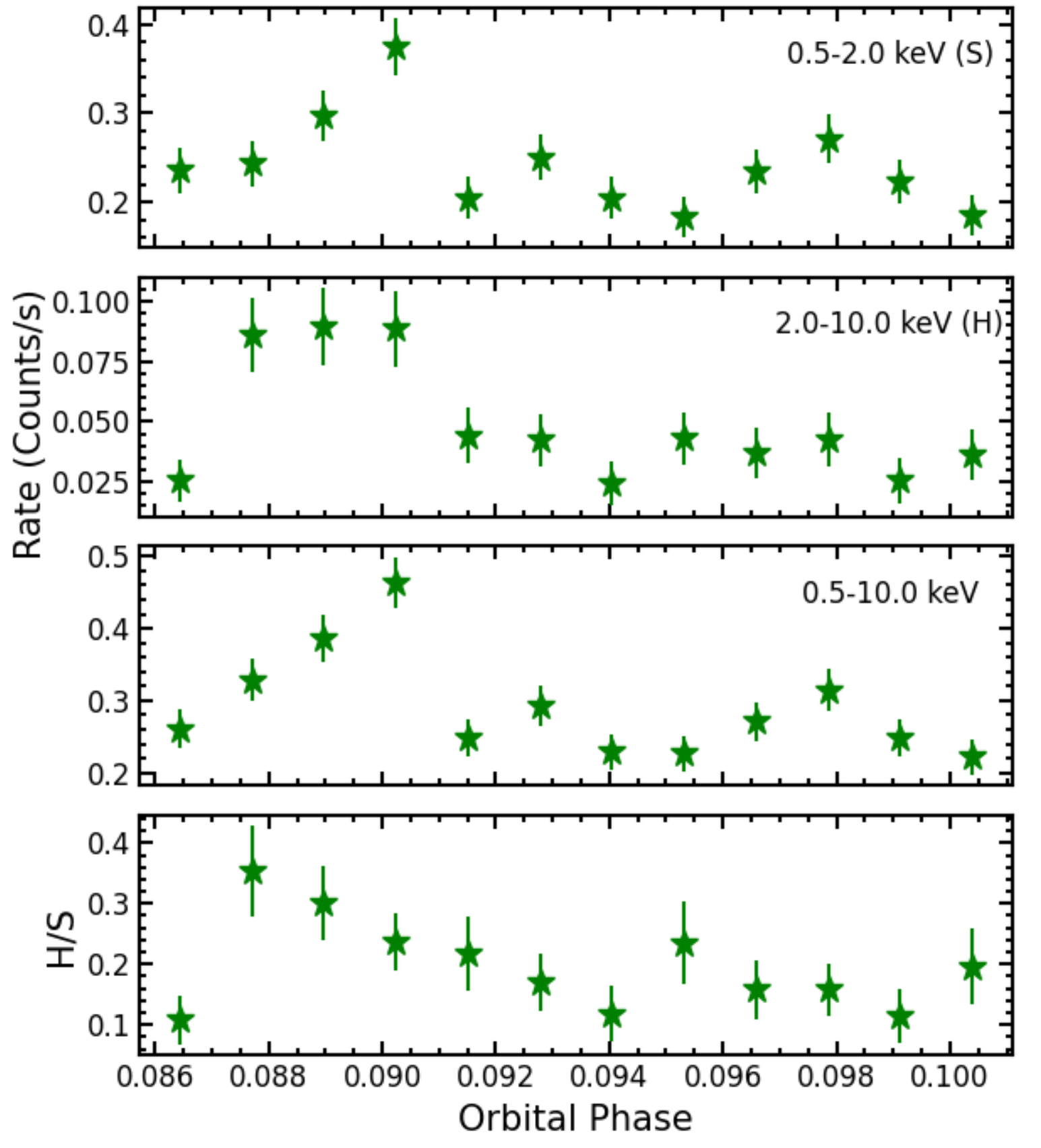}
    \caption{Light curves obtained by reanalysis of the XMM-Newton data of \eps~reported by \citet{naze2014}. For phasing, the reference $\mathrm{HJD_0^{new}}$ is 2456356.085. \textbf{Top three panels:} EPIC-pn count rates as functions of orbital phases over the energy ranges of 0.5--2.0 keV (soft), 2.0--10.0 keV (hard) and 0.5--10.0 keV (total) respectively. \textbf{Bottom:} Hardness ratio defined as the ratio between the count rates over 2.0--10.0 keV to that at 0.5--2.0 keV. Note that the error bars correspond to 1$\sigma$.}
    \label{fig:xmm_count_rate_lc}
\end{figure}

\begin{figure}
    \centering
    \includegraphics[width=0.45\textwidth]{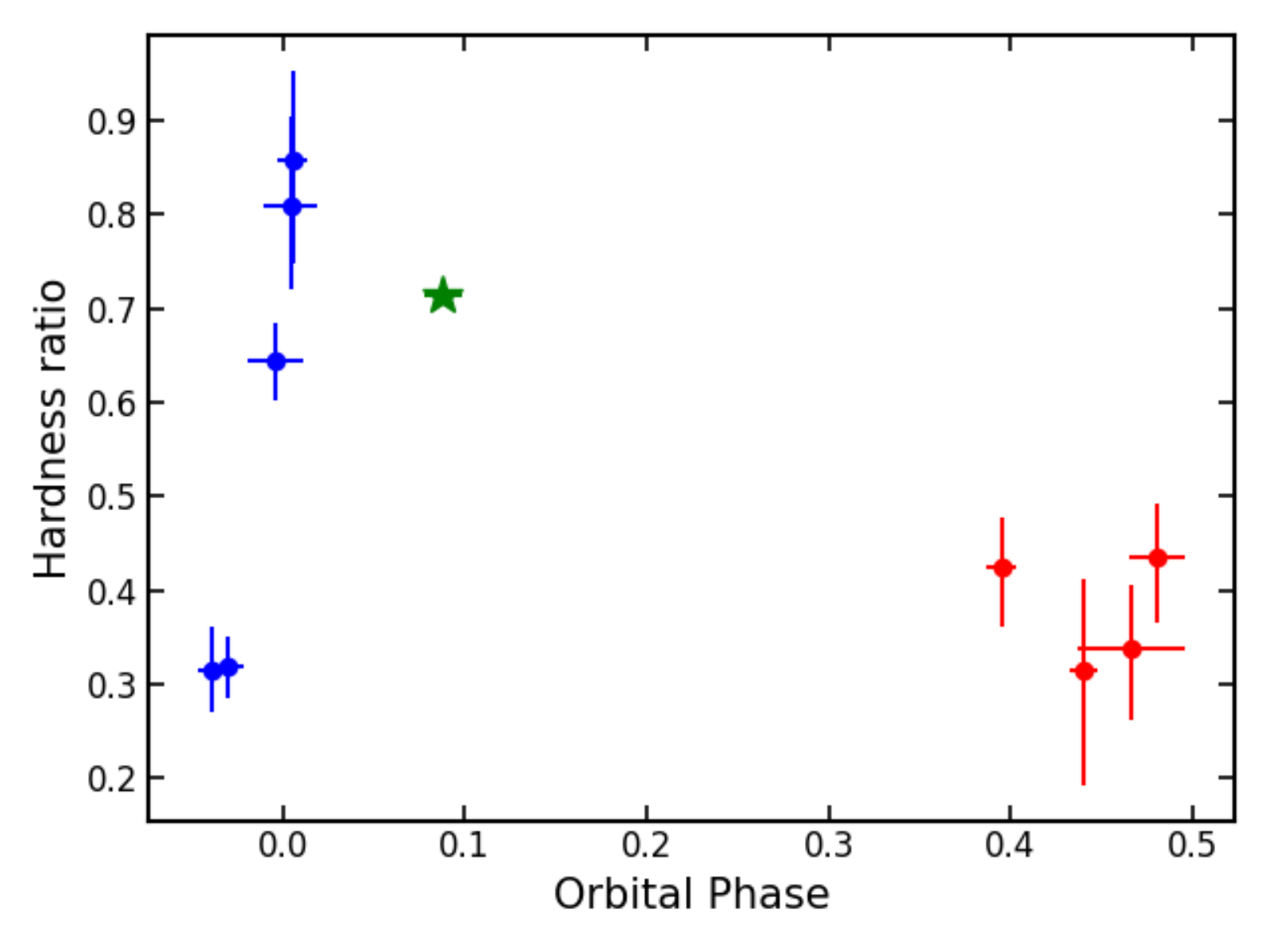}
    \caption{Hardness ratio, defined as the ratio between the fluxes at 1.0--2.0 keV and 0.5--1.0 keV, as a function of orbital phases. The fluxes are corrected for the ISM absorption. The blue and red circles correspond to NICER observations taken at/close to the periastron passage and away from the periastron passage; the green star corresponds to archival XMM-Newton observation of \eps.}
    \label{fig:hardness_ratio}
\end{figure}

\begin{table*}
{\tiny
\caption{Spectral analysis of archival XMM-Newton data of \eps. We fitted the spectra (for full exposure, for times with high count rates, and for times with low count rates, see \S\ref{subsec:past_obs}) with the 4T model with $n_\mathrm{H}$ fixed at the interstellar value. \label{tab:4T_result_xmm}}
\begin{tabular}{c|cccc||c|cccccc}
\hline
 & \multicolumn{4}{c}{Norm $\left(\times 10^{-5}\,\mathrm{cm^{-5}}\right)$} & $\chi^2_\mathrm{red}$ & \multicolumn{6}{c}{Flux $\left(\times 10^{-13}\,\mathrm{erg\,cm^{-2}\,s^{-1}}\right)$}\\
 &  &  &  & & (dof) & \multicolumn{3}{c}{Observed} & \multicolumn{3}{c}{ISM corrected}\\
 & 0.2 keV & 0.6 keV & 1.0 keV & 4.0 keV & & 0.5--1.0 keV & 1.0--2.0 keV & 2.0--10.0 keV & 0.5--1.0 keV & 1.0--2.0 keV & 2.0--10.0 keV\\
\hline\hline
All time & $6.6$ & $2.6$ & $2.7$ & $34.5$ & 1.2(74) & 2.4 & 1.9 & $3.5$ & 2.8 & 2.0 & 3.5\\
& $(5.8-6.6)$ & $(2.0-3.3)$ & $(1.9-3.6)$ & $(32.6-36.5)$ & & (2.3--2.4) & (1.8--2.0) & $(3.3-3.7)$ & (2.7--2.8) & (1.9--2.1) & (3.3--3.7)\\
$t_\mathrm{high}$ & $11.3$ & $2.9$ & $3.0$ & $41.8$ & 1.4(15) & 3.2 & 2.4 & 4.2 & 3.7 & 2.4 & 4.2\\
 & $(8.0-14.4)$ & $(1.1-5.0)$ & $(1.1-5.6)$ & $(34.5-49.0)$ & & (2.9--3.5) & (2.1--2.6) & (3.6--4.9) & (3.4--4.2) & (2.2--2.7) & (3.6--4.9)\\
$t_\mathrm{low}$ & $6.0$ & $2.6$ & $2.4$ & $34.4$ & 1.2(66) & 2.3 & 1.9 & $3.4$ & 2.6 & 2.0 &3.4\\
& $(5.2-6.9)$ & $(1.9-3.3)$ & $(1.6-3.3)$ & $(32.4-36.4)$ & & (2.2--2.3) & (1.8--2.0) & (3.2-3.6) & (2.6--2.7) & (1.9--2.0) & (3.2--3.7)\\
\hline
\end{tabular}
}
\end{table*}

In order to make a more detailed comparison (e.g. in terms of the fluxes over 0.5--1.0 keV and 1.0--2.0 keV) between the archival XMM-Newton data and the new NICER data, and also to maintain a uniformity in the data analysis, we reprocess the data with SAS v.19.1. Light curves in 0.5--2.0\,keV (soft) and 2.0--10.0\,keV (hard) bands were extracted. While the background remains stable over the whole exposure, the light curves clearly reveal that the X-ray emission from the system exhibits variability: the $\chi^2_\mathrm{red}$ for the time-stable model is $\approx 4$ for both soft and hard bands with $p$-values of $\sim 10^{-5}$.
There are two clear "enhancements" and the most prominent feature is that around phase 0.09 (Figure \ref{fig:xmm_count_rate_lc}, top three panels), with a half-width of $\approx 0.005$ cycles (0.02 days), much smaller than the half-width of the enhancement observed at periastron ($\sim 0.05$ cycles or 0.2 days, Figure \ref{fig:lightcurves_4T}). We also show the variation of the hardness ratio (HR), defined as the ratio between the count rates over 2--10 keV to those over 0.5--2.0 keV, as a function of orbital phase, which also exhibits a slight enhancement over the same ranges of phases as that for the enhancement in the count rates. The variation in the HR is, however, less significant than those for the count rates, with a $\chi^2_\mathrm{red}$ of 1.9 and $p$-value of 0.03 for the time-stable model. 


We extracted EPIC-pn spectra over the full exposure as well as for times of low and high count rates, defined by pn count rates below and above 0.18\,cts\,s$^{-1}$ ($t_\mathrm{low}$ and $t_\mathrm{high}$ respectively in Table \ref{tab:4T_result_xmm}) in the 0.3--1.0\,keV band. They were fitted by the same models as presented above, and the results are shown in Figure \ref{fig:spec_xmm}. Our results obtained for the full exposure are consistent with those reported by \citet{naze2014}.
We also find that the spectra taken at low and high flux epochs are similar, within errors, except for the overall luminosity. 

We finally compare the flux over 0.5--1.0 keV and 1.0--2.0 keV for these data with those obtained for the NICER data (since for the NICER data, the flux estimates over 2--10 keV are susceptible to imperfect background modeling) in Figure \ref{fig:hardness_ratio}. We re-define the hardness ratio (HR) as the ISM corrected flux at 1.0--2.0 keV to that at 0.5--1.0 keV. We have excluded the observation 10 as we are unable to obtain a good fit even for the energy range of 0.5--2.0 keV. 
We find that the HR is maximum at the phase of the periastron enhancement. The variation is qualitatively similar to that observed for the flux over 0.5--2.0 keV (Figure \ref{fig:lightcurves_4T}), except for that in the HR plot, the XMM-Newton observation (marked with a star) stands out from the out-of-periastron observations with a value of HR close to that observed at periastron. 

To summarize, we find that the X-ray properties of the system at the phase (and epoch) of the archival XMM-Newton observation are different from those observed by NICER. The overall flux is similar to that observed away from periastron, but the spectrum is harder, as found in the periastron observation. Clearly, re-observing the system is required, with better and higher cadence phase coverage, to confirm the phase-locked nature of the changes and better constrain the exact variability details at all phases.

\section{Discussion}\label{sec:discussion}
We observed in \eps~a statistically significant difference between the X-ray flux at periastron and away from periastron (Table \ref{tab:4T_result_merged_fixed_nH}, Figure \ref{fig:lightcurves_4T}) providing strong evidence for the presence of binary interactions.
Before considering the underlying physical scenarios, we first consider the possibility that the system exhibits X-ray enhancements randomly. In the next subsection, we estimate the probability of such a scenario given our observations. In the subsequent subsection, we will discuss the type of binary interactions that can explain our observations.

\subsection{Estimating the probability of random exhibition of X-ray enhancements by \eps}\label{subsec:random_flaring}
\begin{figure}
    \centering
    \includegraphics[width=0.45\textwidth]{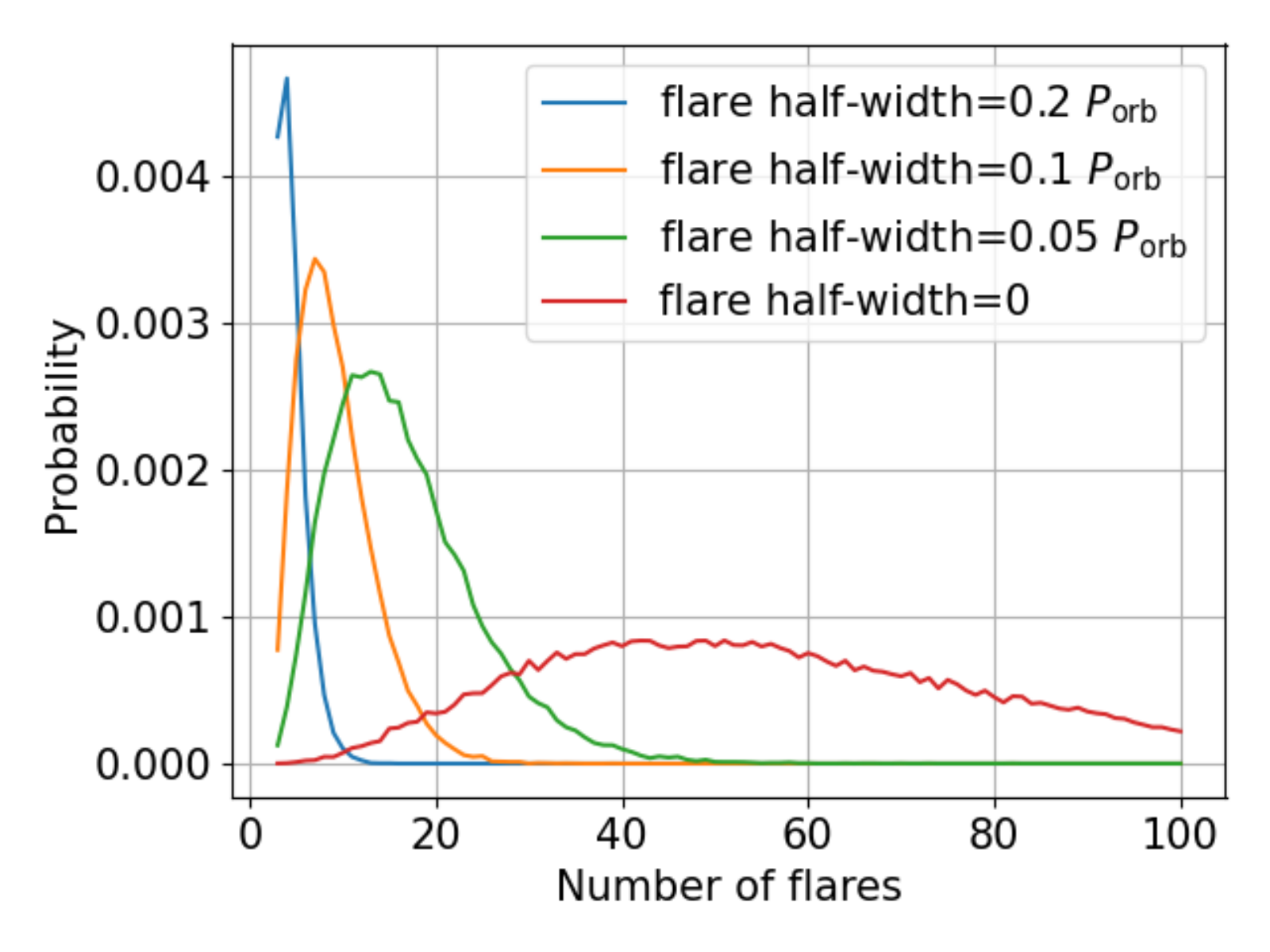}
    \caption{The probability that the observed X-ray enhancements are the results of random flaring by \eps. The $x$-axis shows the total number of flares that we assume to occur within our total observation window. The four curves correspond to four different flare durations. The maximum value of the probablity is $\approx 0.005$.
    }
    \label{fig:FAP}
\end{figure}

Let us assume that the system exhibits X-ray flares randomly without any correlation with orbital phases. Under this scenario, the observation of all three enhancements during our periastron observations would only be a coincidence. We perform a simple test to estimate the plausibility of this situation. 
We consider two parameters: the number of flares $N$ that occur over the total observation window (i.e. 20.48 days; see Table \ref{tab:obs}, column 6), and the flare duration (equivalently, flare half-width). For each combination of $N$ and flare-width, we generate $N$ random flare time-stamps in our whole observing window and check whether these flares fall in the observing windows corresponding to our observations 5, 7 and 9 (and with no flares in the other windows). The assumed duration of each flare is accounted for by extending both boundaries of the time interval corresponding to each observation by that duration. We do this exercise $10^6$ times for each combination of $N$ and flare-width. Each data point in the Figure \ref{fig:FAP} shows the fraction (labelled as `Probability') of these $10^6$ trials that match the observed result (i.e. the flares occur at all three windows corresponding to observations 5, 7 and 9, and no other epochs receive any flare) for each combination.
In all cases, we find that the maximum probability that the observed enhancements are results of random flaring by the system is less than 1\%.
In this context, we would like to mention that no hot magnetic star has been confirmed to exhibit X-ray flaring to the best of our knowledge.


\subsection{Binary interactions in \eps}\label{subsec:binary_interaction}
There are two possible types of interactions relevant here: the first one is the collision between the 
stellar winds from the two stars, and the second one is magnetic reconnection triggered by the anti-aligned configuration of the stellar magnetic fields. 
According to \citet{shultz2015b}, the separation between the two stars are such that their magnetospheres overlap at all orbital phases. This suggests that there could be inter-star magnetospheric interactions at all times, though their strength would vary as a function of orbital phase due to the eccentric orbit, or/and relative spin between the two stars. 

From Figure \ref{fig:lightcurves_4T}, it is clear that although the two observations taken just before periastron have higher X-ray fluxes (over the 0.5--2.0 keV range) than out-of-periastron observations, they display lower X-ray fluxes than those taken at periastron. 
 Thus the enhancement in the X-ray flux is confined to a very narrow orbital phase range of width $\sim 0.05$ cycles, around periastron. A wider and denser coverage of the orbital cycle will be needed to obtain the exact width of that X-ray enhancement and its subsequent evolution. 
Note that the observed enhancement in X-rays is actually seen in observations taken in three different orbital cycles, suggesting that it is a stable characteristics of the system at least over the timescale equal to our total observation duration.

Figures \ref{fig:lightcurves_4T} and \ref{fig:hardness_ratio} reveal that the X-ray properties (flux and hardness ratio) observed for observation 1, which covers the orbital phases 0.95--0.97 (equivalently, $-0.05$ to $-0.03$), are similar to those observed at phases much further away from periastron.
A continuous interaction scenario, where the interaction strength is entirely governed by the binary separation, cannot explain this result. We hence rule out the scenario in which the periastron enhancement is simply a consequence of the fact that the amount of energy released due to inter-star magnetospheric interaction increases as the separation between the two stars decreases. 
The scenario of a wind-wind collision is also ruled out, as it would lead to smooth X-ray variation in orbital phase \citep[e.g.][]{gosset2016}. Besides, no X-ray bright colliding winds have ever been reported from systems with such late-type stars \citep[e.g.][]{rauw2016,naze2011},
which is normal in view of their weak stellar winds.

As mentioned already in the introduction, recurring enhancements in the radio and X-ray light curves at periastron phases have been observed for a few PMS binary systems. Two of the most well-studied such systems are DQ Tau \citep[observed in both X-ray and radio bands, e.g][]{salter2010,getman2011} and V773 Tau \citep[observed only at radio bands, e.g.][]{massi2002,massi2006}.
\citet{getman2011} observed DQ Tau at both periastron (phases 0.95--0.99) and away from periastron (0.66--0.67). They discovered that the X-ray spectrum at periastron is much harder (with a peak plasma temperature of 90 MK, and time-averaged plasma temperature of around 40 MK) than that away from periastron. Qualitatively, this is reminiscent of the case of \eps. 
For both PMS objects, the scenario invoked to explain enhanced flux at/close to periastron is magnetic reconnection triggered by the collision between the two stellar magnetospheres. 
Neither of the two PMS systems is, however, expected to undergo binary magnetospheric interaction at all orbital phases. V773 Tau has a moderately eccentric orbit ($e=0.3$) similar to \eps, but has a much longer orbital period of $\approx 51$ days \citep{welty1995}. DQ Tau, on the other hand, has an orbital period of $\approx 16$ days (shorter than for V773 Tau but longer than that of \eps) and a high eccentricity \citep[$e=0.57$,][]{czekala2016}. Thus in these two systems, the combination of eccentricity and orbital periods makes the combined effects of binarity and magnetism observable only at periastron phases. Another important point is the stability of the enhancements. For both V773 Tau and DQ Tau, the enhancement properties, such as amplitude and orbital phase of enhancement, vary with epoch of observation
\citep[e.g.][]{massi2002,getman2022}, whereas for \eps~, the periastron enhancement appears to be stable. Interestingly, for V773 Tau and DQ Tau also, the observed enhancements at periastron were found to span a very small fraction of the orbital periods \citep[e.g.][]{massi2006,getman2011}. For V773 Tau, \citet{massi2006} invoked unstable magnetic configurations (`helmet streamer') that develop at one star and interact with the corona of the other star at the periastron passage. In this framework, the duration of the flare is determined by the duration of the interaction. For the same system, \citet{adams2011} considered stable anti-aligned dipole configurations (similar to the case of \eps), and proposed that the observed variability in the emission is due to the change in the magnetic energy stored in the system due to the eccentric orbit. Their theoretical analysis predicts a gradual release of magnetic energy, with the maximum emission occurring $\sim 4$ days prior to periastron (for V773 Tau, 4 days is equivalent to 0.08 of the orbital period). This is inconsistent with the much smaller time-width of the enhancements observed for V773 Tau. This apparent discrepancy is resolved by considering the fact that although the deformation of the magnetic field (accumulation of magnetic stress) is continuous and gradual, the relief of that stress resulting in flares can happen over much smaller timescales \citep{adams2011}. 

We suggest that the scenario proposed by \citet{adams2011} for V773 Tau is likely the one responsible for the observed X-ray enhancements from \eps~at periastron. 
Using their Eq. 17, we estimate the maximum power available as a result of magnetospheric interaction to be $\approx 4\times 10^{32}$ erg $\mathrm{s^{-1}}$ \citep[using the stellar parameter values reported by][]{shultz2015b, pablo2019}, corresponding to a flux
$\sim 10^{-10}\,\mathrm{erg\,cm^{-2}\,s^{-1}}$ \citep[using the distance to \eps~as 156 parsec,][]{pablo2019}, which is sufficient to drive the observed X-ray enhancements ($\sim 10^{-12}\,\mathrm{erg\,cm^{-2}\,s^{-1}}$).
The energy released due to magnetic reconnection is also likely to produce non-thermal electrons, which can be probed via radio observations. Thus, future radio observations of \eps~will provide important clues towards understanding the magnetospheric interaction scenario.

Note that in the case of V773 Tau, the rotation periods of the individual stars are known to be smaller than 3 days, i.e., much smaller than the orbital period. For \eps, the rotation periods of individual stars are unknown. Recently, \citet{cherkis2021} showed that both rotational and orbital timescales can be important in determining the timescale of energy release in magnetically coupled stellar binaries.  
The enhancements seen during the archival XMM-Newton observation could also be linked to the same physical phenomenon as the one responsible for the periastron enhancement, and  its occurrence could be a result of relative motion between the magnetospheres due to both rotation about their individual axes and revolution around each other.

\section{Conclusion}\label{sec:conclusion}
In this paper we report evidence of magnetospheric interactions via reconnection between the two components of the only known short-period magnetic massive star binary system \eps. Our observation of the star with the NICER instrument clearly shows that the system produces more X-ray emission at periastron than away from it. The X-ray enhancement is confined to a very narrow orbital phase range of width $\sim 0.05$ cycles. 
In addition, by reprocessing archival X-ray data for the system, we find that there are enhancements with timescales of $\approx 0.005$ orbital cycles, at orbital phases away from periastron, although the persistence of these enhancements remain to be examined.
We conclude that the most favorable scenario is magnetic reconnection that gets triggered due to the relative motion of the magnetospheres at certain orbital configurations. In the future, denser sampling of the orbital cycle will be crucial to understand this unique binary system.


\section*{Acknowledgements} 
We thank the referee for the useful comments and suggestions that help us to improve the manuscript.
BD acknowledges support from the Bartol Research Institute. 
BD thanks Patrick Stanley for useful help in TESS data analysis.
BD thanks Surajit Mondal for useful discussions.
BD and VP acknowledge support by NASA NICER Guest Observer Program 80NSSC21K0134.
YN acknowledges support from the Fonds National de la Recherche Scientifique (Belgium), the European Space Agency (ESA) and the Belgian Federal Science Policy Office (BELSPO) in the framework of the PRODEX Programme (contract linked to XMM-Newton).
MFC acknowledges support by NASA under award number 80GSFC21M0002.
DHC acknowledges support from the NASA Chandra grants TM4-15001B and AR6-17002A.
ADU is supported by NASA under award number 80GSFC21M0002.
MAL acknowledges support from NASA’s astrophysics division.
MES acknowledges support from the Annie Jump
Cannon Fellowship, supported by the University of Delaware and endowed by the Mount Cuba Astronomical Observatory. 
AuD acknowledges support by the National Aeronautics and Space
Administration under Grant No. 80NSSC22K0628 issued through the
Astrophysics Theory Program and Chandra Award Numbers TM-22001 and
GO2-23003X, issued by the Chandra X-ray Center, which is operated by the
Smithsonian Astrophysical Observatory for and on behalf of the National
Aeronautics Space Administration under contract NAS8-03060.
GAW acknowledges Discovery Grant support from the Natural Sciences and Engineering Research Council (NSERC) of Canada. 

\section*{Data availability}
The X-ray data used in this work are available from the HEASARC archive (\url{https://heasarc.gsfc.nasa.gov/db-perl/W3Browse/w3browse.pl}), and the TESS data are available from the MAST archive (\url{https://mast.stsci.edu/portal/Mashup/Clients/Mast/Portal.html}). The analyzed data are available upon request.



\bibliographystyle{mnras}
\bibliography{das} 


\appendix

\section{The 2T model}\label{app_sec:2T}
\begin{figure*}
    \centering
    \includegraphics[width=0.48\textwidth]{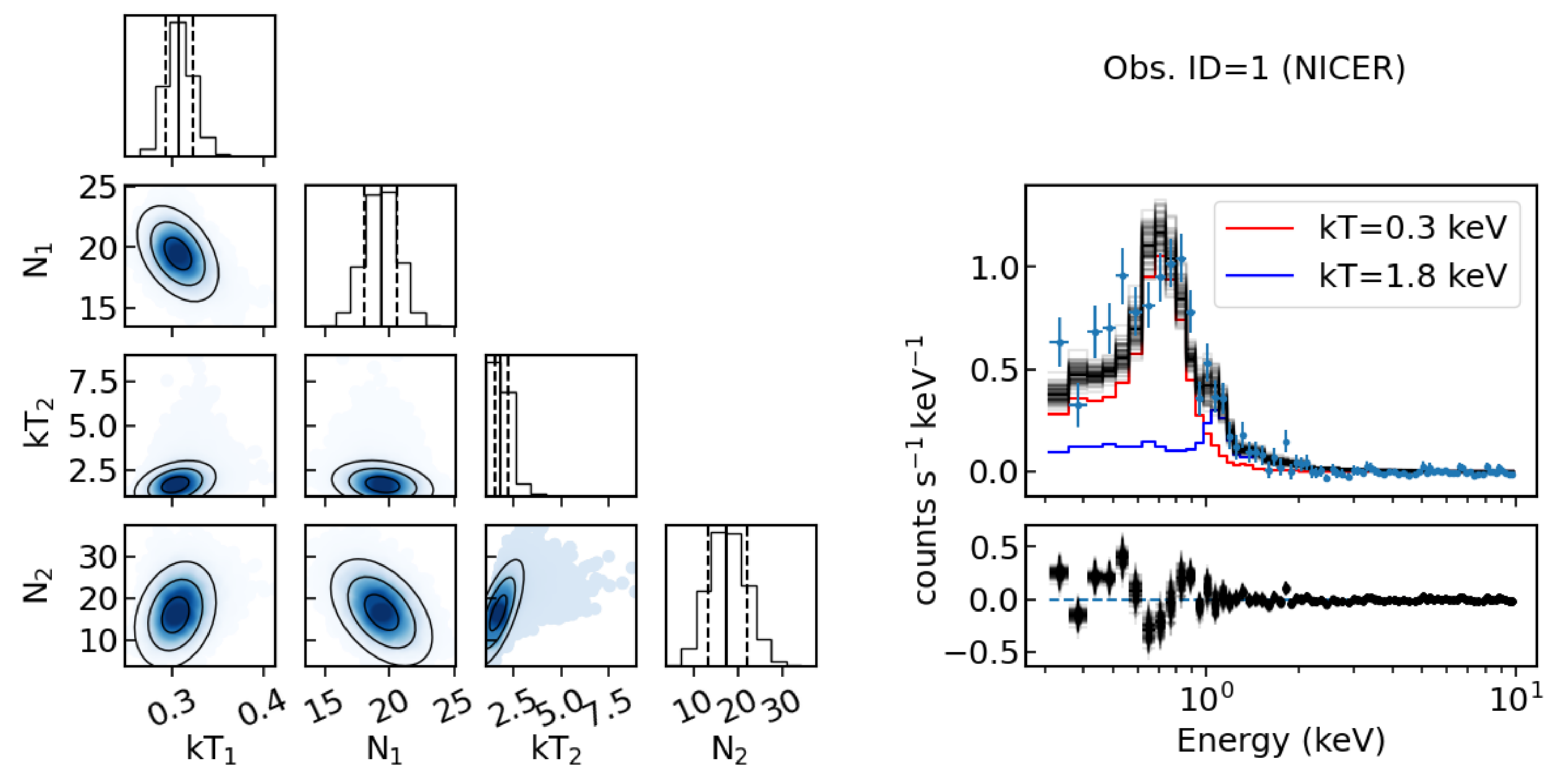}\hspace{0.5cm}
    \includegraphics[width=0.48\textwidth]{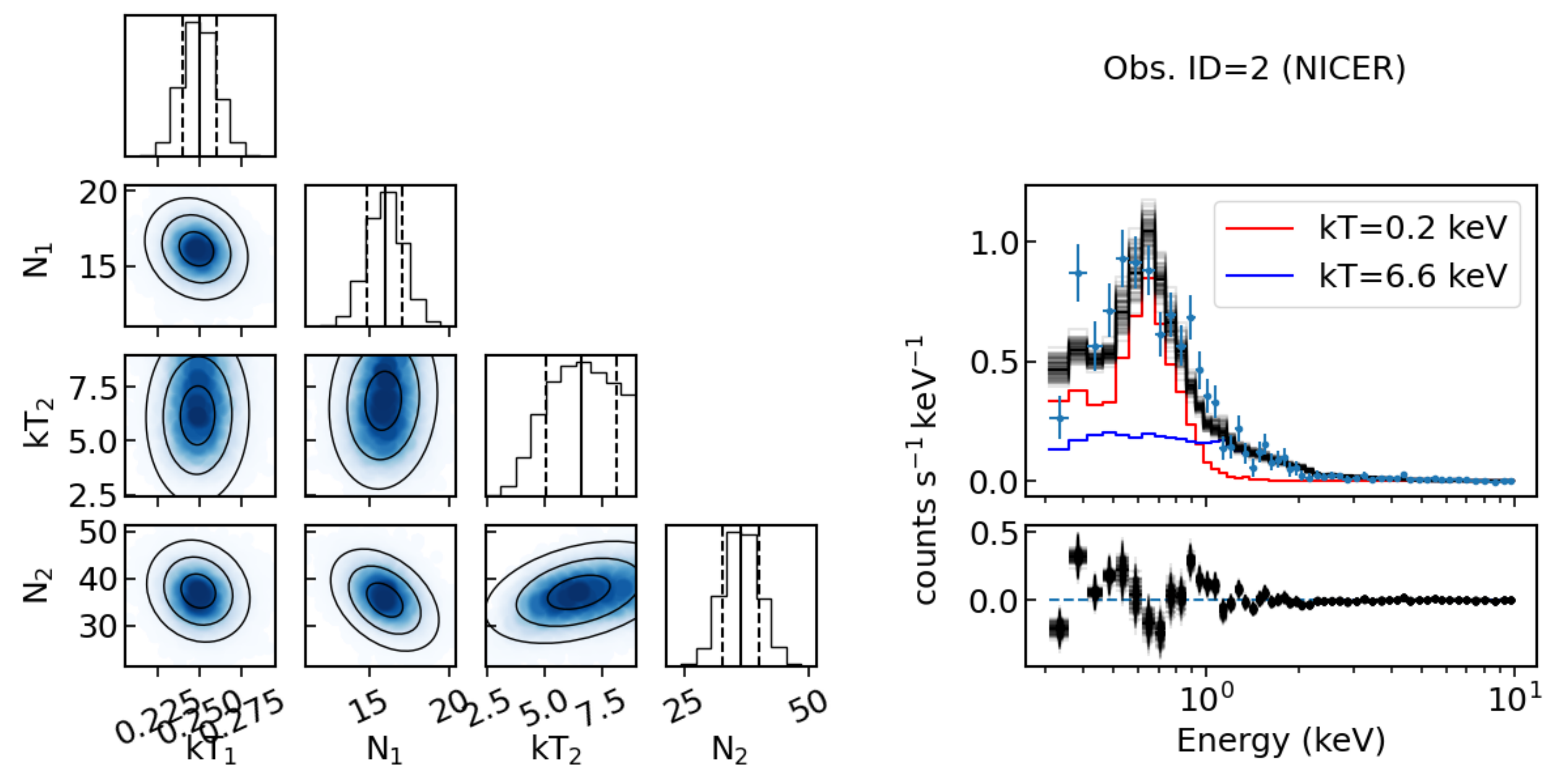}
    \includegraphics[width=0.48\textwidth]{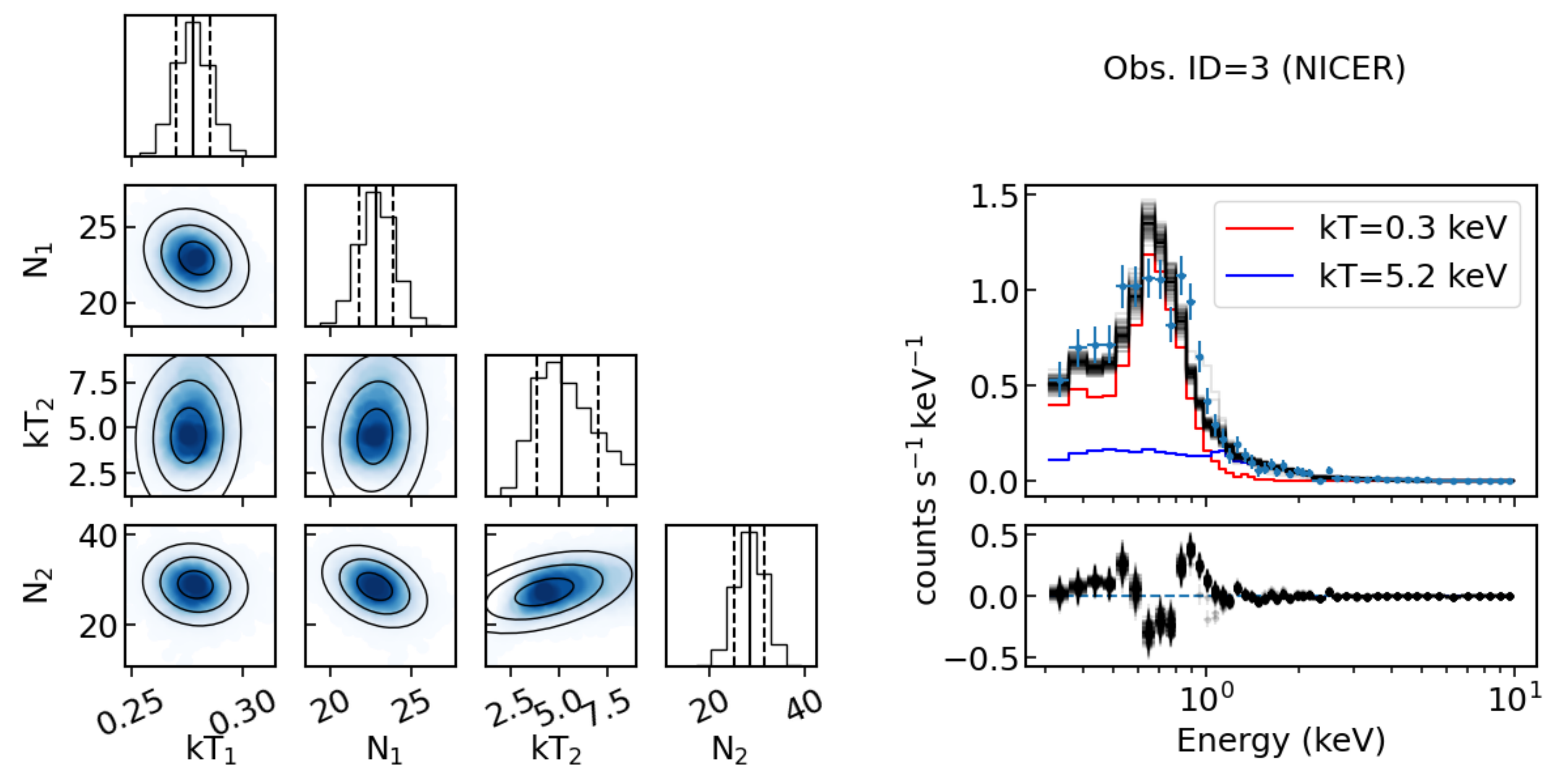}\hspace{0.5cm}
    \includegraphics[width=0.48\textwidth]{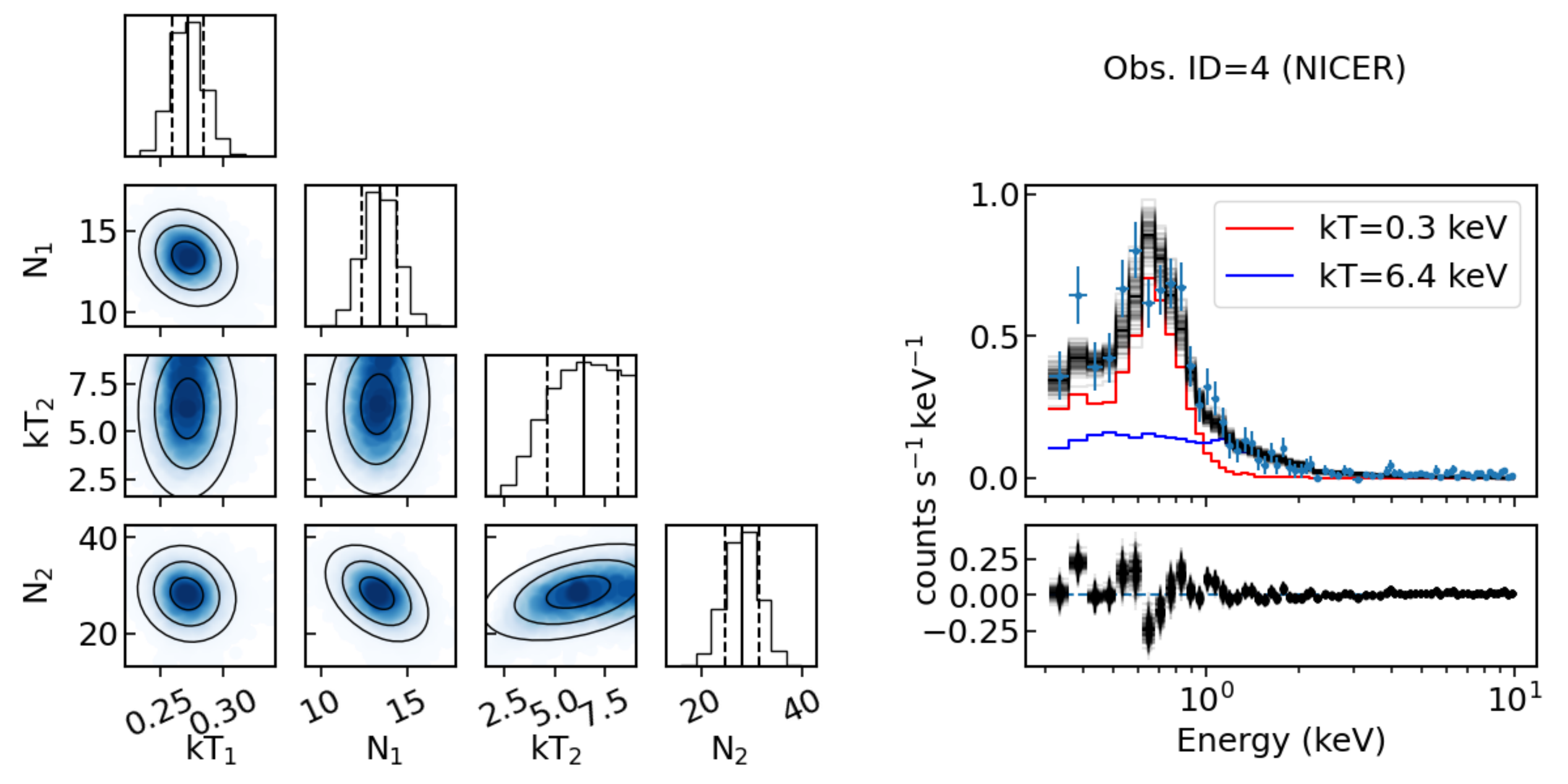}
    \includegraphics[width=0.48\textwidth]{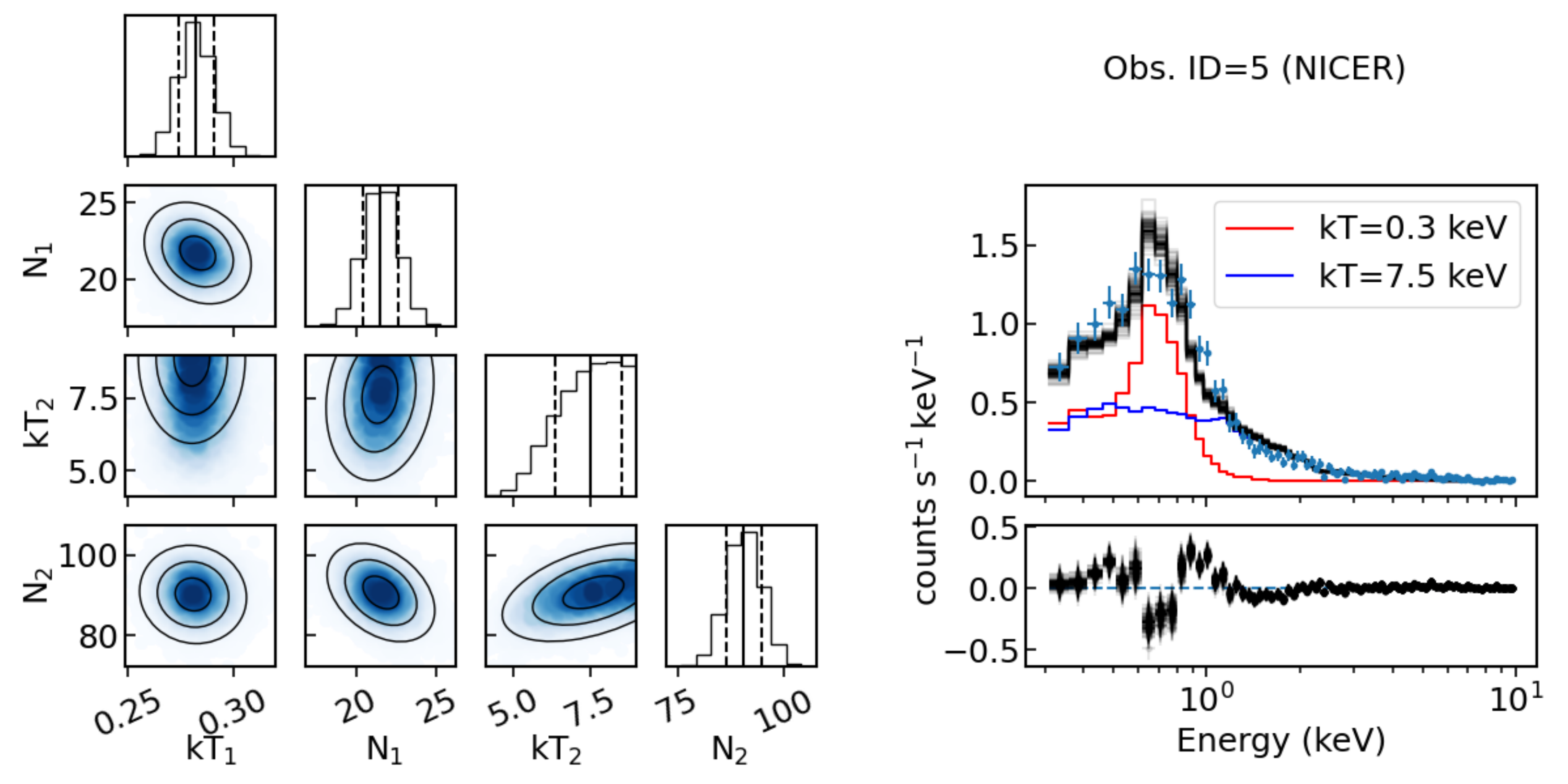}\hspace{0.5cm}
    \includegraphics[width=0.48\textwidth]{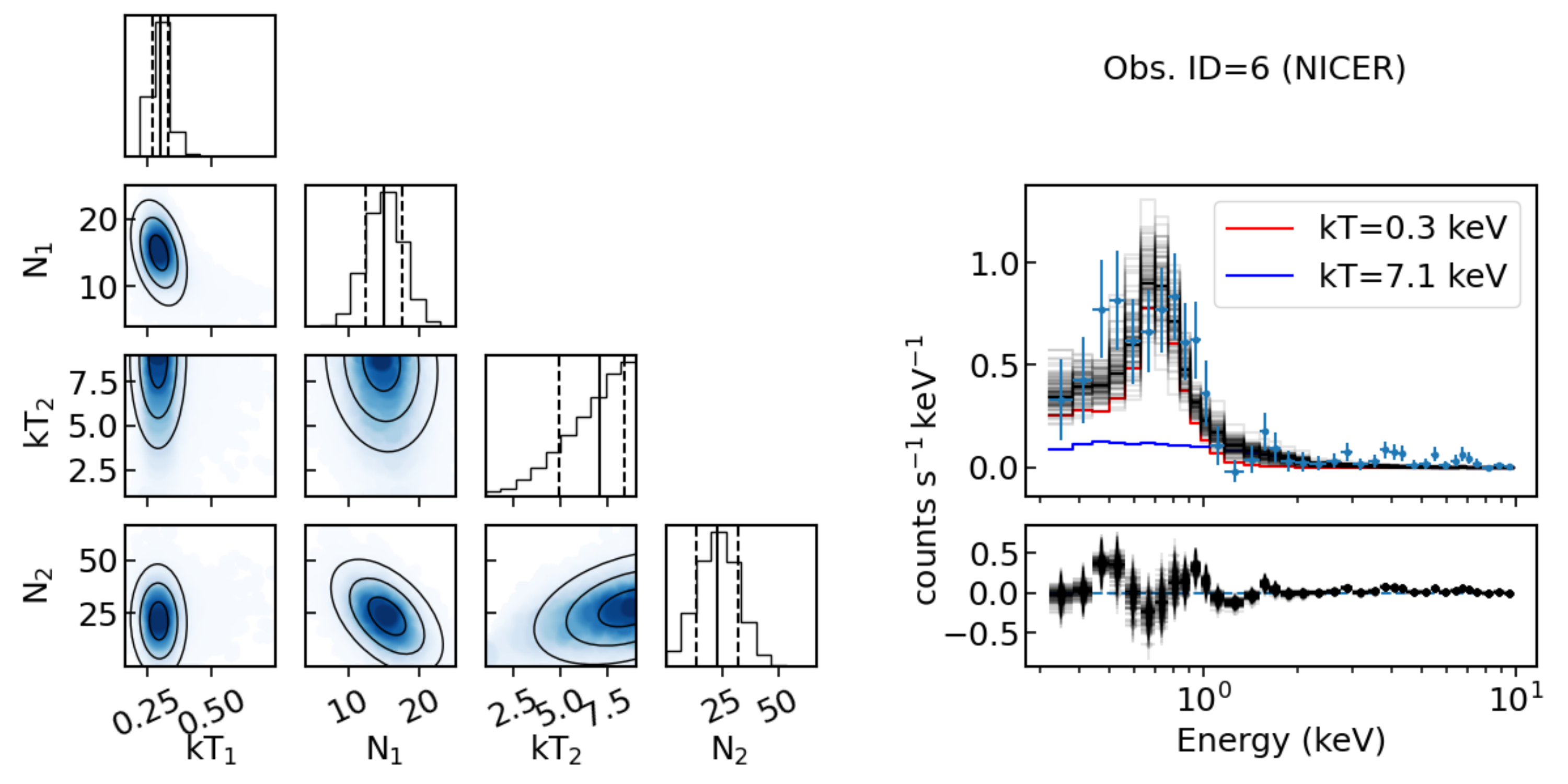}
    \includegraphics[width=0.48\textwidth]{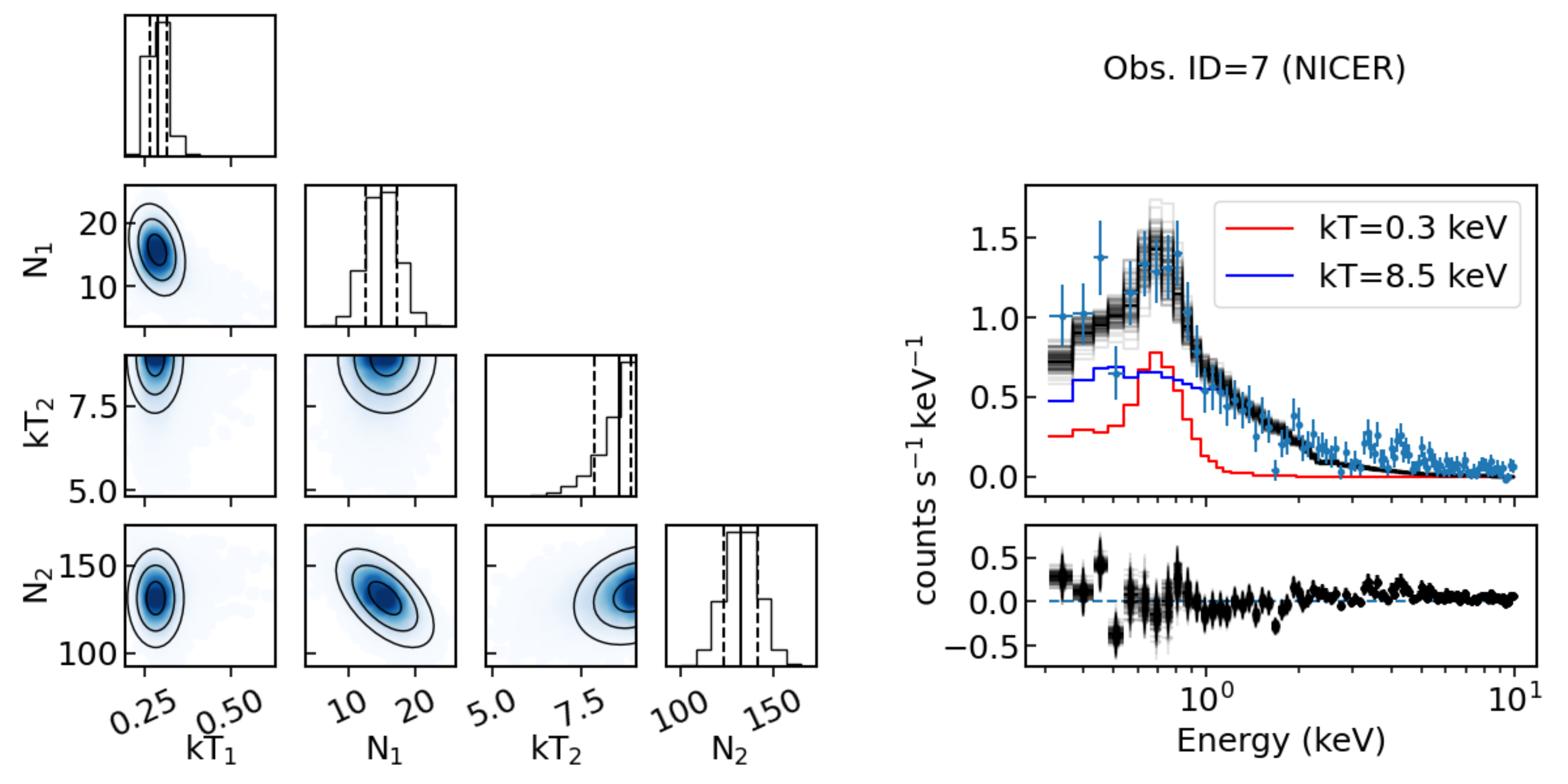}\hspace{0.5cm}
    \includegraphics[width=0.48\textwidth]{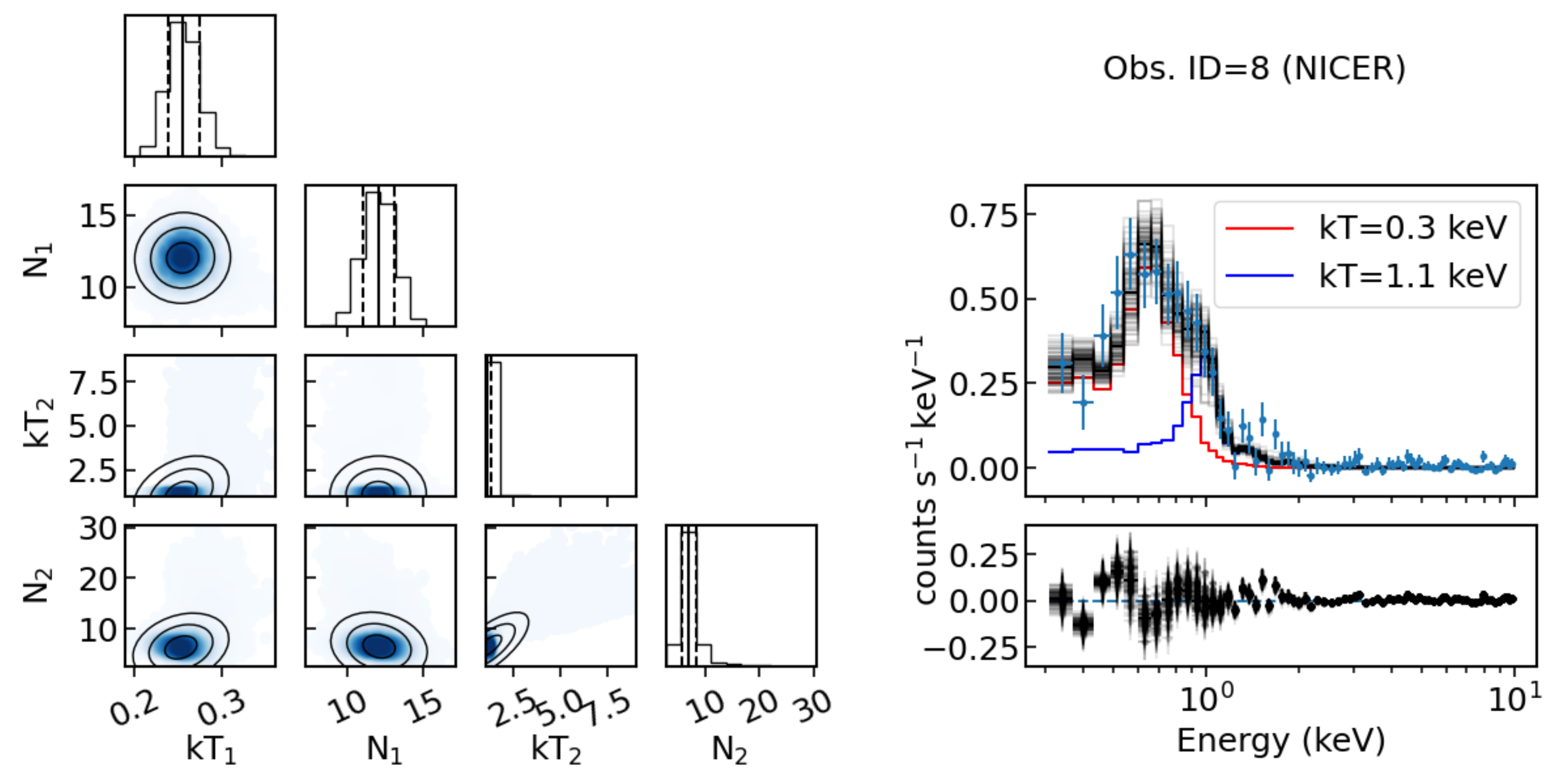}
    \includegraphics[width=0.48\textwidth]{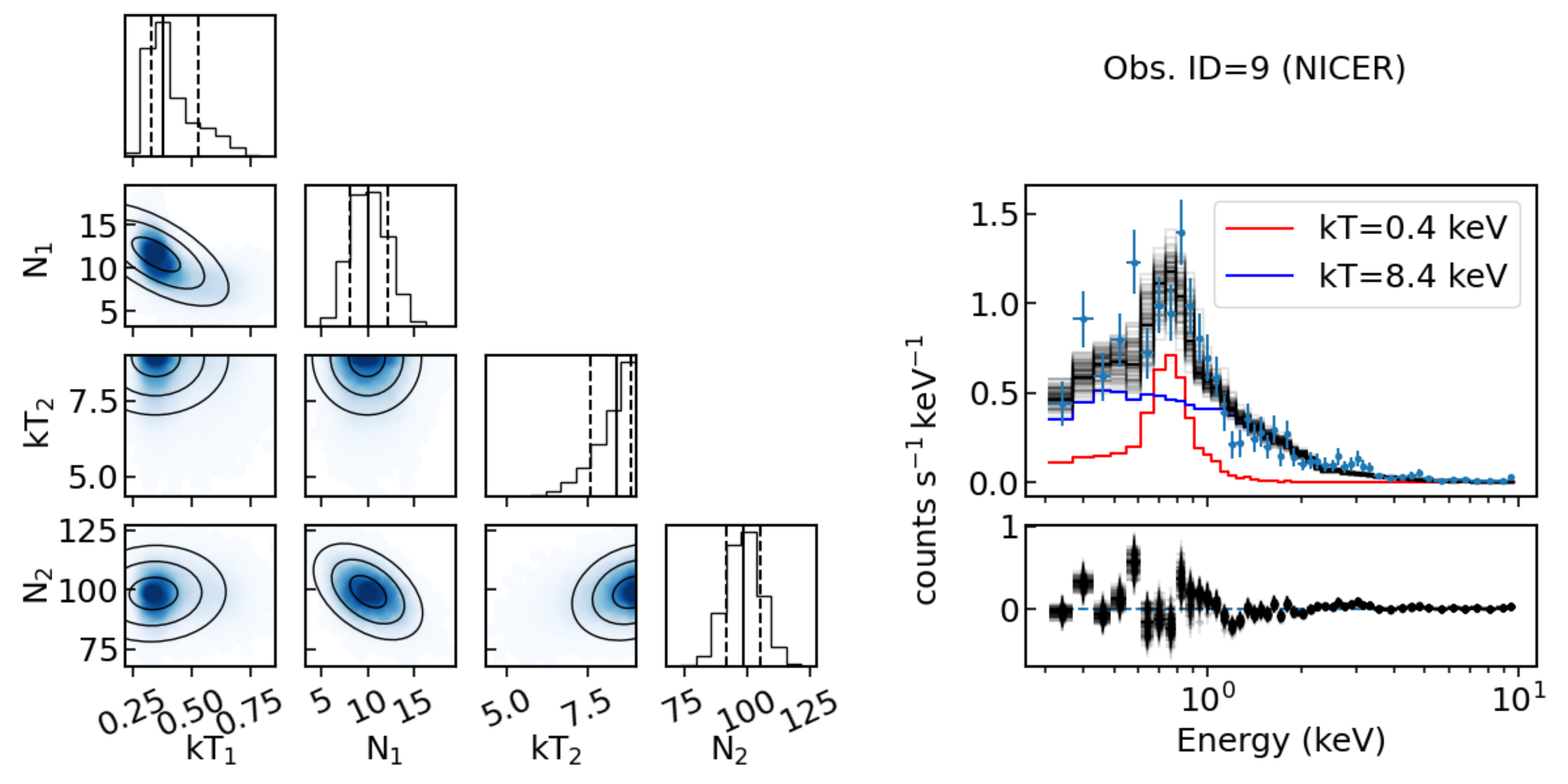}\hspace{0.5cm}
    \includegraphics[width=0.48\textwidth]{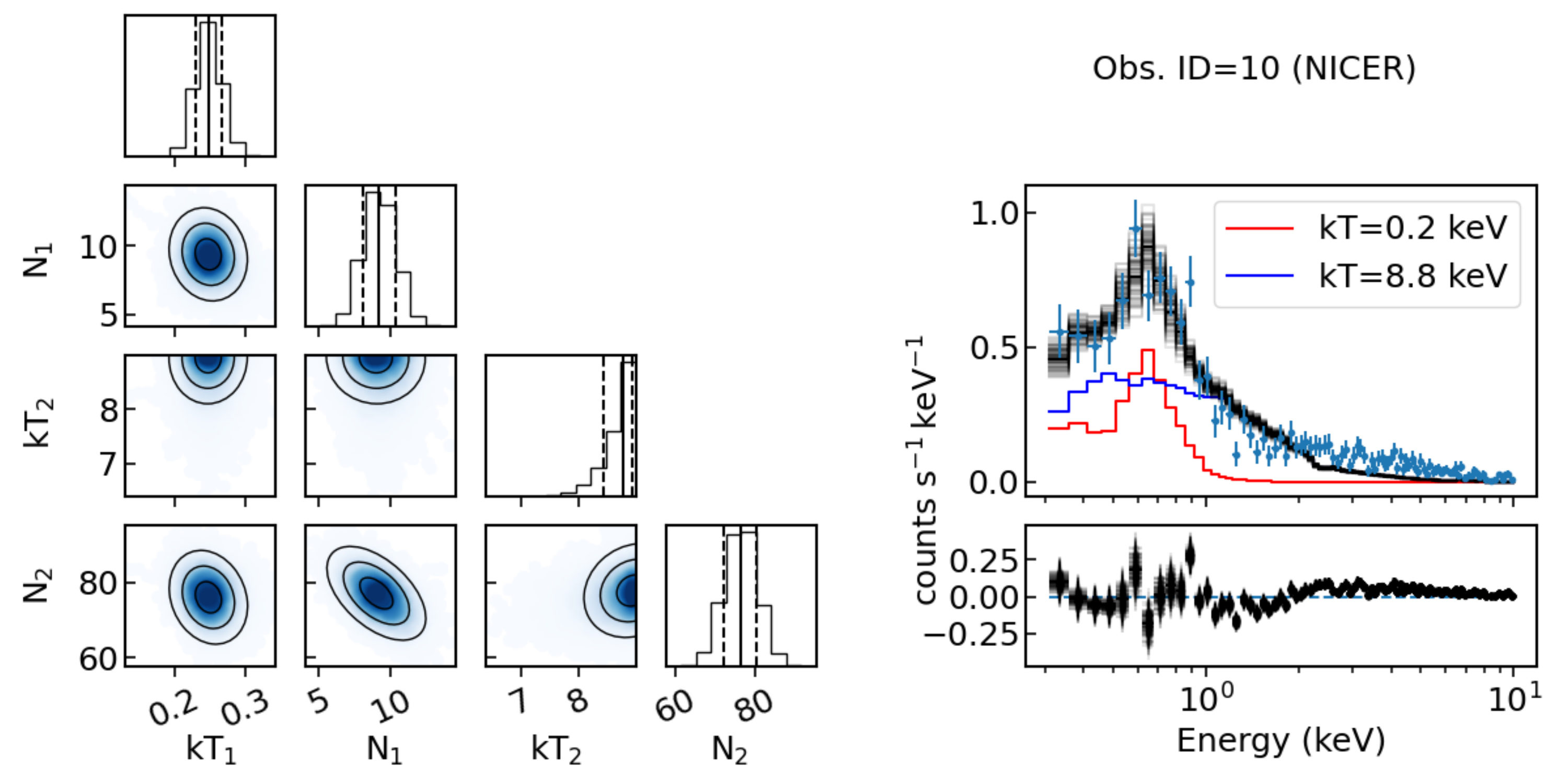}
    \caption{Spectral analysis of individual observations for the 2T model with $n_\mathrm{H}$ fixed at $0.03\times 10^{22}$\,cm$^{-2}$. The Obs. IDs correspond to the IDs listed in Table \ref{tab:obs}. The solid vertical lines on the histograms mark the median values, and the vertical dashed lines mark the 68\% confidence intervals. $\mathrm{N}_i$s ($i=1,2$) are proportional to the norms at the two plasma temperatures given by $\mathrm{kT_1}$ and $\mathrm{kT_2}$ respectively, with $\mathrm{N}_i=\mathrm{Norm}_i\times 10^5$. See \S\ref{app_sec:2T} for details.}
    \label{fig:2T_fixed_nH_mcmc_spectra}
\end{figure*}

The results of the fitting are listed in Table \ref{tab:2T_result}. During the fitting process, we provided an initial guess values of temperatures as 0.3 and 3.0 keV \citep[the temperatures obtained by][using the same model]{naze2014}, and also set a minimum and maximum values as 0.1--0.9 keV and 1.0--9.0 keV for the cooler and the hotter plasma components respectively. We fix $n_\mathrm{H}$ to $0.03\times 10^{22}$\,cm$^{-2}$. Although we obtained reasonable reduced $\chi^2$ for most of the spectra, it is often at the cost of inferring extremely high temperature for the hotter plasma component which is unphysical. In addition, the MCMC analysis (Figure \ref{fig:2T_fixed_nH_mcmc_spectra}) shows that for many of the spectra, the temperature of the hotter component cannot be constrained within the limit provided (1.0--9.0 keV).

Due to the above issues, we decided not to use this model for further analysis.

\begin{table*}
{\tiny 
\caption{The spectral fitting results for individual observations using the 2T model with $n_\mathrm{H}$ fixed at interstellar value (\S\ref{sec:results}). The energy range used during the fitting process is 0.3--10.0 keV. For each observation number, the first row gives the median values and the second row gives the 68\% confidence interval from the posterior probability distribution marginalized for a given parameter.\label{tab:2T_result}}
\begin{tabular}{c|cccc|c|cccccc}
\hline
ID & $\mathrm{kT_1}$ & $\mathrm{Norm_1}$ & $\mathrm{kT_2}$ & $\mathrm{Norm_2}$ & $\chi^2_\mathrm{red}$ & \multicolumn{6}{c}{Flux $\left(\times 10^{-13}\,\mathrm{erg\,cm^{-2}\,s^{-1}}\right)$}\\
  &  &  &  & & & \multicolumn{3}{c}{Observed} & \multicolumn{3}{c}{ISM corrected}\\
 & (keV) & ($\times 10^{-5}\mathrm{cm^{-5}}$) & (keV) & ($\times 10^{-5}\mathrm{cm^{-5}}$) & & 0.5--1.0 keV & 1.0--2.0 keV & 2.0--10.0 keV & 0.5--1.0 keV & 1.0--2.0 keV & 2.0--10.0 keV\\
 \hline\hline
1 & 0.31 & 19.3 & 1.8 & 17.4 & 1.4(74) & 3.3 & 1.4 & 0.8 & 3.9 & 1.5 & 0.8\\ 
& (0.29--0.32) & (18.0--20.6) & 1.5--2.2 & (13.2--21.9) &  & (3.2--3.4) & (1.3--1.6) & (0.5--1.2) & (3.7--4.0) & (1.4--1.7) & (0.5--1.2)\\
2 & 0.25 & 16.0 & 6.6 & 36.3 & 1.4(49) & 2.8 & 1.5 & 4.6 & 3.3 & 1.6 & 4.6\\
& (0.24--0.26) & (14.8--17.1) & (5.1--8.1) & (32.7--39.9) & & (2.7--2.9) & (1.4--1.6) & (3.9--5.2) & (3.2--3.5) & (1.5--1.7) & (3.9--5.3)\\
3 & 0.28 & 22.8 & 5.2 & 28.3 &  1.9(43) & 3.7 & 1.4 & 3.3 & 4.4 & 1.5 & 3.3\\
& (0.27--0.28) & (21.7--23.9) & (3.8--7.1) & (25.2-31.3) & & (3.6--3.9) & (1.4--1.5) & (2.5--4.1) & (4.3--4.6) & (1.4--1.6) & (2.5--4.1)\\
4 & 0.27 & 13.4 & 6.4 & 28.1 & 1.4(64) & 2.4 & 1.2 & 3.7 & 2.8 & 1.3 & 3.7\\
& (0.26--0.28) & (12.4--14.4) & (4.6--8.1) & (24.7--31.4) & & (2.3--2.5) & (1.1--1.4) & (2.9--4.2) & (2.7--3.0) & (1.2--1.4) & (2.9--4.2)\\
5 & 0.28 & 21.5 & 7.5 & 90.5 & 1.8(87) & 4.8 & 3.6 & 12.2 & 5.7 & 3.8 & 12.3\\
& (0.27--0.29) & (20.4--22.6) & (6.3--8.5) & (86.5--94.6) & & (4.7--5.0) & (3.5--3.8) & (11.3--13.0) & (5.5--5.8) & (3.6--3.9) & (11.4--13.1)\\
6 & 0.30 & 15.0 & 7.1 & 23.1 & 1.3(33) & 2.6 & 1.1 & 2.8 & 3.1 & 1.2 & 2.8\\
& (0.27--0.33) & (12.4--17.6) & (4.9--8.4) & (13.8--31.8) & & (2.4--2.9) & (0.8--1.4) & (1.3--4.1) & (2.8--3.4) & (0.8--1.5) & (1.3--4.1) \\
7 & 0.29 & 14.9 & 8.5 & 132.6 & 2.0(89) & 4.8 & 5.0 & 18.5 &  5.6 & 5.2 & 18.5\\
& (0.27--0.31) & (12.5--17.3) & (7.8--8.9) & (123.4--142.0) & & (4.5--5.0) & (4.7--5.3) & (17.2--19.8)& (5.3--5.8) & (4.9--5.5) & (17.2--19.9)\\
8 & 0.26 & $1.2$ & 12.0 & 6.6 & 0.8(68) & 2.1 & 0.7 & 0.2 & 2.5 & 0.8 & 0.2\\
& (0.24--0.27) & (11.0--13.1) & (1.0--1.3) & (5.5--8.1) & & (2.0--2.3) & (0.6--0.8) & (0.1--0.2) & (2.3--2.7) & (0.7--0.9) & (0.1--0.2)\\
9 & 0.38 & 10.1 & 8.4 & 98.2 & 1.9(44) & 3.6 & 3.9 & 13.7 & 4.2 & 4.0 & 13.7\\
& (0.33--0.53) & (8.1--12.2) & (7.6--8.8) & (91.6--105.1) & & (3.4--3.8) & (3.7--4.1) & (12.7--14.7) & (4.0--4.5) & (3.8--4.3) & (12.7--14.7)\\
10 & 0.25 & 9.2 & 8.8 & 76.4 & 3.4(86) & 2.7 & 2.8 & 10.8 & 3.2 & 2.9 & 10.8\\
& (0.23--0.26) & (8.1--10.3) & (8.4--8.9) & (72.2-80.5) & & (2.6--2.8) &  (2.7--3.0) & (10.2--11.40 & (3.0--3.3) & (2.8--3.1) & (10.2--11.5)\\
\hline
\end{tabular}
}
\end{table*}

\section{MCMC results and spectra}\label{sec:mcmc_results}
This section shows the MCMC plots.

\begin{figure}
    \centering
    \includegraphics[trim={0.2cm 0cm 14cm 0cm},clip,width=0.235\textwidth]{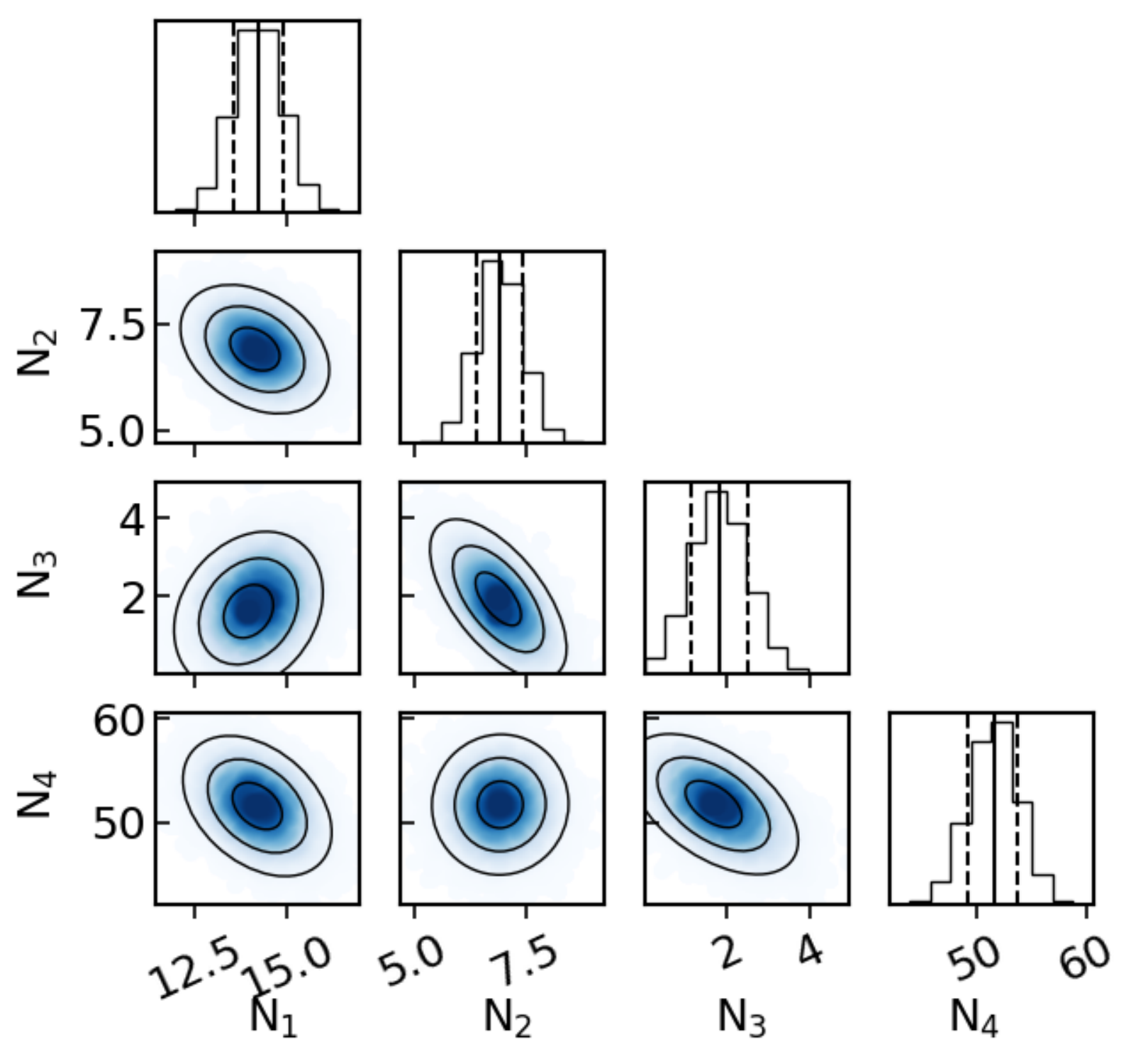}
    \includegraphics[trim={0.2cm 0cm 14cm 0cm},clip,width=0.235\textwidth]{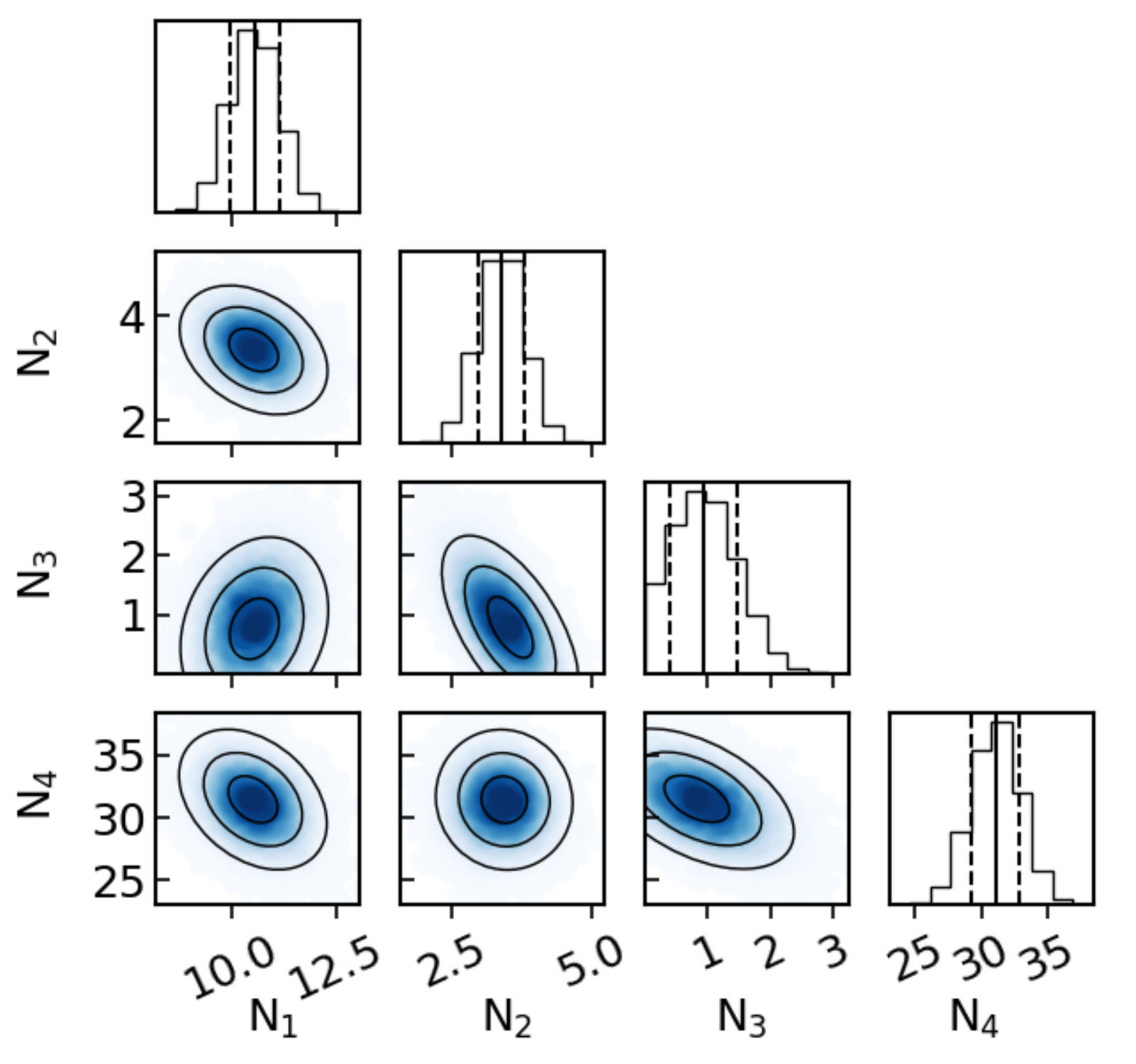}
    
    \vspace{0.5cm}

    \includegraphics[trim={0.2cm 0cm 14cm 0cm},clip,width=0.235\textwidth]{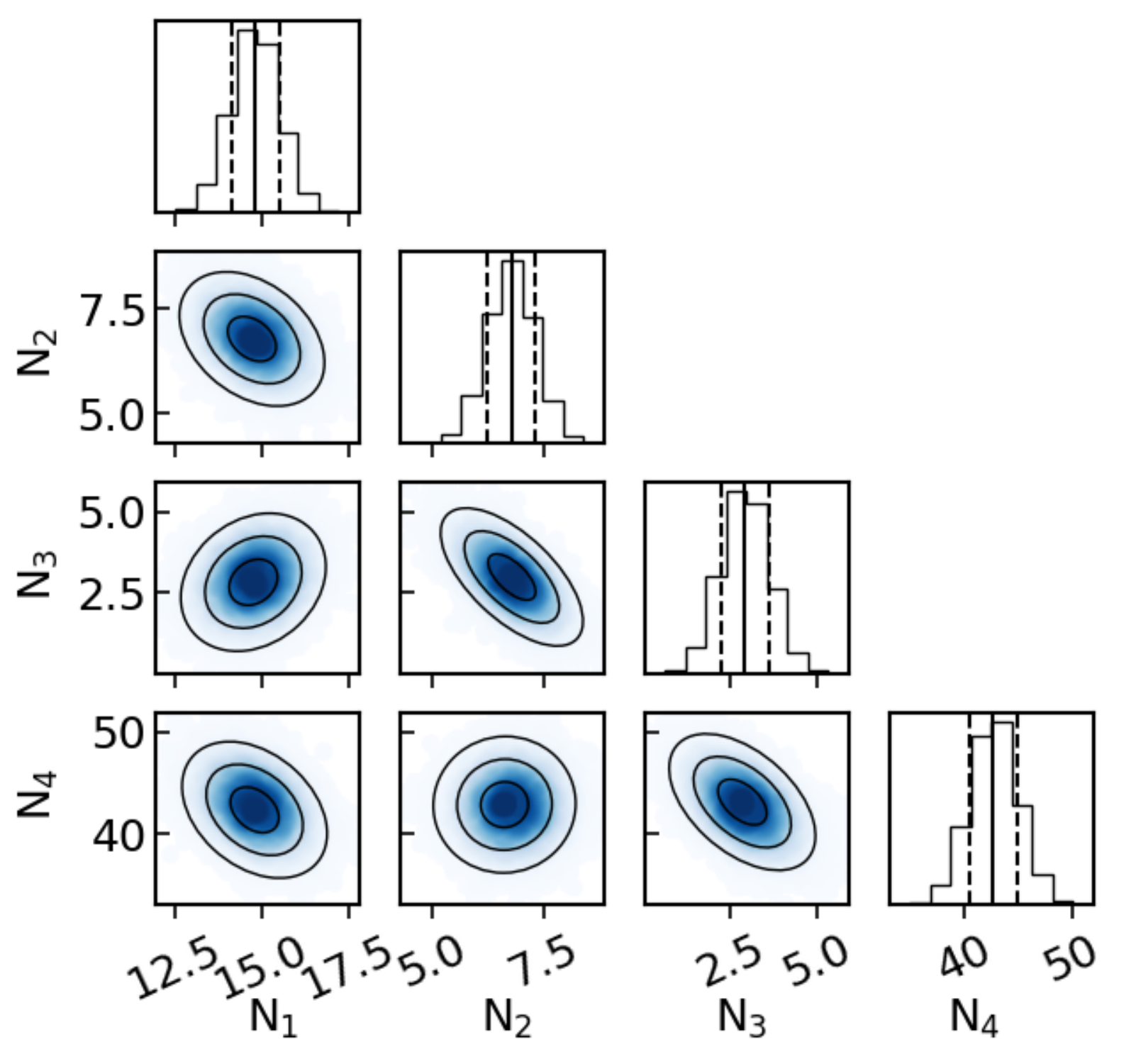}
    \includegraphics[trim={0.2cm 0cm 14cm 0cm},clip,width=0.235\textwidth]{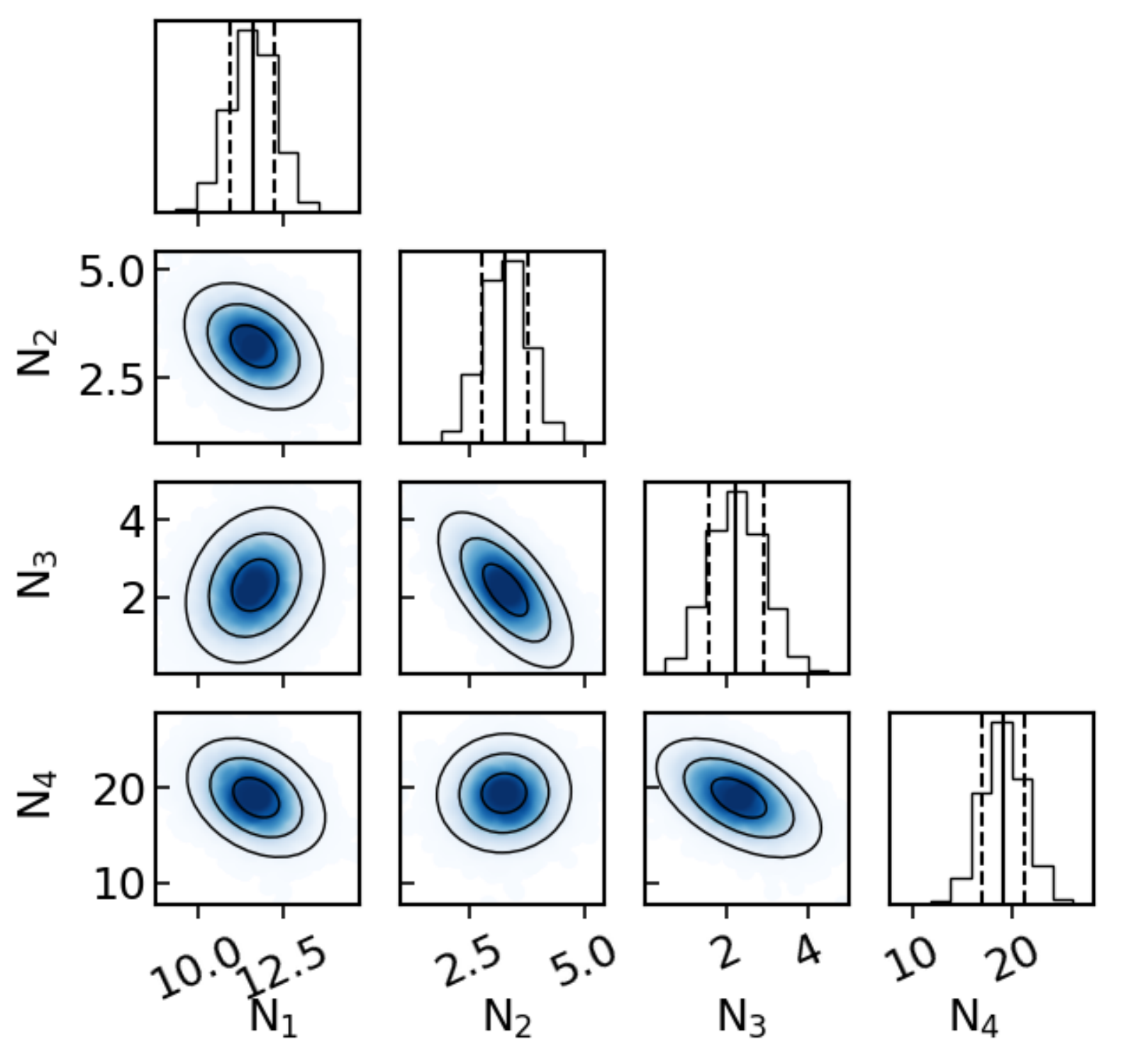}
    \caption{Corner plots corresponding to Figure \ref{fig:merged_spectra}. The top left panel corresponds to average periastron observation obtained by merging observations 1, 3, 5, 7 and 9; the top right panel corresponds to average out-of-periastron observation obtained by merging observations 2, 4, 6, 8 and 10; the bottom left panel corresponds to average periastron observation obtained by merging all the periastron observations excluding 7; and the bottom right panel corresponds to average out-of-periastron observation obtained by merging all the out-of-periastron observations excluding 10. The model used is the 4T model with $n_\mathrm{H}$ fixed at its interstellar value. $\mathrm{N}_i$s ($i=1,2$) are proportional to the normalization factors (norms) at 0.2, 0.6, 1.0 and 4.0 keV respectively, with $\mathrm{N}_i=\mathrm{Norm}_i\times 10^5$.
    \label{fig:merged_spectra_corner_plots}}
\end{figure}

\begin{figure*}
    \centering
    \includegraphics[width=0.48\textwidth]{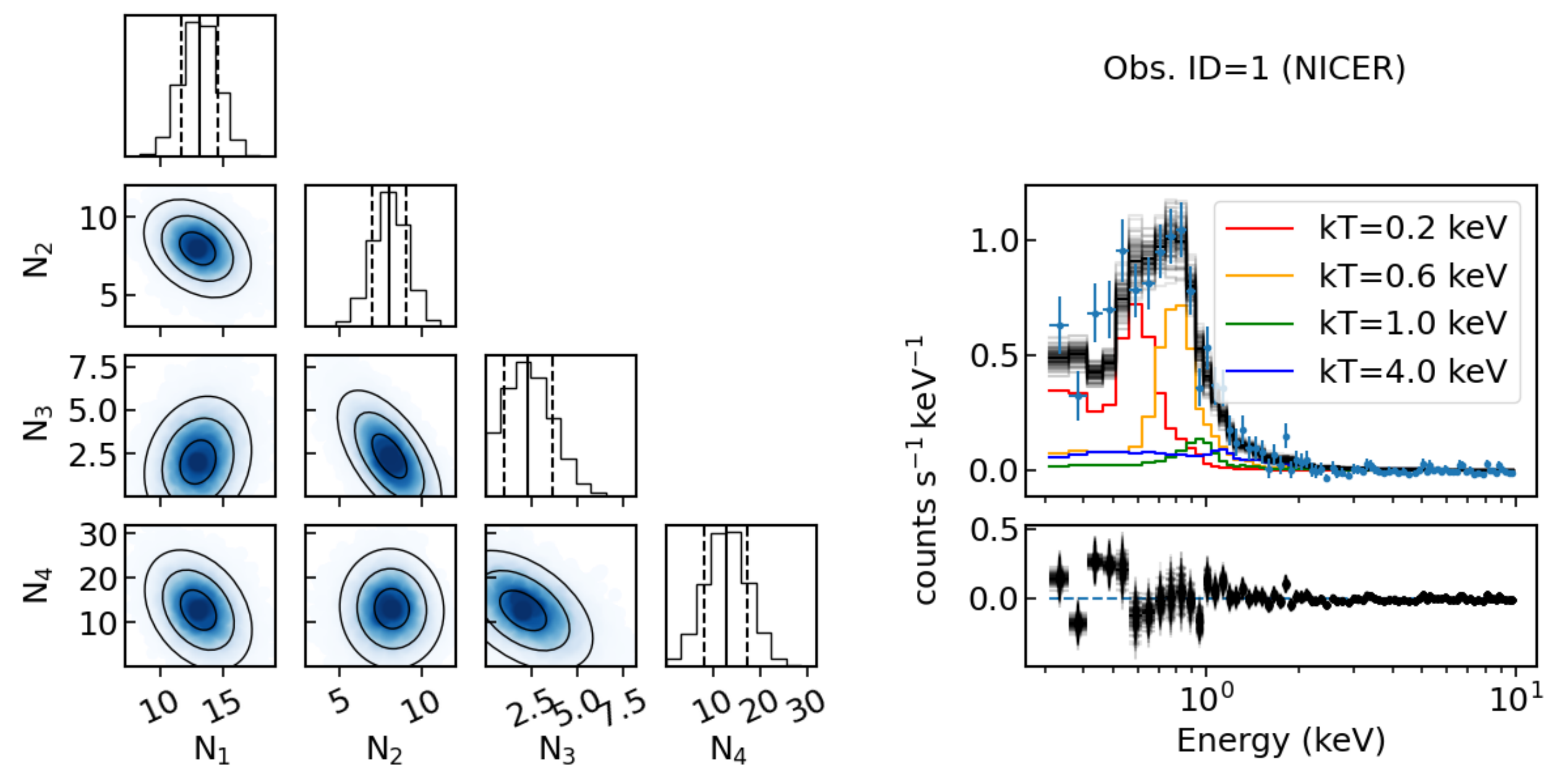}\hspace{0.5cm}
    \includegraphics[width=0.48\textwidth]{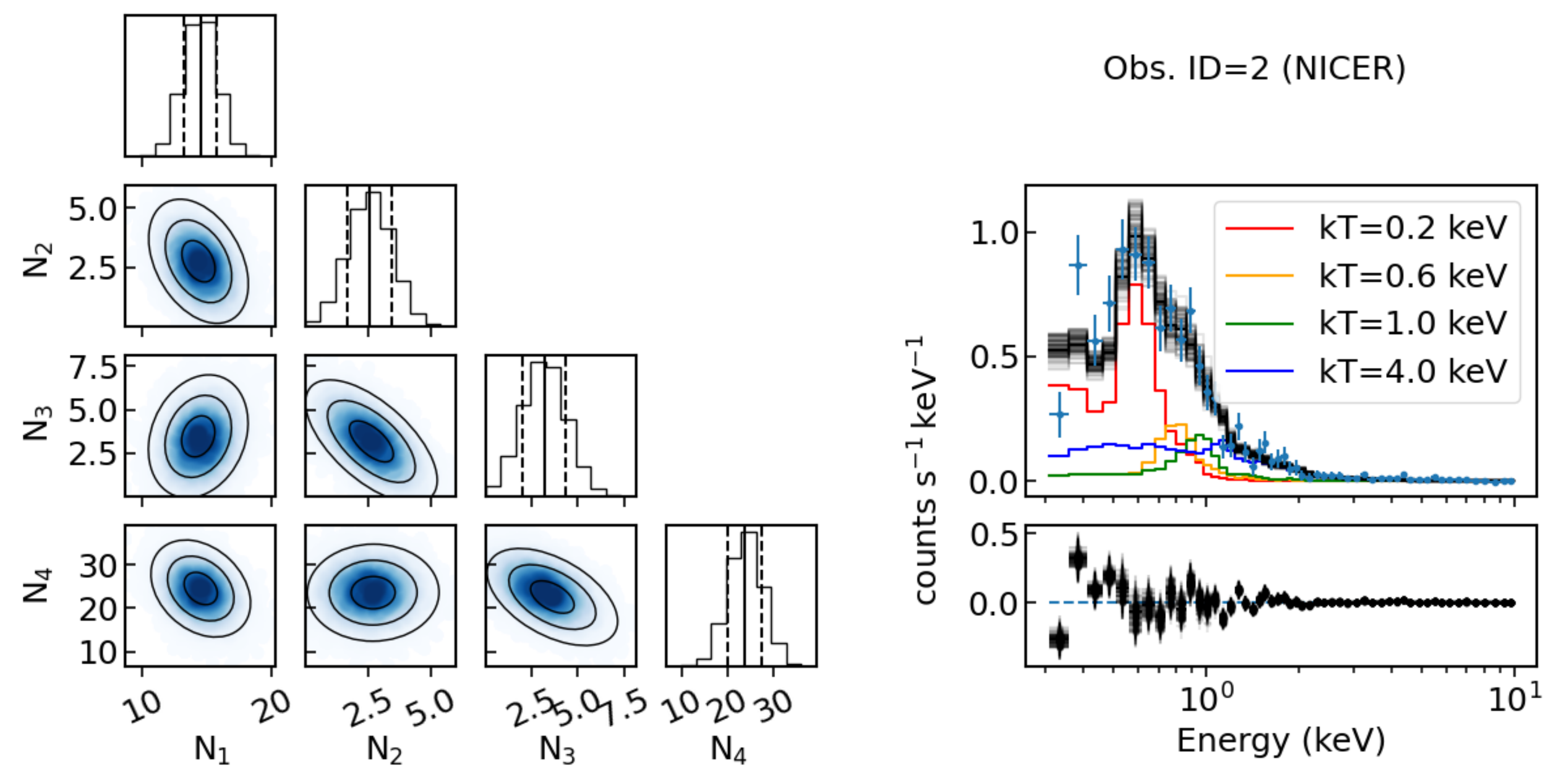}
    \includegraphics[width=0.48\textwidth]{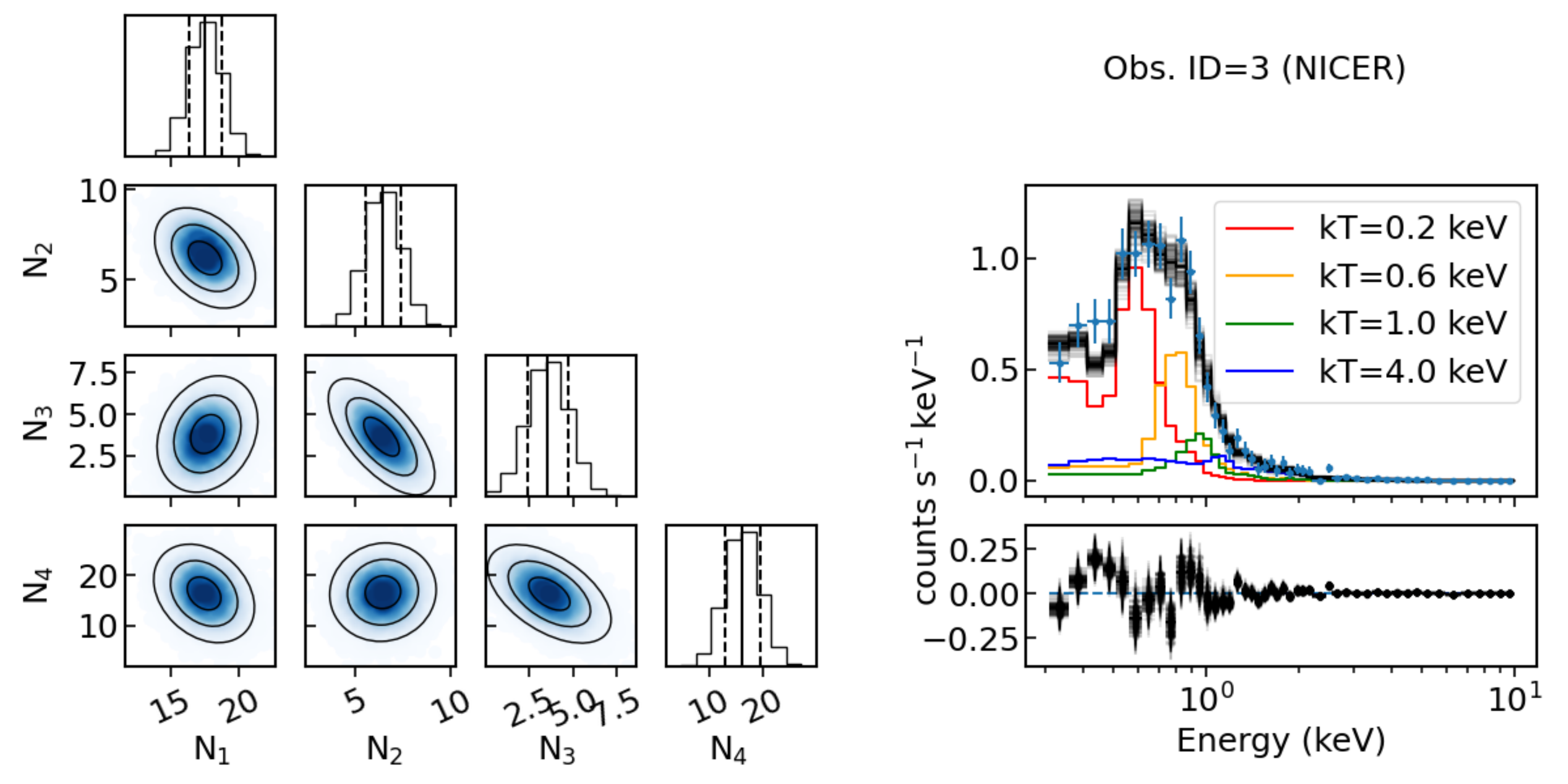}\hspace{0.5cm}
    \includegraphics[width=0.48\textwidth]{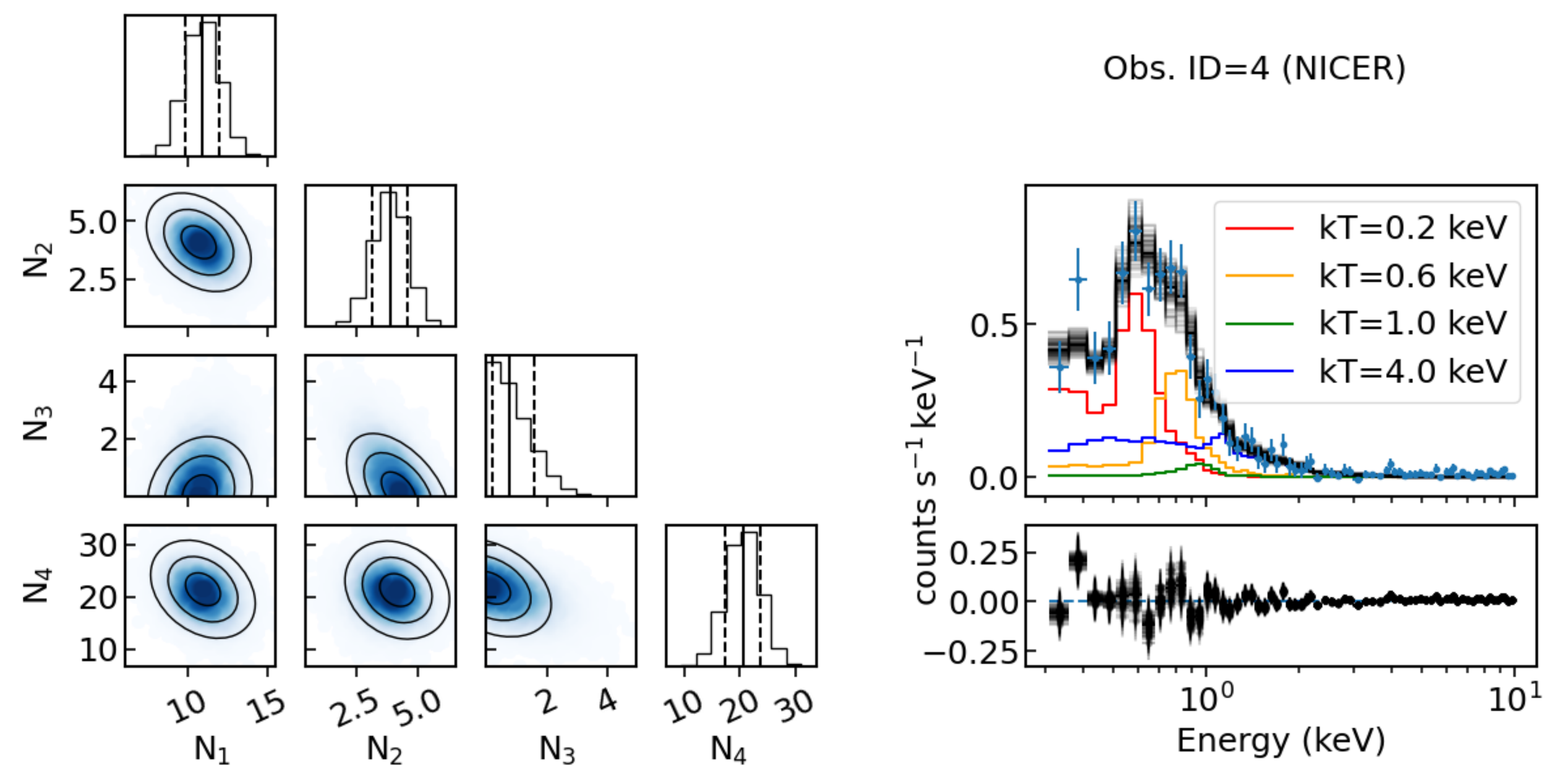}
    \includegraphics[width=0.48\textwidth]{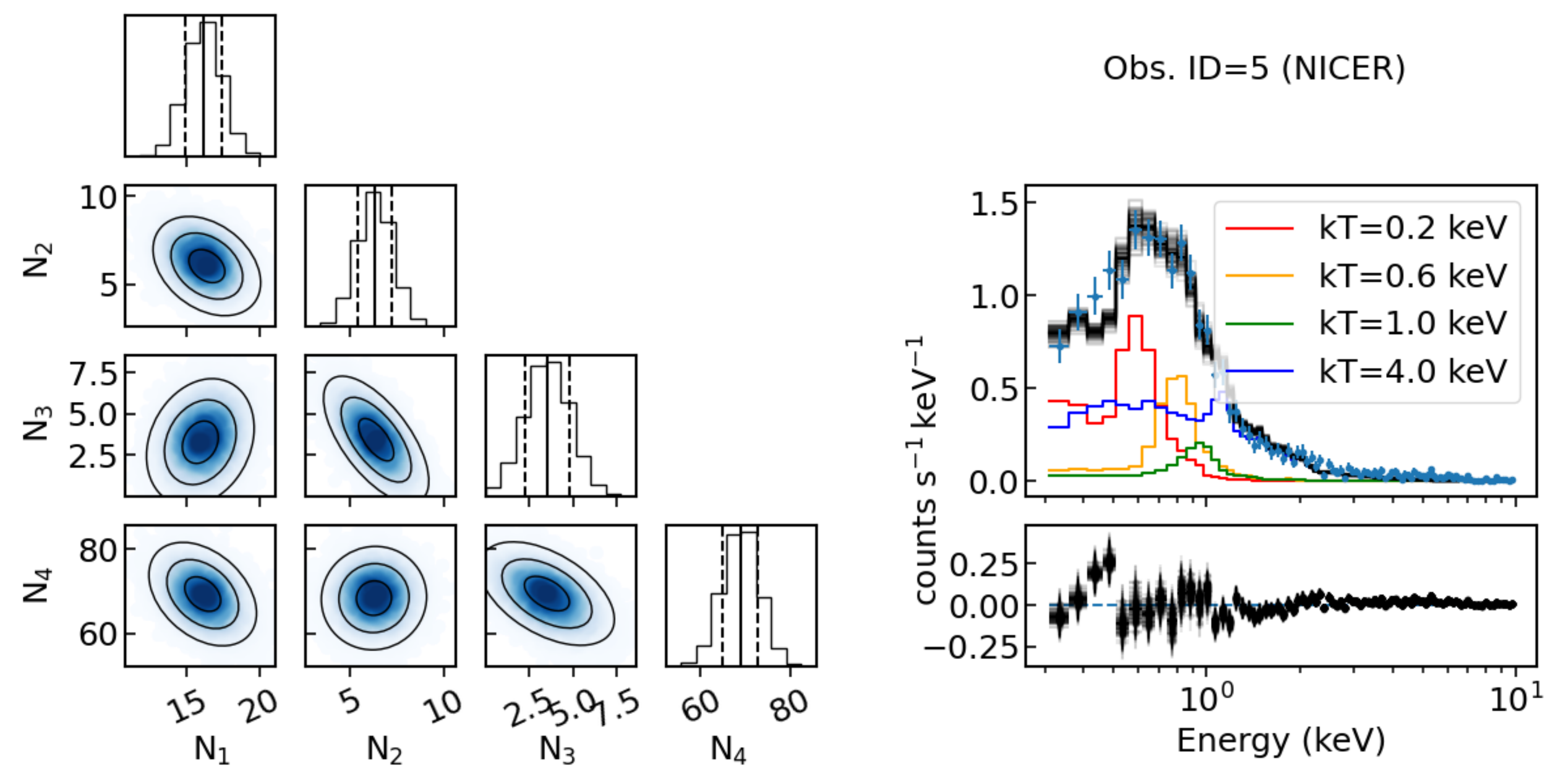}\hspace{0.5cm}
    \includegraphics[width=0.48\textwidth]{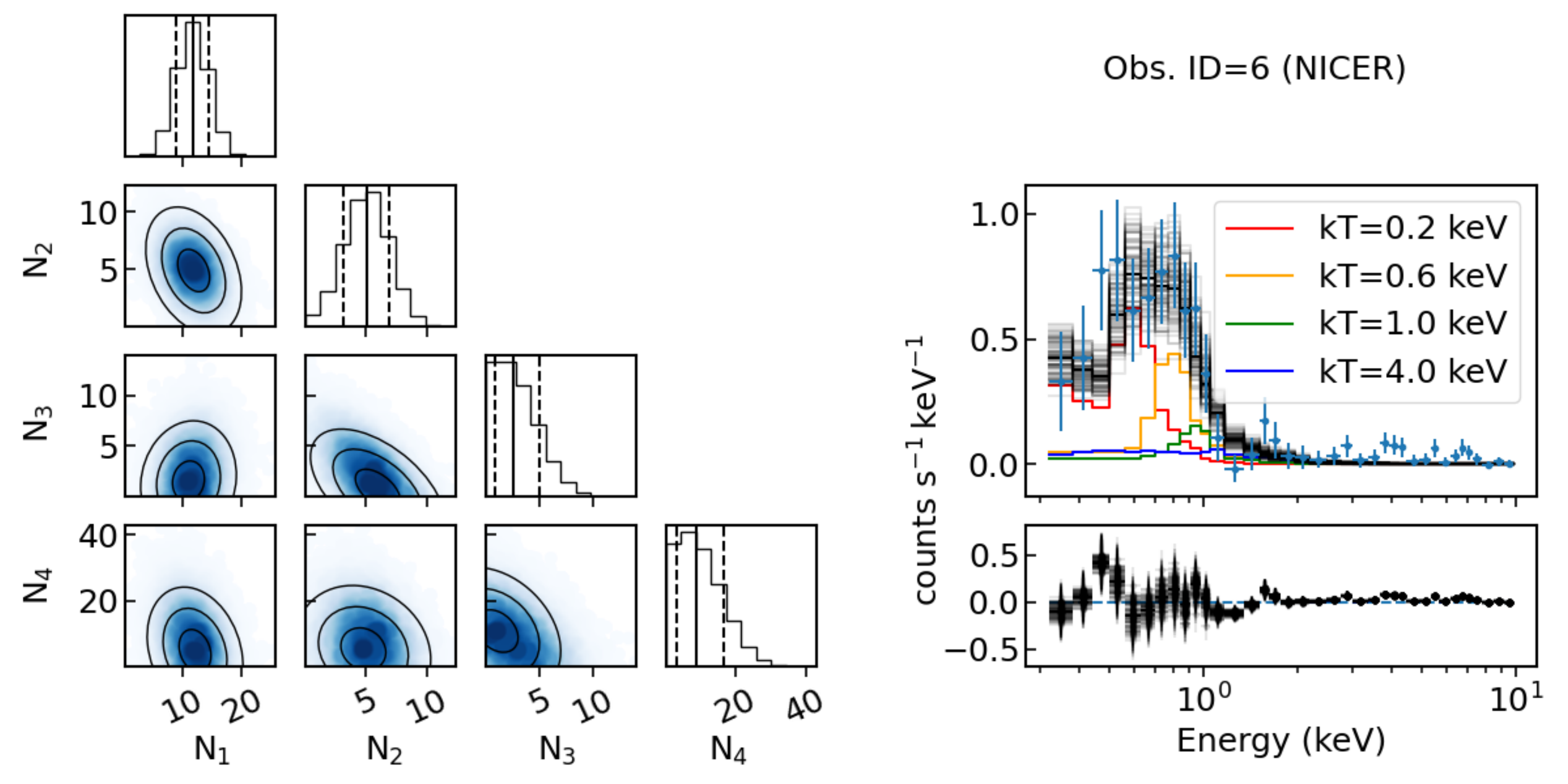}
    \includegraphics[width=0.48\textwidth]{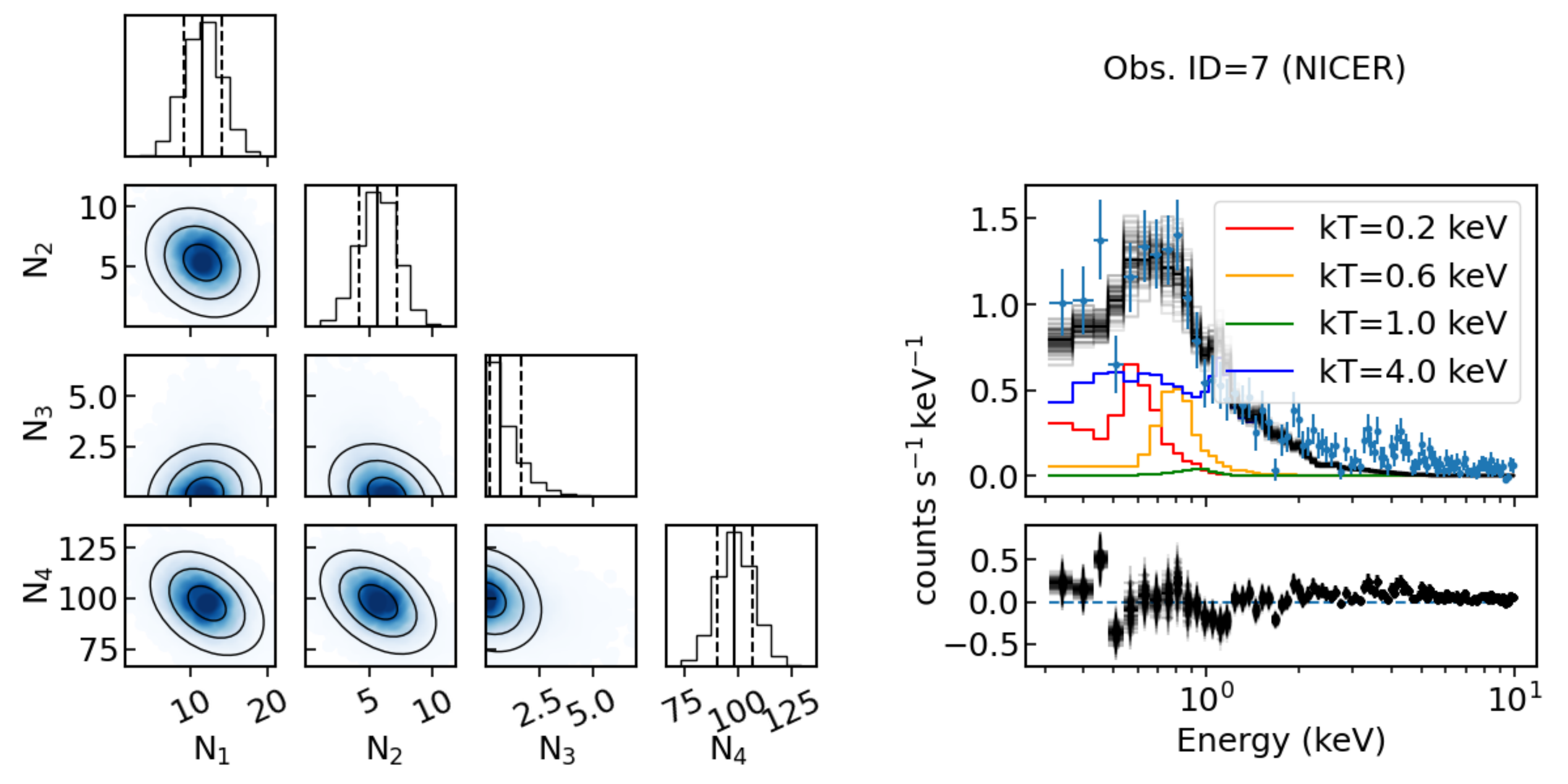}\hspace{0.5cm}
    \includegraphics[width=0.48\textwidth]{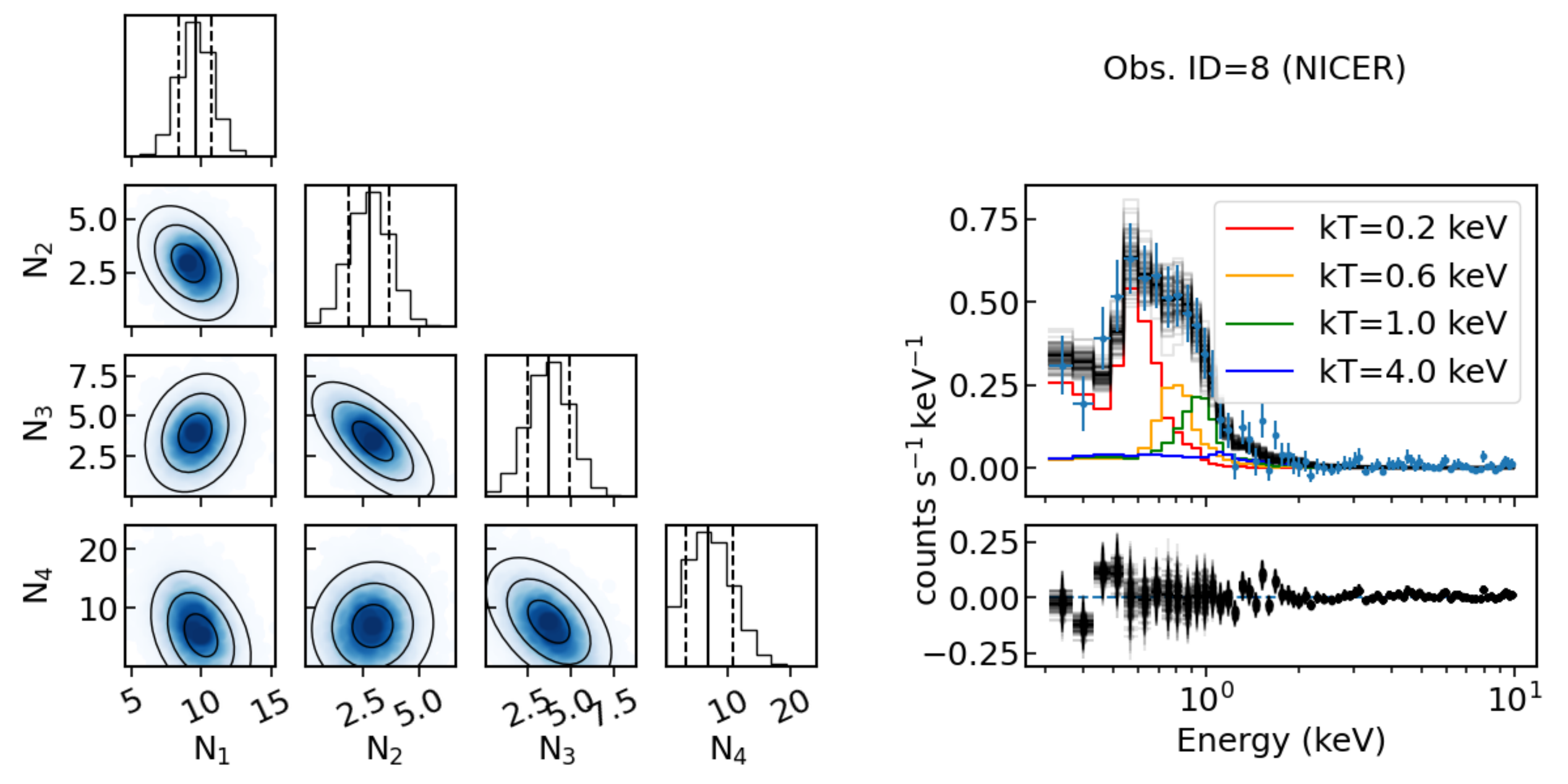}
    \includegraphics[width=0.48\textwidth]{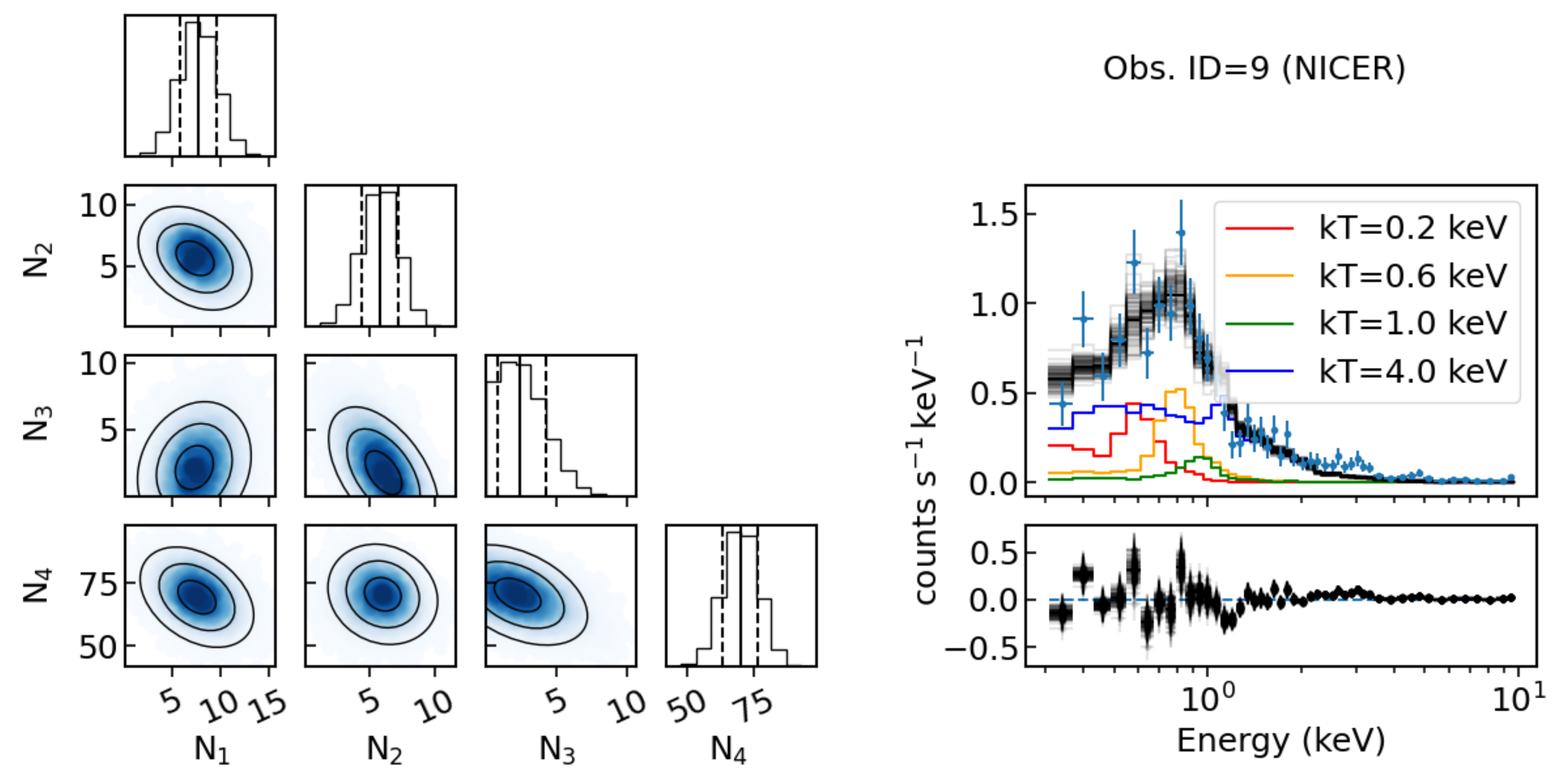}\hspace{0.5cm}
    \includegraphics[width=0.48\textwidth]{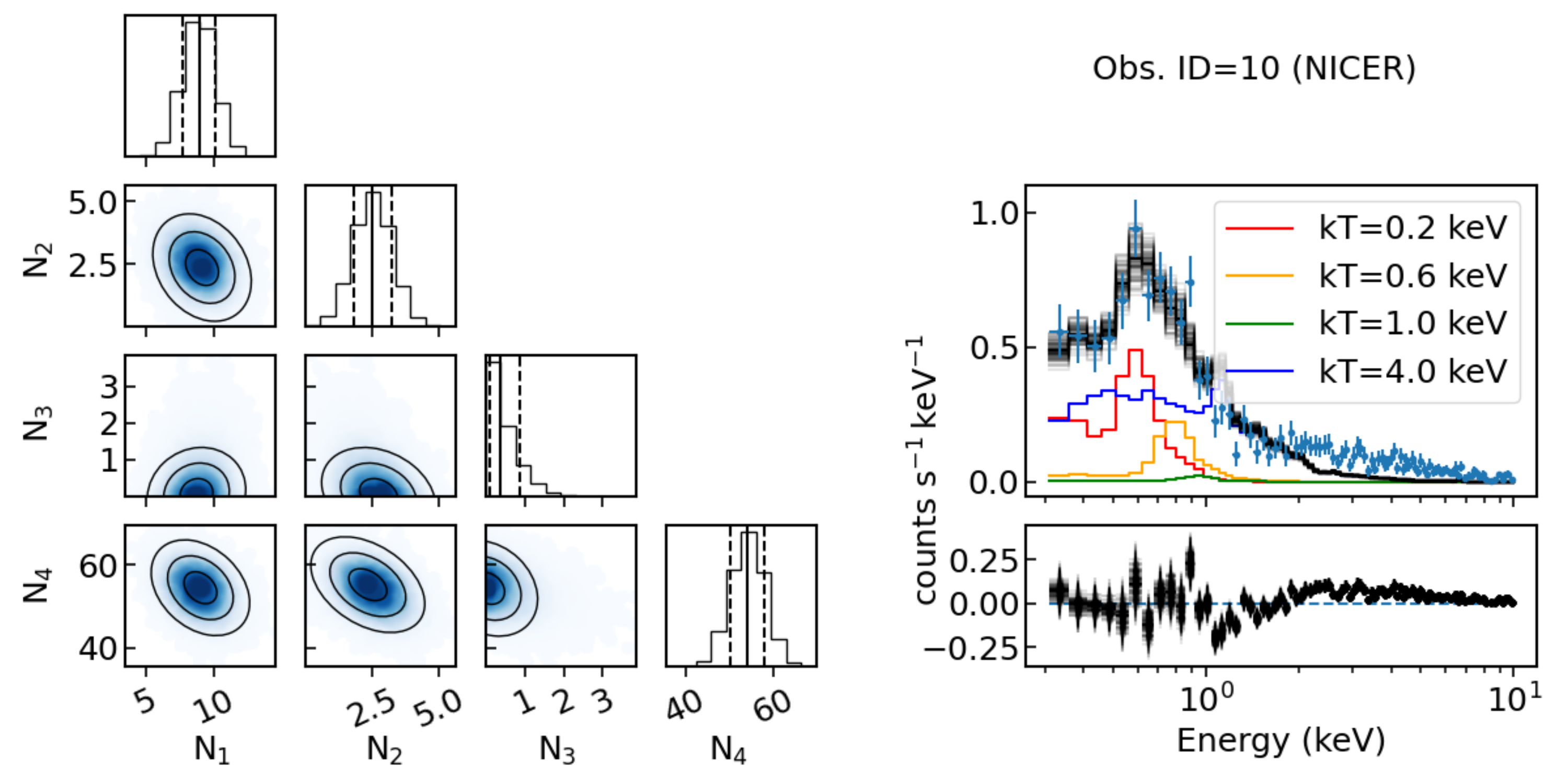}
    \caption{Spectral analysis of individual observations. The model fitted is the 4T model with $n_\mathrm{H}$ fixed at $0.03\times 10^{22}$\,cm$^{-2}$. The Obs. IDs correspond to the IDs listed in Table \ref{tab:obs}. The solid vertical lines on the histograms mark the median values, and the vertical dashed lines mark the 68\% confidence intervals. $\mathrm{N}_i$s ($i=1,2,3,4$) are proportional to the norms at 0.2, 0.6, 1.0 and 4.0 keV respectively, with $\mathrm{N}_i=\mathrm{Norm}_i\times 10^5$. See \S\ref{subsec:result_individual_obs} for details.}
    \label{fig:4T_fixed_nH_mcmc_spectra}
\end{figure*}

\begin{figure}
    \centering
    \includegraphics[width=0.48\textwidth]{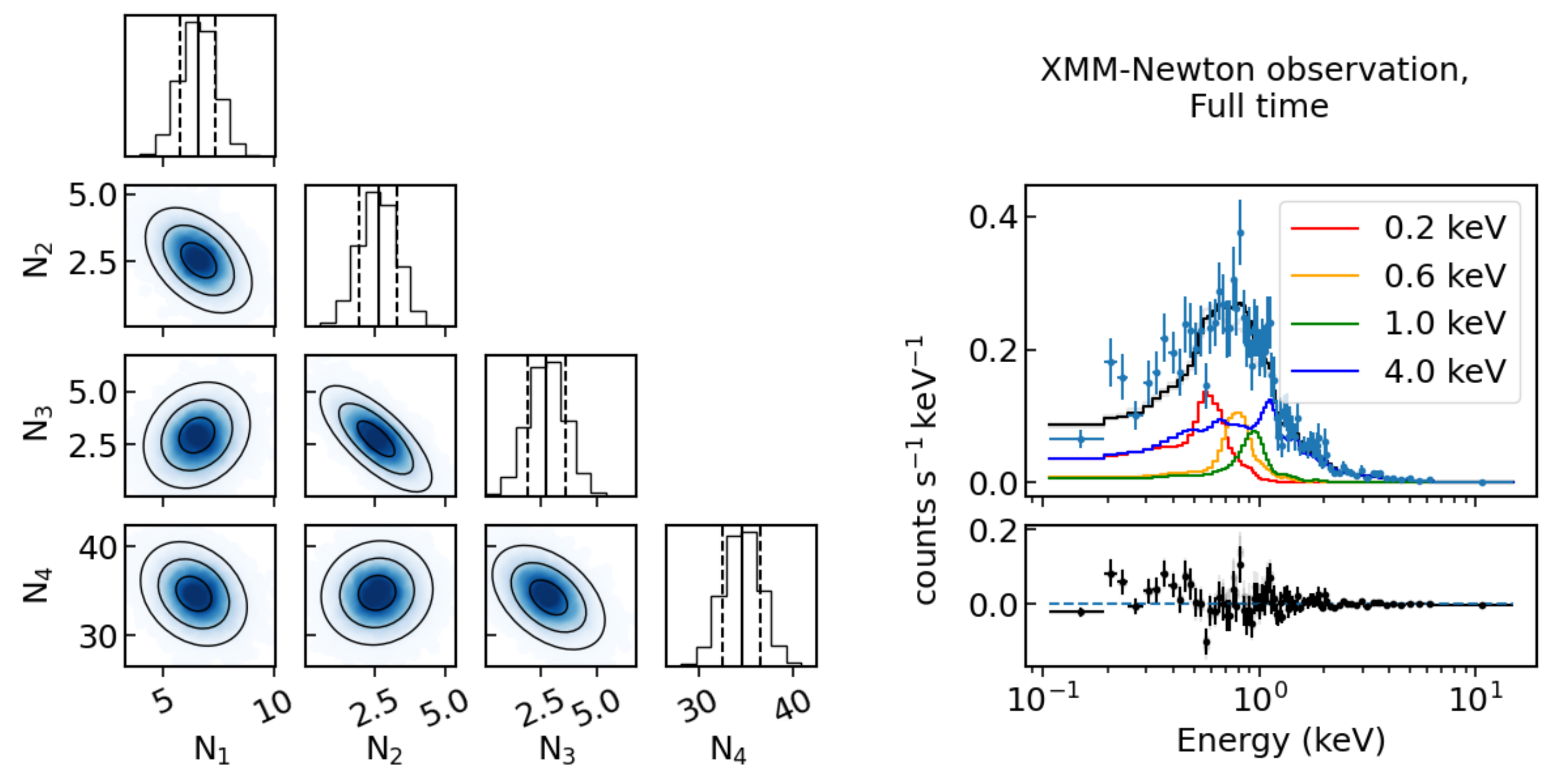}
    
    \vspace{0.5cm}

    \includegraphics[width=0.48\textwidth]{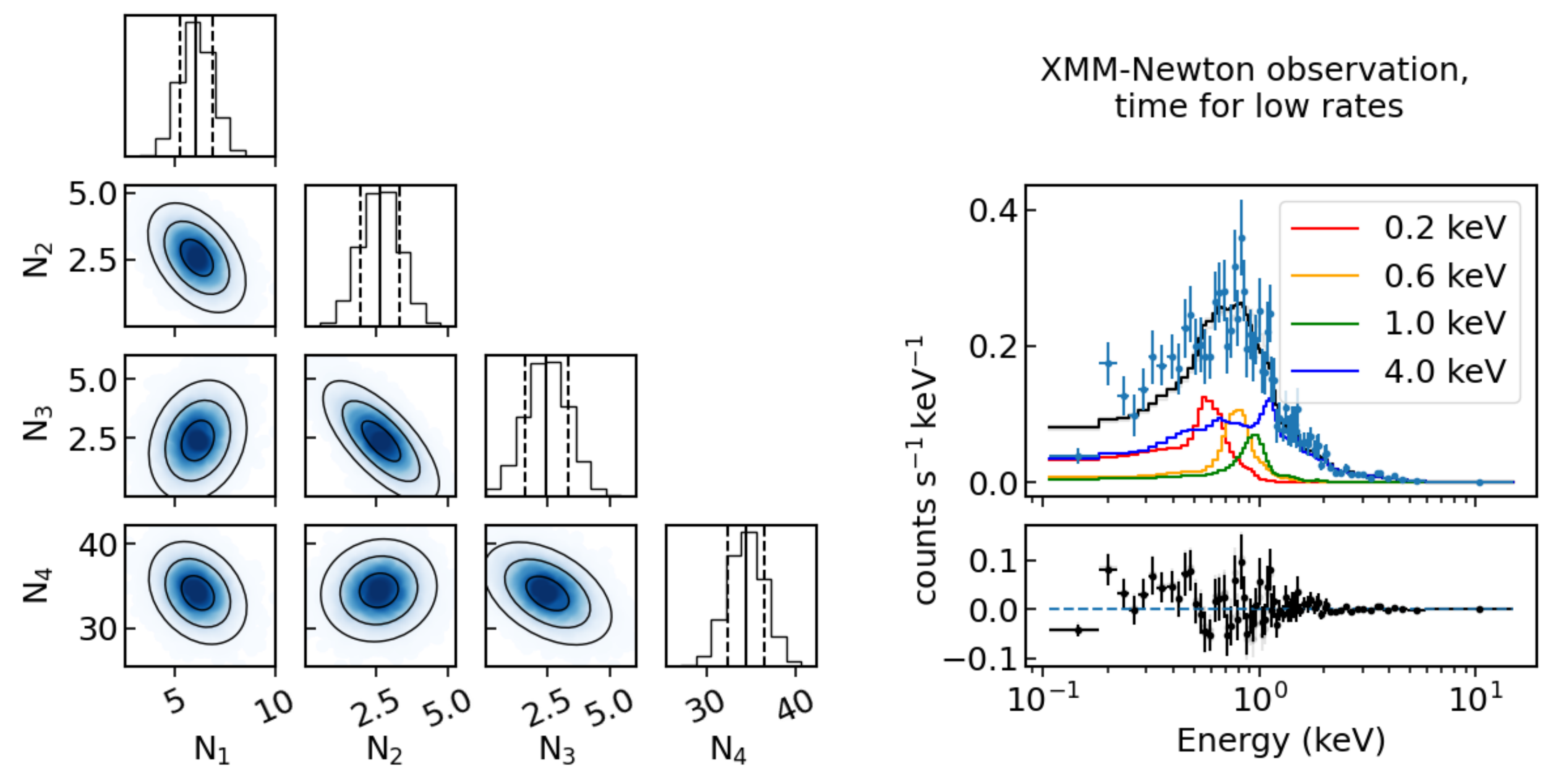}
    
    \vspace{0.5cm}
    \includegraphics[width=0.48\textwidth]{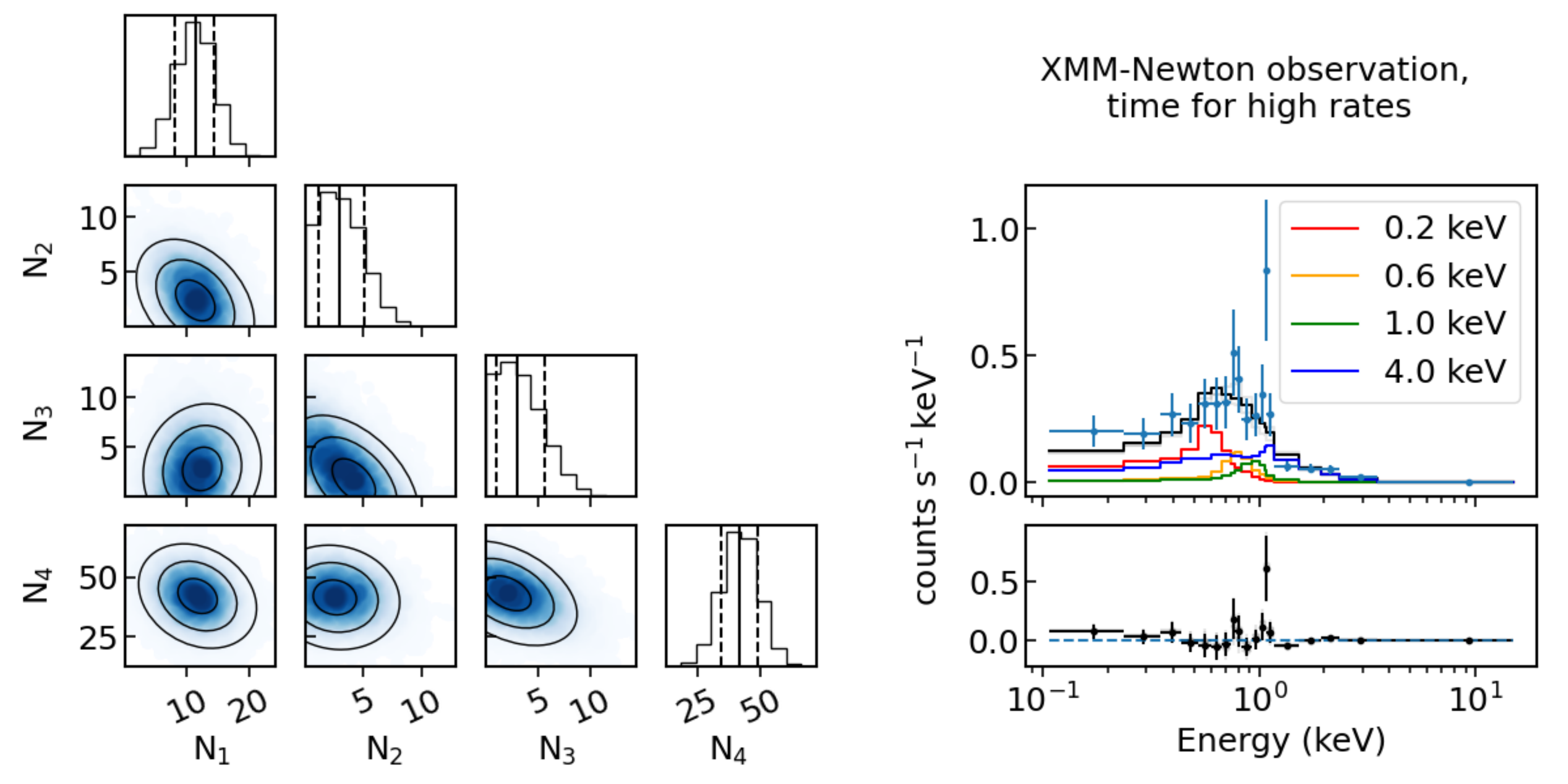}
    \caption{Results from spectral analysis of the XMM-Newton observation of \eps~first reported by \citet{naze2014}. $\mathrm{N_1}$, $\mathrm{N_2}$, $\mathrm{N_3}$ and $\mathrm{N_4}$ have the same meaning as that described in the caption of Figure \ref{fig:4T_fixed_nH_mcmc_spectra}. See \S\ref{subsec:past_obs} for details.}
    \label{fig:spec_xmm}
\end{figure}

\section{Detailed investigation of the observations 7 and 10}\label{sec:obs_7_10}
\begin{figure}
    \centering
    \includegraphics[width=0.3\textwidth]{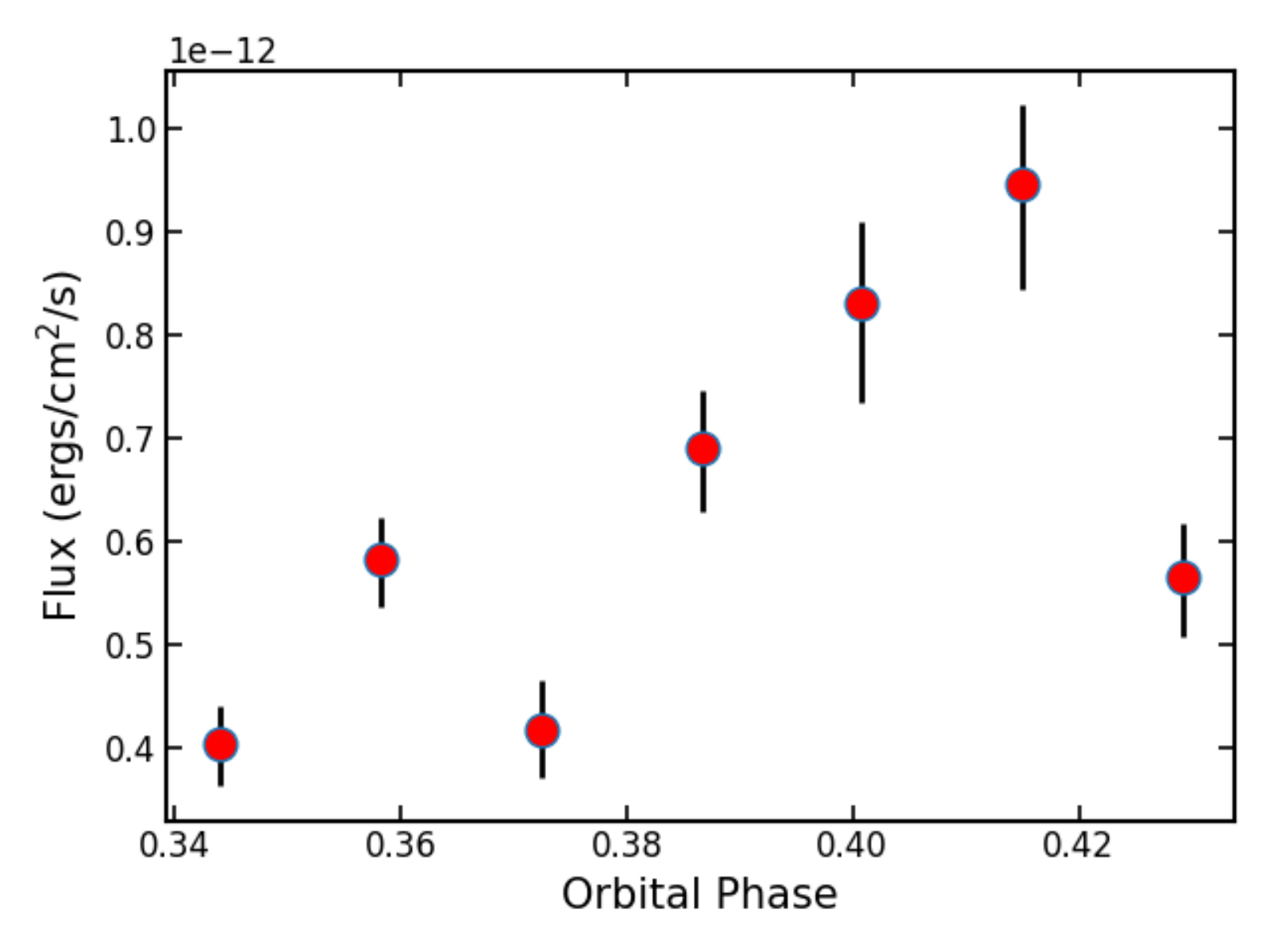}
    \includegraphics[width=0.3\textwidth]{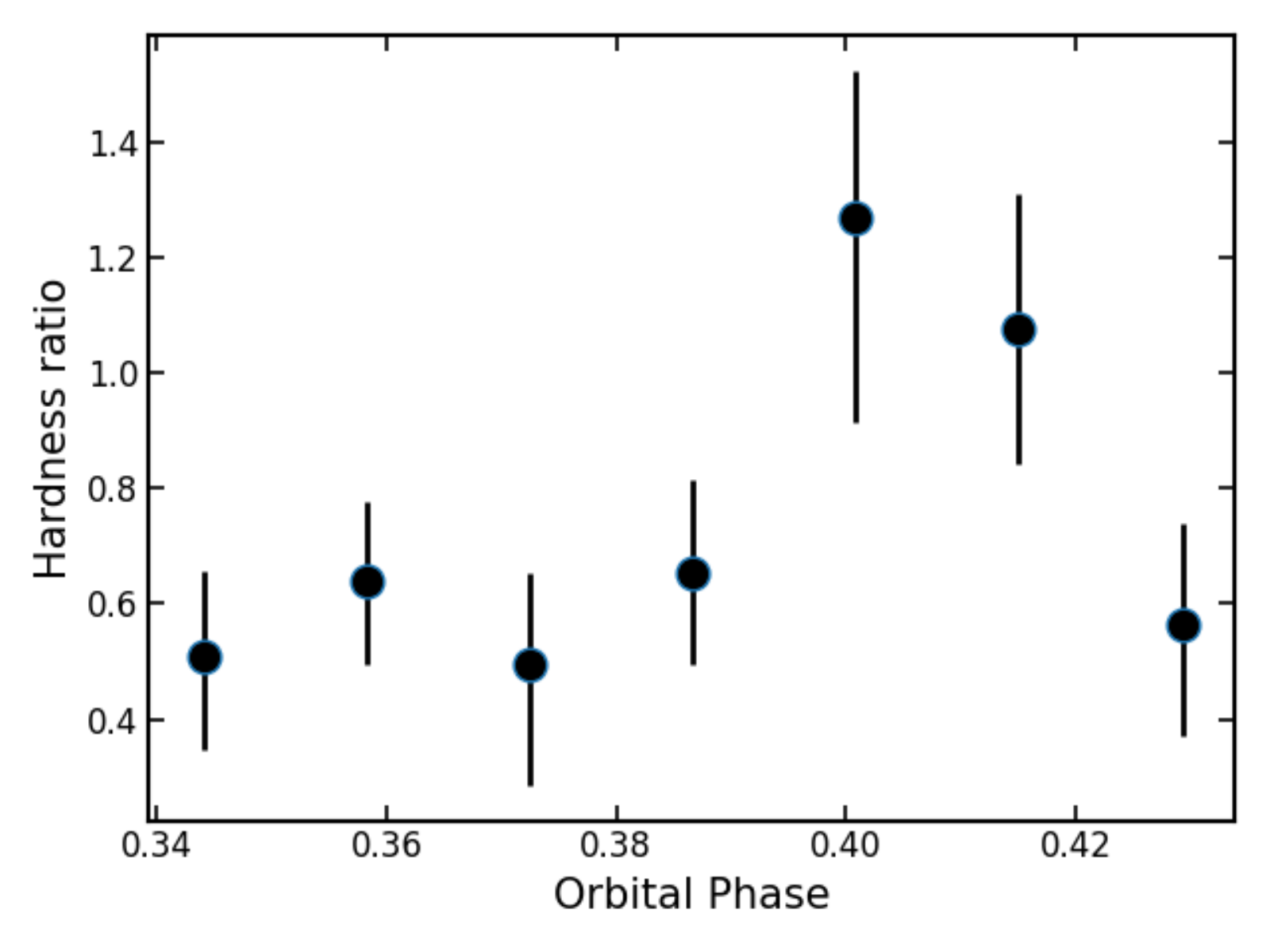}
    \caption{\textit{Top:} The variation of flux over 0.5--2.0 keV (using the 4T model with $n_\mathrm{H}$ fixed at its interstellar value) within observation 10 (\S\ref{sec:obs_7_10}). The markers represent the median values and the error bars correspond to the 68\% confidence intervals (MCMC analysis).
    \textit{Bottom:} The corresponding variation of the hardness ratio defined as the ratio of flux over 1--2 keV to that over 0.5--1.0 keV.}
    \label{fig:1001}
\end{figure}

\begin{table*}
{\tiny
\caption{Spectral analysis of the individual GTIs of observations 7 and 10. We fitted the spectra with the 4T model with $n_\mathrm{H}$ fixed at its interstellar value.\label{tab:gti_analysis_IDs_7_10}}
\begin{tabular}{c|cccc||c|cccccc}
\hline
GTI & \multicolumn{4}{c}{Norm $\left(\times 10^{-5}\,\mathrm{cm^{-5}}\right)$} & $\chi^2_\mathrm{red}$ & \multicolumn{6}{c}{Flux $\left(\times 10^{-13}\,\mathrm{erg\,cm^{-2}\,s^{-1}}\right)$}\\
 &  &  &  & & & \multicolumn{3}{c}{Observed} & \multicolumn{3}{c}{ISM corrected}\\
No. & 0.2 keV & 0.6 keV & 1.0 keV & 4.0 keV & & 0.5--1.0 keV & 1.0--2.0 keV & 2.0--10.0 keV & 0.5--1.0 keV & 1.0--2.0 keV & 2.0--10.0 keV\\
\hline\hline
\multicolumn{12}{c}{ID 7}\\
\hline
1 & 14.8 & 6.5 & 1.9 & 37.0 & 1.2(55) & 4.1 & 2.2 & 3.6 & 4.8 & 2.3 & 3.7\\
& (11.7--17.9)  & (4.4--8.5) & (0.6--4.1) & (26.3--47.3) & & (3.7--4.4) & (1.8--2.6) & (2.7--4.7) & (4.4--5.1) & (1.9--2.7) & (2.7--4.7)\\
2 & 7.0 & 3.6 & 0.7 & 171.6 & 4.4(58) & 5.6 & 7.6 & 17.0 & 6.6 & 7.9 & 17.0\\
 & (3.4--11.0) & (1.6--5.9) & (0.2--1.8) & (157.6--184.9) & & (5.2--6.0) & (7.1--8.1) & (15.5--18.3) & (6.1--7.0) & (7.3--8.4) & (15.6--18.4)\\
\hline 
\multicolumn{12}{c}{ID 10}\\
\hline 
1 &  9.9 & 2.6 & 1.2 & 23.1 & 1.4(21) & 2.3 & 1.3 & 2.3 & 2.7 & 1.4 & 2.3\\
& (7.5--12.3)  & (1.2--4.0) & (0.3--2.6) & (16.1--29.9) & & (2.0--2.5) & (1.0--1.5) & (1.6--3.0) & (2.4--3.0) & (1.1--1.6) & (1.6--3.0)\\
2 & 7.9 & 4.2 & 4.0 & 35.8 & 1.7(18) & 3.0 & 2.2 & 3.6 & 3.6 & 2.3 & 3.6\\
 & (5.3--10.5) & (2.2--6.0) & (1.5--6.7) & (27.4--44.0) & & (2.8--3.4) & (1.9--2.4) & (2.8--4.4) & (3.2--3.9) & (2.0--2.5) & (2.8--4.4)\\
3 & 9.9 & 2.7 & 1.2 & 23.0 & 2.1(19) & 2.4 & 1.3 & 2.4 & 2.8 & 1.4 & 2.4\\
& (6.9--12.9) & (1.2--4.4) & (0.4--2.8) & (14.5--31.5) & & (2.0--2.7) & (1.0--1.6) & (1.5--3.1) & (2.4--3.2) & (1.1--1.7) & (1.5--3.1)\\
4 & 17.9 & 0.7 & 0.7 & 57.3 & 4.8(23) & 3.5 & 2.6 & 5.6 & 4.2 & 2.8 & 5.7\\
& (14.2--21.6) & (0.2--1.6) & (0.2--1.7) & (48.0--66.6) & & (3.2--3.9) & (2.2--3.0) & (4.8--6.5) & (3.8--4.7) & (2.3--3.1) & (4.8--6.5)\\
5 & 1.1 & 1.7  & 2.1 & 96.3 & 4.5(21) & 3.2 & 4.5 & 9.6 & 3.7 & 4.7 & 9.7\\
& (0.3--2.8) & (0.5--3.7) & (0.6--4.8) & (81.3--111.0) & & (2.8--3.5) & (3.9--5.1) & (8.1--11.1) & (3.2--4.0) & (4.1--5.3) & (8.2--11.1) \\
6 & 5.4 & 3.5 & 1.4 & 101.4  & 4.5(14) & 3.9 & 4.7 & 10.0 & 4.6 & 4.9 & 10.0\\
& (2.0--9.6) & (1.3--6.2) & (0.4--3.4) & (86.6--115.7)  & & (3.6--4.4) & (4.2--5.3) & (8.5--11.4) & (4.2--5.1) & (4.3--5.5) & (8.5--11.4)\\
7 & 12.0 & 3.4 & 2.8 & 32.0 & 1.4(13) & 3.1 & 2.0 & 3.3 & 3.6 & 2.0 & 3.3\\
& (8.1--15.9) & (1.4--5.5) & (0.9--5.6) & (21.1--42.7) & & (2.7--3.5) & (1.6--2.4) & (2.4--4.4) & (3.2--4.2) & (1.6--2.4) & (2.4--4.4)\\
\hline
\end{tabular}
}
\end{table*}
Finally, we take a deeper look at the two observations (IDs 7 and 10) that we decided to exclude while constructing the average periastron and average out-of-periastron observations in \S\ref{subsec:result_merged_events}. From Table \ref{tab:4T_result}, observations 10 and 7 are, respectively, the exposures with the highest and the second highest $\chi^2_\mathrm{red}$ values.
We first consider observation 10 for which the $\chi^2_\mathrm{red}$ for the 4T model is the highest (Table \ref{tab:4T_result}). This is the only observation for which the spectrum fitted with the 4T model has a $\chi^2_\mathrm{red}$ greater than 2 even for the energy range below 2 keV. To understand the possible reason, we performed spectral analysis (4T model, $n_\mathrm{H}$ fixed at interstellar value) for each of the seven GTIs of the observation. To do this exercise, we calculated the background spectra for the individual GTIs using \texttt{nibackgen3C50}. The spectra (over 0.3--10.0 keV) were then fitted with the 4T model (again with fixed $n_\mathrm{H}$). The results of the spectral analysis are given in Table \ref{tab:gti_analysis_IDs_7_10}, and the variation of the flux over 0.5--2.0 keV with orbital phases are shown in the top panel of Figure \ref{fig:1001}. Note that although the median $\chi^2_\mathrm{red}$ is larger than 2 for GTIs 3, 4, 5 and 6 for the energy range 0.3--10.0 keV, it is smaller than 2 for all but GTI 4 for the energy range 0.5--2.0 keV.

Figure \ref{fig:1001} appears to suggest that there is an enhancement within the orbital phase range spanned by this observation. This enhancement is also reflected in the variation of the hardness ratio (bottom panel of Figure \ref{fig:1001}), which is consistent with the periastron enhancement as well as the enhancements found in the archival XMM-Newton observation. However, the orbital phase of the enhancement is partially covered by observations 4 and 6 (Table \ref{tab:obs}, also see Figure \ref{fig:lightcurves_4T}) and we do not find any hint of enhancement in these two observations. This shows that the enhancement within observation 10 is not a persistent characteristic of the system. There are two possibilities: the enhancement represents an X-ray flare from the system (or, from an invisible companion), or it appeared due to an additional background contribution not accounted for by the 3C50 background model. With the present data, we cannot rule out any of these possibilities. Further monitoring of the system will be needed to understand the origin of the observed flux enhancement at a phase much away from the periastron.

\begin{figure}
    \centering
    \includegraphics[width=0.3\textwidth]{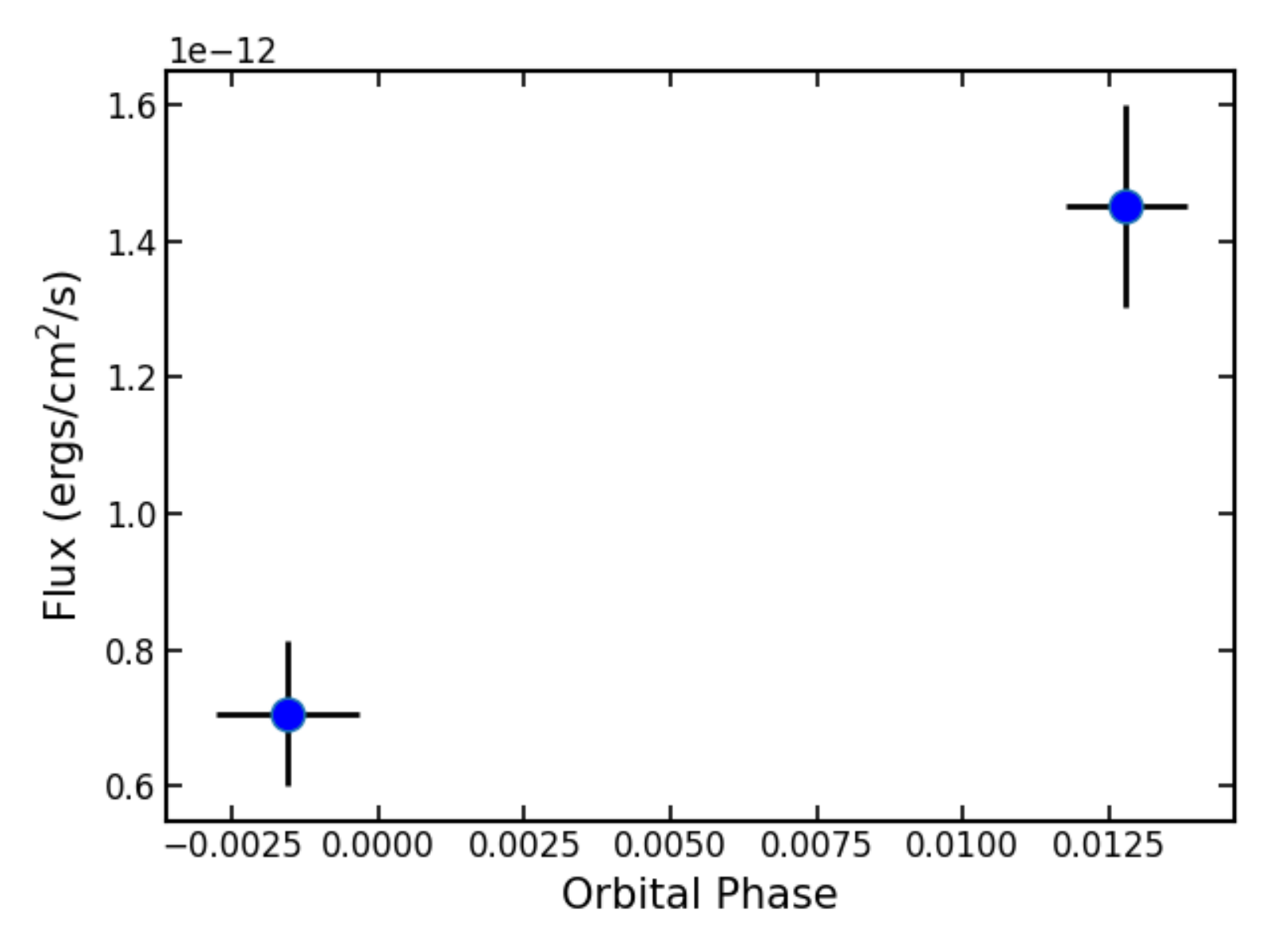}
    \includegraphics[width=0.3\textwidth]{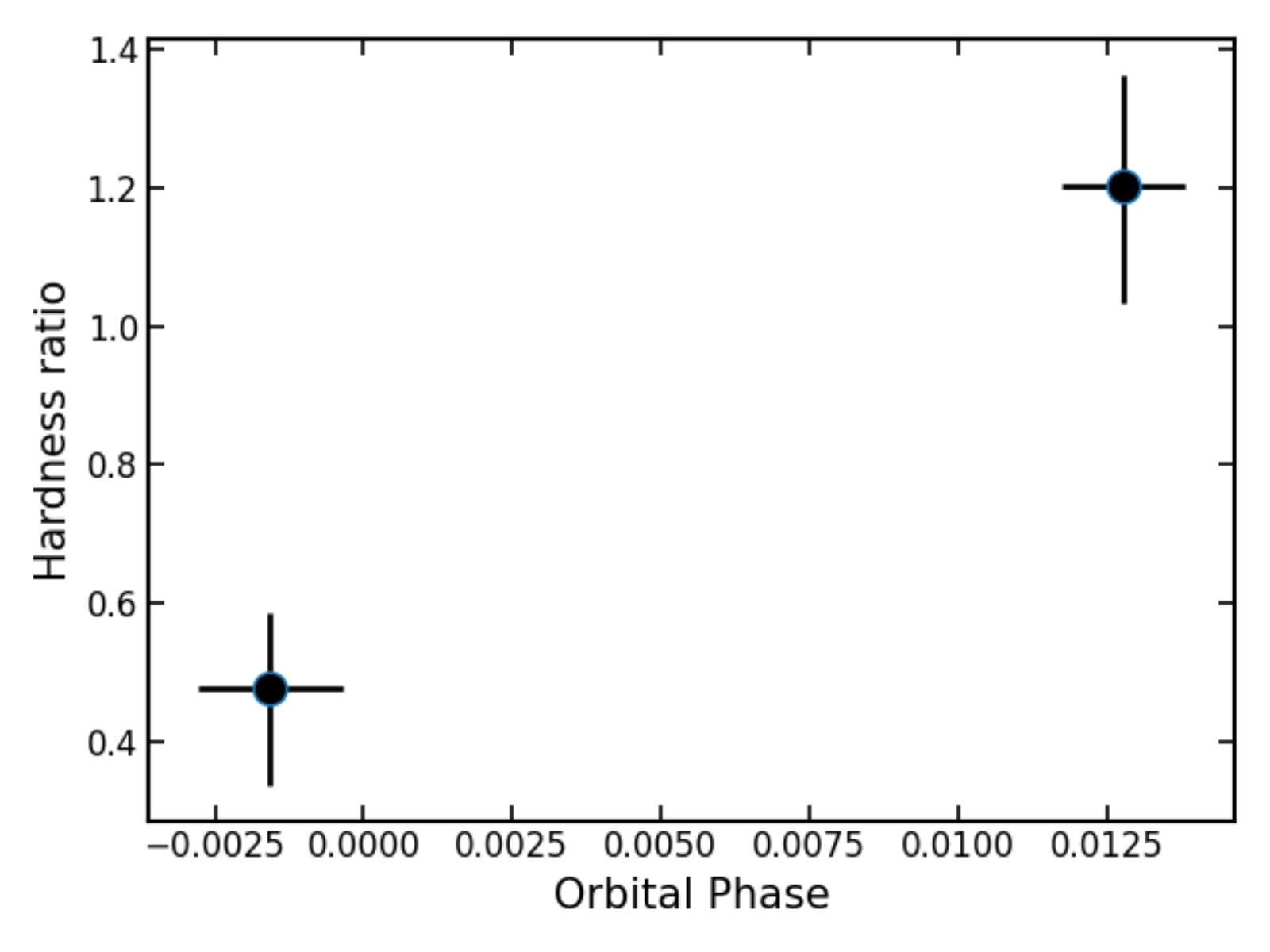}
    \caption{Same as Figure \ref{fig:1001}, but for the GTIs within observation 7.
    }
    \label{fig:0701}
\end{figure}

We performed a similar analysis for observation 7 (Table \ref{tab:gti_analysis_IDs_7_10} and Figure \ref{fig:0701}), and arrived at the same conclusion that the observed variation over small orbital phase ranges could be due to the use of an imperfect background model. It is, however, to be noted that the difference between the fluxes (0.5--2.0 keV) at the two GTIs for spectrum 7 is significantly higher than that observed among the GTIs of spectrum 10. In addition, the observed trend in the variation of the X-ray flux is consistent with that shown by other periastron observations (Figure \ref{fig:lightcurves_4T}). Thus, the observed difference between the two GTIs of spectrum 7 is physically more plausible than that for spectrum 10.

\section{Tests to check the robustness of our primary result}\label{sec:test}
\begin{figure}
    \centering
    \includegraphics[width=0.3\textwidth]{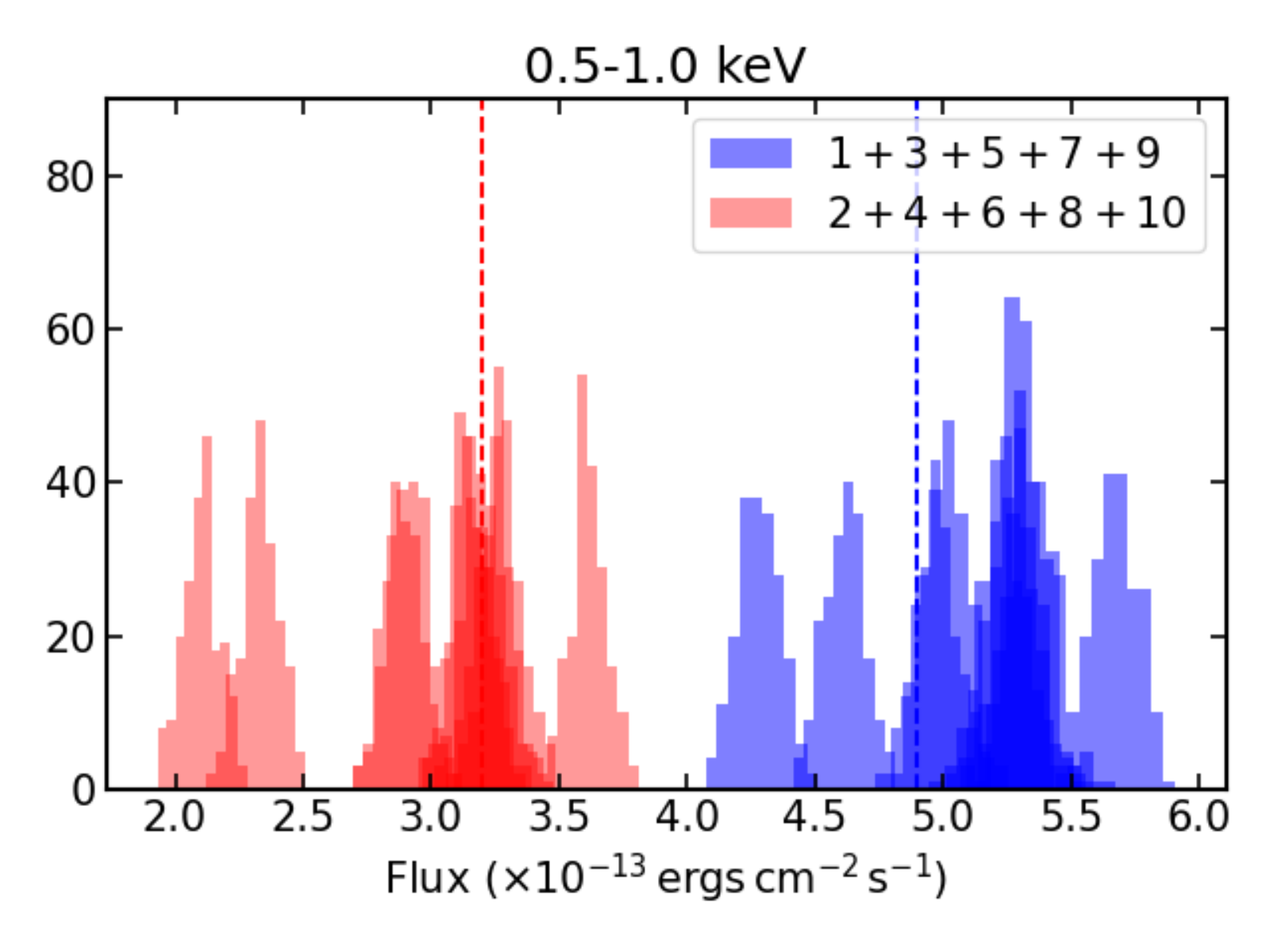}
    \includegraphics[width=0.3\textwidth]{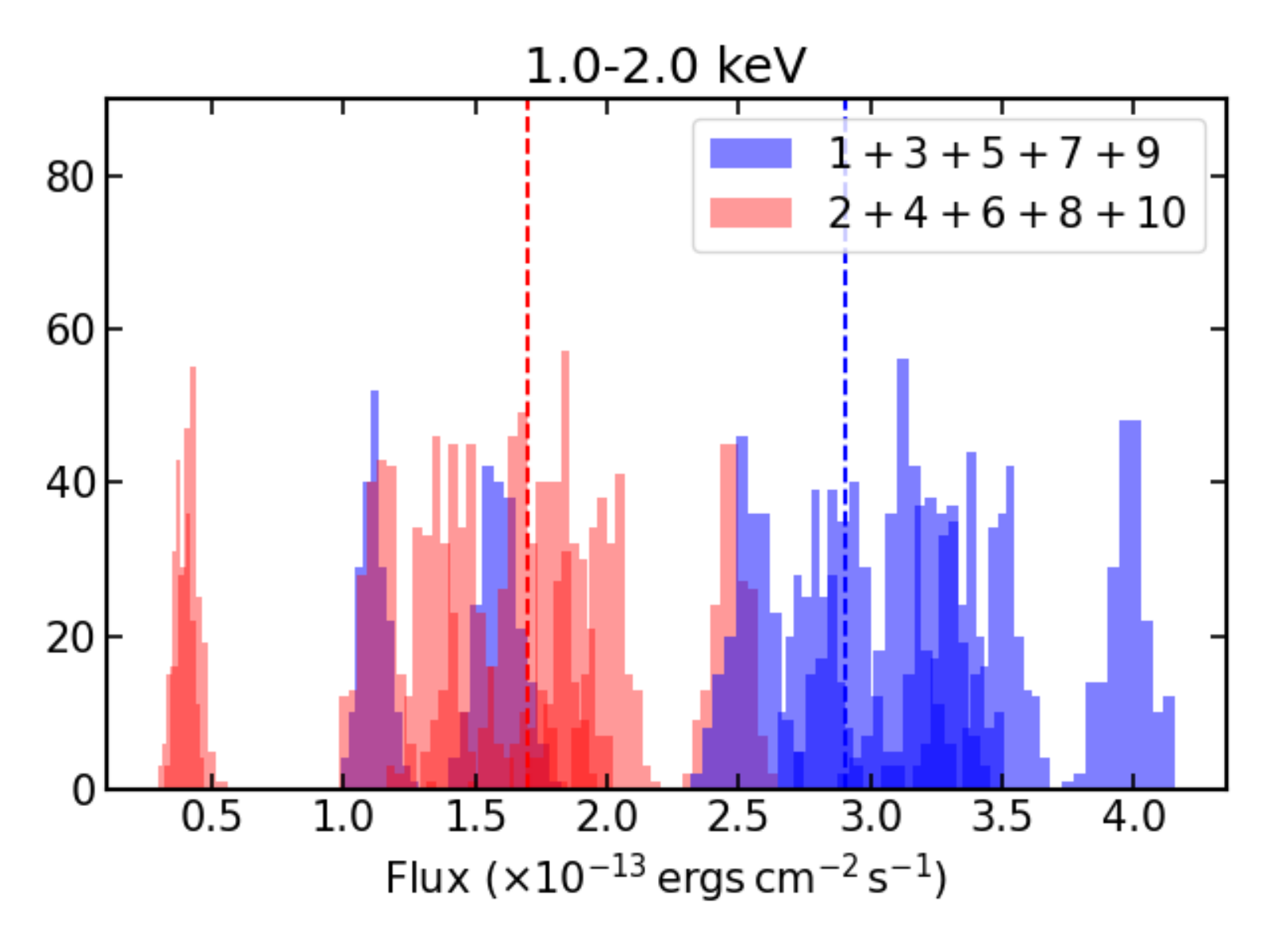}
    \includegraphics[width=0.3\textwidth]{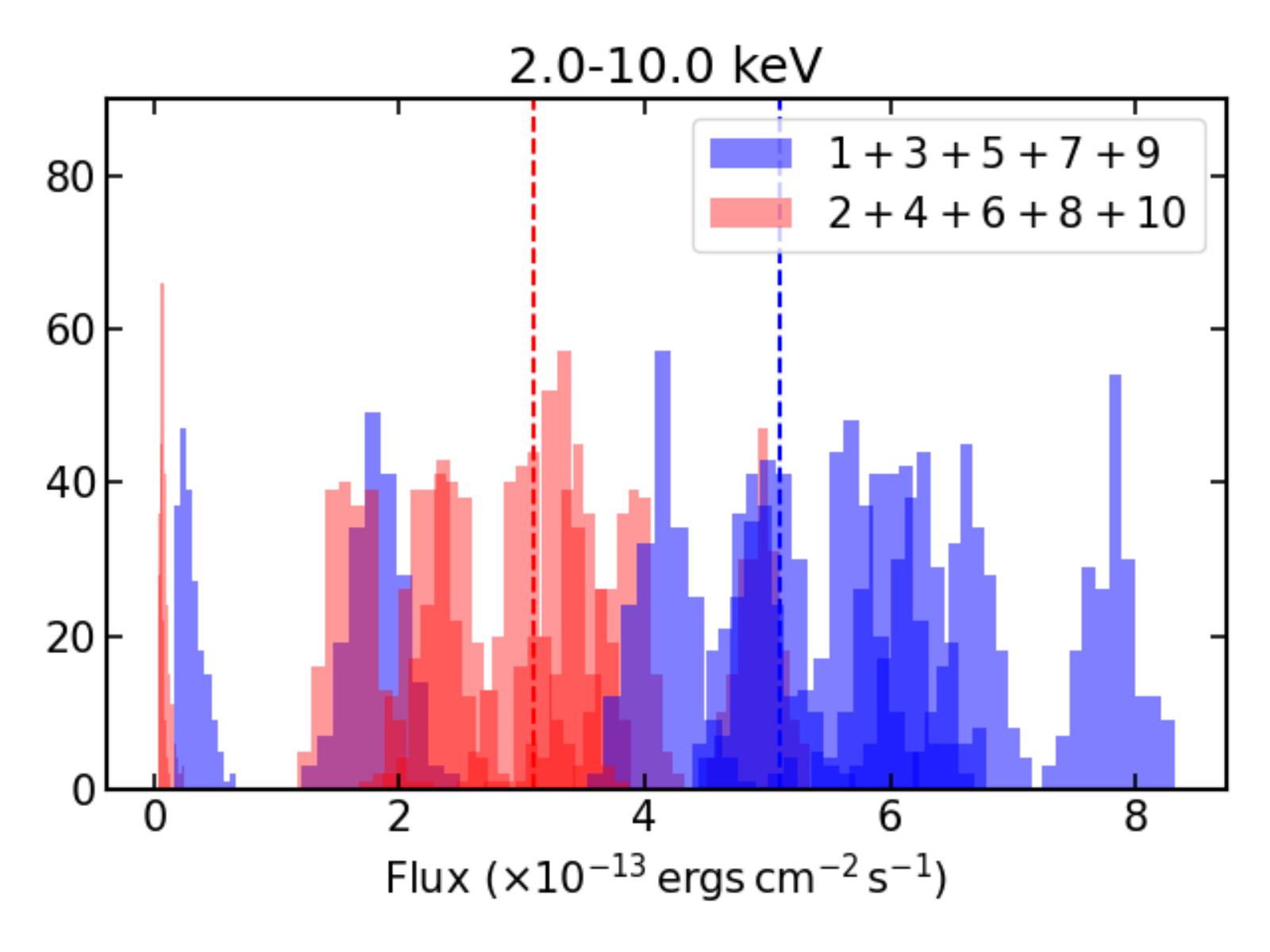}
    \caption{Effect of using different background spectra on the estimated flux for the merged periastron (red) and merged out-of-periastron (blue) observations. The legends show the observation IDs used to obtain the merged spectra. The different bars correspond to different background spectra. The vertical dashed lines mark the flux obtained by using the `true' background spectra. Note that here `true' background for a given observation implies that the background spectrum was calculated using that particular X-ray observation (3C50 model). The three panels are for the three energy bins: 0.5--1.0 keV (top), 1.0--2.0 keV (middle) and 2.0--10.0 keV (bottom). See \S\ref{sec:test} for details.}
    \label{fig:bkg_check}
\end{figure}

The key result of this work is that the system \eps~produces more X-ray emission at periastron than that away from periastron. However, the NICER data, based on which this conclusion is derived, suffer from an important limitation, which is that the background estimations are based on modelling, rather than using a subtraction of the background directly observed around the X-ray source for an imaging instrument. This is especially important for a relatively faint source like \eps~(average count rates are $\sim 1\,\mathrm{count\,s^{-1}}$). In order to understand whether the observed difference in X-ray flux is robust against this limitation, we performed a spectral analysis (4T model with fixed $n_\mathrm{H}$) for the merged periastron (IDs: $1+3+5+7+9$) and merged out-of-periastron ($2+4+6+8+10$) observations considering each of the ten background spectra obtained for our ten observations using the 3C50 model. In Figure \ref{fig:bkg_check}, we plot the posterior distribution of the flux values (for all ten background spectra) obtained from the MCMC analysis for the three energy bins: 0.5--1.0 keV (top), 1.0--2.0 keV (middle) and 2.0--10.0 keV (bottom). The red and blue histograms respectively represent the results for periastron and out-of-periastron. As expected, the flux values are found to vary significantly with the use of different background models. Despite that, we find that over 0.5--1.0 keV, the periastron flux is always higher than that away from periastron. Over 1.0--2.0 keV, the periastron flux is higher than that away from periastron except for a few combinations of background spectra. The difference between the two type of observations is most affected by the variation of background for the highest energy bin (2--10 keV), which we have not used for further analysis. Note that, for a given background spectrum, the (merged) periastron flux is always much higher than that away from periastron over 0.5--2.0 keV. Finally, based on the three panels in Figure \ref{fig:bkg_check}, we do not have any evidence for the opposite case.

Thus we conclude that our key result of observing higher X-ray flux at periastron is robust against any limitation on the part of our chosen background model. 

\bsp	

\label{lastpage}
\end{document}